\documentclass{aa}
\usepackage{rotating}
\usepackage{graphicx}
\usepackage{txfonts}
\usepackage{natbib}
\usepackage{epsfig}

\begin{document}

\title{X-ray and optical spectroscopy of the massive young open cluster IC~1805\thanks{Based on observations collected with {\it XMM-Newton}, an ESA science mission with instruments and contributions directly funded by ESA member states and the USA (NASA), and with the TIGRE telescope (La Luz, Mexico).}\fnmsep \thanks{Table\,\ref{Xcat} is only available in electronic form at the CDS via anonymous ftp to cdsarc.u-strasbg.fr (130.79.128.5) or via http://cdsarc.u-strasbg.fr/viz-bin/qcat?J/A+A/}}
\author{G.\ Rauw \and Y.\ Naz\'e\thanks{Research Associate FRS-FNRS (Belgium)}}
\offprints{G.\ Rauw}
\mail{rauw@astro.ulg.ac.be}
\institute{Space sciences, Technologies and Astrophysics Research (STAR) Institute, Universit\'e de Li\`ege, All\'ee du 6 Ao\^ut, 19c, B\^at B5c, 4000 Li\`ege, Belgium}
\date{}
\abstract{Very young open clusters are ideal places to study the X-ray properties of a homogeneous population of early-type stars. In this respect, the IC~1805 open cluster is very interesting as it hosts the O4\,If$^+$ star HD~15570 thought to be in an evolutionary stage intermediate between a normal O-star and a Wolf-Rayet star.}{Such a star could provide a test for theoretical models aiming at explaining the empirical scaling relation between the X-ray and bolometric luminosities of O-type stars.}{We have observed IC~1805 with XMM-Newton and further collected optical spectroscopy of some of the O-star members of the cluster.}{The optical spectra allow us to revisit the orbital solutions of BD+60$^{\circ}$~497 and HD~15558, and provide the first evidence of binarity for BD+60$^{\circ}$~498. X-ray emission from colliding winds does not appear to play an important role among the O-stars of IC~1805. Notably, the X-ray fluxes do not vary significantly between archival X-ray observations and our XMM-Newton pointing. The very fast rotator BD+60$^{\circ}$~513, and to a lesser extent the O4\,If$^+$ star HD~15570 appear somewhat underluminous. Whilst the underluminosity of HD~15570 is only marginally significant, its amplitude is found to be compatible with theoretical expectations based on its stellar and wind properties. A number of other X-ray sources are detected in the field, and the brightest objects, many of which are likely low-mass pre-main sequence stars, are analyzed in detail.} {}
\keywords{Stars: early-type -- open clusters and associations: individual (IC~1805) -- Stars: individual (HD~15558, HD~15570, HD~15629, BD+60$^{\circ}$~497, BD+60$^{\circ}$~498, BD+60$^{\circ}$~499, BD+60$^{\circ}$~501, BD+60$^{\circ}$~513) -- X-rays: stars}
\authorrunning{Rauw \& Naz\'e}
\titlerunning{The X-ray view of IC~1805}
\maketitle

\section{Introduction}
X-ray emission of massive O-type stars was discovered in the late seventies with the {\it EINSTEIN} satellite \citep{Harnden}. X-rays in single O-stars are generally thought to arise from a distribution of hydrodynamic shocks produced by the so-called line deshadowing instability \citep[LDI, e.g.][]{Feldmeier} that affects the radiatively driven winds of these objects. Additional X-ray emission can arise in magnetically confined stellar winds of O-stars with a strong enough magnetic field \citep[e.g.][and references therein]{BabelMontmerle,ud-Doula,YNBfield,ud-DoulaYN}, or in interacting wind binary systems \citep[e.g.][and references therein]{SBP,PP,GRYN}. 

Soon after the discovery of the X-ray emission of O-type stars with the {\it EINSTEIN} satellite, it was realised that their X-ray luminosity $L_{\rm X}$ scales with their bolometric luminosity $L_{\rm bol}$ \citep[e.g.][]{Sciortino}. This scaling relation was confirmed and refined with large samples of O-type stars observed with {\it ROSAT} \citep{Berghoefer}, and most recently {\it XMM-Newton} \citep{NGC6231,YN} and {\it Chandra} \citep{Carina,CygOB2}. A very different situation holds for presumably single Wolf-Rayet stars, where no clear dependence between X-ray and bolometric luminosities exists \citep{Wessolowski,Ignace,OskinovaWR}. Some classes of Wolf-Rayet stars remain below the detection threshold of the current generation of X-ray observatories \citep[e.g.][]{Oskinova,Gosset2}. 

\citet{Owocki} attempted to theoretically explain the origin of the empirical $L_{\rm X}/L_{\rm bol}$ scaling relation of O-stars. They predicted that X-ray emission produced by LDI shocks should scale as $L_{\rm X} \propto  L_{\rm bol}^{1.7}$ or $L_{\rm X} \propto L_{\rm bol}^{3.4}$ for radiative and adiabatic shocks, respectively. These authors also argued that the shocks in the winds of most O-star are most likely radiative, although the wind density remains sufficiently low in most cases to prevent the wind absorption from playing a significant role. \citet{Owocki} considered that turbulence in the radiatively cooling post-shock gas efficiently mixes cold and hot material. To reproduce the observed linear $L_{\rm X}$ versus $L_{\rm bol}$ relation, these authors adopted a scaling of the hot gas volume filling factor with some ad hoc power $m \simeq 0.2 - 0.4$ of the ratio between cooling length and position in the wind. Following their formalism, the X-ray luminosity should scale as $(\frac{\dot{M}}{v_{\infty}})^{1-m}$ (where $\dot{M}$ and $v_{\infty}$ are the mass-loss rate and the terminal velocity of the stellar wind) over most of the spectral range of O-type stars, where the LDI shocks that produce the X-rays are radiative and the winds themselves are optically thin.  Such a relation between the X-ray flux (corrected for the absorption of the interstellar medium, ISM) and the theoretical value of $\frac{\dot{M}}{v_{\infty}}$ was indeed found for the O-type stars observed within the {\it Chandra} Cyg\,OB2 legacy survey \citep{CygOB2}. The \citet{Owocki} scenario furthermore predicts a change in the behaviour of the $L_{\rm X}$ versus $L_{\rm bol}$ relation at the high-luminosity, high mass-loss rate end of the O-star domain, where the winds should progressively become optically thick. \citet{Gayley} extended the \citet{Owocki} formalism and proposed a first unified approach to estimate the X-ray emergence efficiency for winds of non-magnetic hot stars. He predicted a smooth evolution of the X-ray emergence efficiency over the full range of wind strengths from early B-stars into Wolf-Rayet stars.

The most extreme O-stars are so-called O\,If$^+$ stars. These objects have spectral properties intermediate between those of normal O-stars and those of WN-type Wolf-Rayet stars \citep{Willis,Conti95,DeBecker09} and are therefore considered to be transition objects. Possible indications for a difference in the  $L_{\rm X}/L_{\rm bol}$ ratio for O\,If$^+$ stars were reported by \citet{DeBecker13}, who found for the O4\,If$^+$ and O5\,f$^+$ stars HD~16691 and HD~14947 $\log{L_{\rm X}/L_{\rm bol}} \sim -7.1$ or $-7.5$ depending on the adopted value of their $L_{\rm bol}$. However, because these two stars are rather isolated, that is, are located outside a stellar cluster, a proper $L_{\rm X}/L_{\rm bol}$ benchmark is lacking against which a genuine X-ray underluminosity might be established. Whilst the X-ray emissions of O-type stars inside a specific cluster exhibit only a very small scatter about the mean $L_{\rm X}/L_{\rm bol}$ ratio, values ranging between $-6.5$ and $-7.2$ have been reported for $\log{L_{\rm X}/L_{\rm bol}}$ in different populations. Some of the dispersion around the $L_{\rm X}/L_{\rm bol}$ relation inferred from large samples of O-stars (mixing objects from different clusters and field stars) may be due to environmental effects, different approaches in the data analysis \citep{YN,Carina}, or even a poor determination of the spectral properties when the interstellar absorption is high \citep{HM1,M17}. Therefore, an X-ray over- or underluminosity is hard to demonstrate for an isolated O-star. 

The best way to probe the $L_{\rm X}/L_{\rm bol}$ ratio in the higher luminosity regime, is to observe a cluster hosting an O\,If$^+$ star, along with a significant population of normal O-stars, which allows homogeneously building the local  $L_{\rm X}/L_{\rm bol}$ relation of normal O-stars. The only Galactic young open cluster hosting a large sample of O-stars and including an O4\,If$^+$ star (HD~15570) is IC~1805 at a distance near 2\,kpc \citep{Megeath}. 

Within a radius of 12\arcmin\ around the central O + O binary HD~15558, this cluster hosts about 40 early-type stars in the spectral range O4 -- B2, including 8 O-stars with spectral types from O9.5\,V to O4\,If$^+$ \citep{Massey,SH99}. IC~1805 is not only an interesting place to study the X-ray properties of the population of massive stars, but is also an important site to study star formation. IC~1805 lies at the centre of a large superbubble inside the W4 region \citep[][and references therein]{Megeath}. The massive stars have created an H\,{\sc ii} region that is surrounded by a cloud of molecular (CO) and atomic (H\,{\sc i}) gas. The cluster is embedded in a bubble of hot plasma produced by the winds of the O-stars, which yields a diffuse X-ray emission detected with {\it Chandra} \citep{Townsley}. IC~1805 and the entire W4 complex have been the theatre of multiple episodes of star formation over the past 10 -- 20\,Myr and low-mass star formation is still ongoing \citep{Megeath,Vilnius,sun16}, although photoevaporation by the harsh UV-radiation fields of the massive cluster members led to the destruction of optically thick accretion disks around most of the intermediate mass stars \citep{Wolff}. 
  
In this paper we discuss a new X-ray observation of the massive star population of IC~1805 along with some support optical spectroscopy. Our observational material is described in Sect.\,\ref{obs}. The results of our optical monitoring are provided in Sect.\,\ref{overview}, and our X-ray data are analysed and discussed in Sect.\,\ref{xrays}. The summary of our results is given in Sect.\,\ref{conclusion}. 

\section{Observations \label{obs}}
\subsection{{\it XMM-Newton}\label{obsxmm}}
{\it XMM-Newton} \citep{Jansen} observed IC~1805 for 49\,ks in August 2014 (PI Rauw, ObsID 0740020101, see Table\,\ref{journalX}). The raw data were reduced with SAS v14.0.0 using calibration files available in November 2015 and following the recommendations of the {\it XMM-Newton} team\footnote{http://xmm.esac.esa.int/sas/current/documentation/threads/ }. The EPIC \citep{pn,MOS} cameras were operated in the full-frame mode and the medium filter was used to reject optical and UV light. During reduction, these data were filtered to keep only best-quality data ({\sc{pattern}} of 0--12 for EPIC-MOS and 0--4 for EPIC-pn data). We note that no background flares due to soft protons affected the observation.

Source detection was performed on the three EPIC datasets using the task {\it edetect\_chain} on the 0.4--2.0 (soft) and 2.0--10.0 (hard) energy bands for a log-likelihood of 10. This task searches for sources using a sliding box and determines the final source parameters from point spread function (PSF) fitting. The final count rates correspond to equivalent on-axis, full PSF count rates. The task was run both with and without considering the possibility of extended sources and simultaneous fit of up to five neighbouring sources. The results were similar in both cases. A total of 191 sources were found, nine of them appearing potentially problematic (e.g.\ because of their position in or near a CCD gap, or in the wings of the PSF of a brighter source). The full catalogue of sources is given in Table\,\ref{Xcat}. The properties of the brightest X-ray sources are analysed in Sect.\,\ref{xrays}, whilst a full correlation with optical catalogues is performed in \citet{sun16}. 

We then extracted spectra using the task {\it especget} for the eight detected massive stars and for 17 other sources having at least 400 EPIC counts. For the source regions, we used circular regions with radii between 7 and 25\arcsec\ (depending on crowding in the area) and centred on the Simbad coordinates of the massive stars or on the best-fit positions found in the source detection step for the other sources. For the background regions, circular regions of at least 18\arcsec\ radius were chosen in a region devoid of sources and as close as possible to the targets. Dedicated ARF and RMF responses, which are used to calibrate the flux and energy axes, respectively, were also calculated. The EPIC spectra were grouped with the SAS command {\it specgroup} to obtain an oversampling factor of five and to ensure that at least a signal-to-noise ratio of 3 (i.e.\ a minimum of ten counts) was reached in each spectral bin of the background-corrected spectra.

\begin{table}
\caption{Journal of X-ray observations of IC~1805\label{journalX}}
\begin{tabular}{c c c c c}
\hline 
Observatory & Inst.       & Duration & JD(start) & JD(end) \\
            &             &  (ks)    & $-2\,440\,000$ & $-2\,440\,000$ \\
\hline
{\it ROSAT} & PSPC-B      &  8.5     & 8856.966 & 8858.827 \\
{\it Chandra} & ACIS-I    & 82.1     &14065.465 &14066.416 \\
{\it XMM-Newton} & EPIC   & 48.7     &16895.380 &16895.943 \\
\hline
\end{tabular}
\end{table}

EPIC light curves of the brightest sources were extracted for time bins of 1\,ks and for the total 0.4--10.\,keV energy band in the same regions as the spectra. They were further processed by the task {\it epiclccorr}, which corrects for loss of photons due to  vignetting, off-axis angle, or other problems such as bad pixels. In addition, to avoid very large errors and poor estimates of the count rates, we discarded bins displaying effective exposure times lower than 50\% of the time bin length. Our previous experience with {\it XMM-Newton} has shown us that including such bins degrades the results. As the background is much fainter than the source, in fact too faint to provide a meaningful analysis, three sets of light curves were produced and analysed individually: the raw source+background light curves, the background-corrected light curves of the source, and finally the light curves of the background region alone. The results found for the raw
and background-corrected light curves of the source are indistinguishable. As for $\zeta$\,Pup \citep{zetaPup2}, the same set of tests was applied to all cases. We first performed a $\chi^2$ test for three different null hypotheses (constancy, linear variation, quadratic variation), and also compared the improvement of the $\chi^2$ when increasing the number of parameters in the model (e.g.\ linear trend versus constancy) by means of Snedecor F tests (nested models, see Sect.\ 12.2.5 in \citealt{lin76}).

\subsection{Archival {\it ROSAT} and {\it Chandra} data\label{obsacis}}
IC~1805 was also observed during 80\,ks with {\it Chandra} in November 2006 \citep[PI Townsley, ObsID 7033, see Table\,\ref{journalX},][]{Feigelson,Townsley}. We processed these data using the CIAO v.4.7 software (with CALDB 4.6.9). Using {\it specextract}, we extracted the spectra and generated source-specific response matrices for six massive stars (BD+60$^{\circ}$~498, BD+60$^{\circ}$~499, BD+60$^{\circ}$~501, HD~15558\footnote{The ACIS spectrum of this star is subject to moderate pile-up, see \citet{Townsley}.}, HD~15570, and HD~15629) as well as six bright {\it XMM-Newton} sources (see Sect.\,\ref{other}) with at least 100 net ACIS counts as reported in \citet{Townsley}. To this aim, we used circular regions of radius between 2 and 15\arcsec, depending on the off-axis angle and crowding\footnote{To evaluate crowding, we used the source list of \citet{Townsley}. BD+60$^{\circ}$~499 appears to have a close companion, which is blended with the O-star emission in {\it XMM-Newton} data, therefore we extracted the spectra of BD $+60^{\circ}$~499 and its companion to evaluate the contamination in the {\it XMM-Newton} data. The small separation of the two sources forced the use of ellipses, rather than circles, to extract their spectra.}. Circular background regions as close as possible to the targets were used and a similar grouping as for the {\it XMM-Newton} data was applied.

Finally, we also retrieved an archival {\it ROSAT} observation (PI Pauldrach, ObsID 201263, see Table\,\ref{journalX}) obtained in August 1992. These data were processed using the {\it xselect} software. Spectra for five O-stars (BD+60$^{\circ}$~497, BD+60$^{\circ}$~501, HD~15558, HD~15570, and HD~15629) were extracted using circular extraction regions of radius 42\arcsec. The background was evaluated over a circular source-free region of 75\arcsec\ radius. We used the standard response matrix {\tt pspcb\_gain2\_256.rmf} and generated source-specific ancillary response files using the {\it pcarf} command.

\subsection{Optical spectroscopy}
In support of the {\it XMM-Newton} observation, we collected optical spectroscopy for several O-type stars of IC~1805 with the refurbished HEROS spectrograph at the 1.2~m TIGRE telescope \citep{Schmitt} at La Luz Observatory (Guanajuato, Mexico). The HEROS echelle spectra have a resolving power of 20\,000 and cover the full optical range, although with a small gap near 5800\,\AA. The HEROS data were reduced with the corresponding reduction pipeline \citep{Mittag,Schmitt}. Telluric absorptions in the He\,{\sc i} $\lambda$\,5876 and H$\alpha$ regions were corrected with the {\it telluric} command within IRAF using the list of telluric lines of \citet{Hinkle}. The spectra were normalized self-consistently using a series of carefully chosen continuum windows. Journals of the observations for each star, along with the radial velocities that we have determined are given in Appendix\,\ref{Journalopt} (see Tables\,\ref{RV15558} -- \ref{RV513}).

\section{Optical monitoring of the O-stars in IC~1805 \label{overview}}
To assess the $L_{\rm X}/L_{\rm bol}$ relation of single O-type stars in IC~1805, it is important to derive accurate spectral types and evaluate the contamination of our sample by colliding wind binaries. So far, investigations of the multiplicity of O-type stars in IC\,1805 only revealed two binaries: the long-period binary HD~15558 \citep[O5.5\,III(f) + O7\,V, P$_{\rm orb}$ = 442\,days, e = 0.4;][]{GM,DeBecker06}, and the short-period system BD+60$^{\circ}$~497 \citep[O6.5\,V((f)) + O8.5-9.5\,V((f)), P$_{\rm orb}$ = 3.96\,days;][]{RDB}. In this section, we briefly review the properties of the O-star population within the field of view of our X-ray observations and, whenever appropriate, we revise them using our HEROS spectra.

Membership of the stars in the field of IC~1805 was investigated by \citet{Vasilevskis} based on proper motion. The corresponding membership probabilities were subsequently revised by \citet{Sanders}. Except for HD~15558, all O-stars considered here have a membership probability exceeding 50\% \citep{Sanders}. For some of the stars, proper motion membership probabilities were also derived from {\it Hipparcos} data \citep{Baumgardt}. Unfortunately, for the few stars in common between the studies of \citet{Sanders} and \citet{Baumgardt}, the results often differ significantly\footnote{For instance, HD~15558 has a cluster membership probability of 2\% according to \citet{Sanders} and of 94.8\% according to \citet{Baumgardt}.}. Nevertheless, in what follows, we assume that all the O-type stars belong to IC~1805. 

\subsection{HD~15558}
The spectrum of HD~15558 displays a number of remarkable features \citep[see discussion by][]{DeBecker06}. The combined spectral type of the system was recently given as O4.5\,III(f) by \citet{Sota}. \citet{DeBecker06} tentatively proposed an O5.5\,III(f) + O7\,V spectral classification for the binary system. 

\citet{GM} presented the first single-lined spectroscopic binary (SB1) orbital solution of this eccentric long-period ($\sim 440$\,days) spectroscopic binary system. The SB1 orbital parameters were improved by \citet{DeBecker06}. These authors also identified weak spectroscopic signatures of the secondary star in some spectral lines, which led to a first double-lined spectroscopic binary (SB2) orbital solution. The latter yielded very high minimum masses for the primary star ($m_1\,\sin^3{i} > 150$\,M$_{\odot}$). However, this result needs to be confirmed, as the luminosities of HD~15558 (assuming the star to be a member of IC~1805) would be surprisingly low for such a massive star. \citet{DeBecker06} suggested that HD~15558 might be a triple system with a short-period binary moving around the O5\,III star in a wide orbit of $\sim 440$\,days. 

\begin{figure}[thb]
\begin{center}
\resizebox{8cm}{!}{\includegraphics{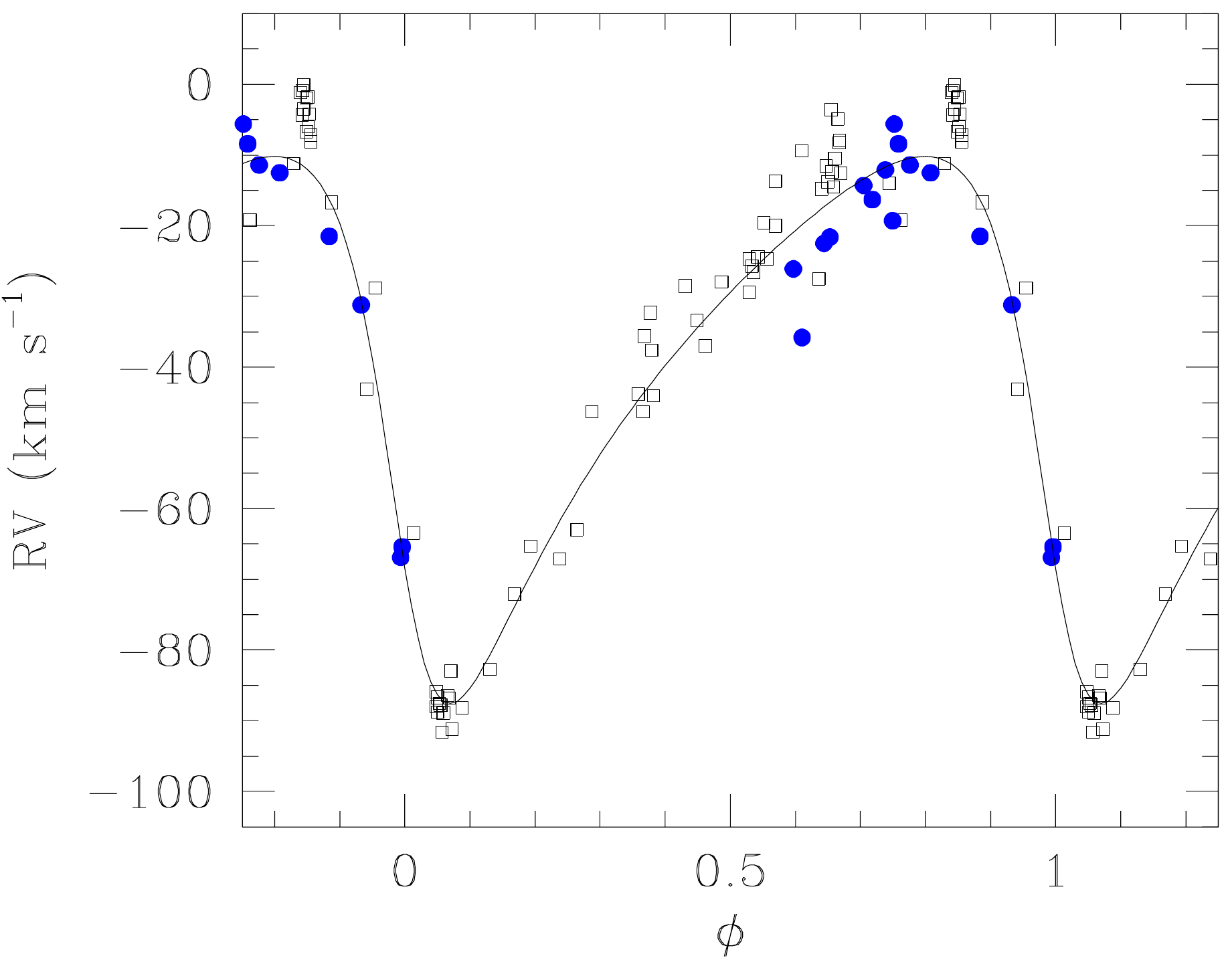}}
\end{center}
\caption{New SB1 orbital solution of HD~15558, using He\,{\sc ii} $\lambda$\,4542 RVs from \citet[][open squares]{DeBecker06} and from our new HEROS data (filled circles, see Table\,\ref{RV15558}). The corresponding orbital elements are given in Table\,\ref{orbit15558}.\label{solorbit15558}}
\end{figure}

We have analysed 16 new HEROS spectra of HD~15558 taken between August 2013 and October 2014. Whilst this dataset is not sufficient to search for a possible signature of the putative close binary system, it allows us to improve the SB1 orbital solution and to establish accurate orbital phases corresponding to the X-ray observations. We measured the radial velocities (RVs) of the He\,{\sc ii} $\lambda$\,4542 line on our HEROS spectra (Table\,\ref{RV15558}) and combined these new RVs with those listed by \citet{DeBecker06}. Using the Fourier period search technique of \citet{HMM} and \citet{Gosset}, we found a best estimate of the orbital period of $447.25 \pm 4.10$\,days for 87 RV data points spread over 4880\,days. We then used the Li\`ege Orbital Solution Package (LOSP) code \citep{SGR}, which is an improved version of the code originally proposed by \citet{WHS}. 
The resulting orbital elements are listed in Table\,\ref{orbit15558} and the new orbital solution is shown in Fig.\,\ref{solorbit15558}. We note that the new mass function is lower than in the solution of \citet{DeBecker06}, which obviously also leads to a reduction of the masses of the individual components. More observations are needed to determine whether HD~15558 is a triple system and to search for the secondary (and possibly tertiary) spectral signatures. 

\begin{table}[htb]
\caption{Revised SB1 orbital solution of HD~15558 based on RVs of the He\,{\sc ii} $\lambda$\,4542 line (see Table\,\ref{RV15558}).\label{orbit15558}} 
\begin{center}
\begin{tabular}{c c}
\hline
Element & Value \\
\hline
$P_{\rm orb}$ (days) & $445.76 \pm 0.42$ \\
$e$                & $0.42 \pm 0.02$ \\
$T_0$ (HJD)        & $2\,456\,692.47 \pm 3.71$ \\
$K$ (km\,s$^{-1}$)  & $38.7 \pm 1.1$ \\
$\gamma$ (km\,s$^{-1}$) & $-40.7 \pm 0.8$ \\
$\omega$ ($^{\circ}$)& $120.2 \pm 4.3$ \\
$a\,\sin{i}$ (R$_{\odot}$)& $309.4 \pm 9.8$ \\
$f(m)$ (M$_{\odot}$) & $2.00 \pm 0.19$ \\
\hline
\end{tabular}
\end{center}
\end{table}  

As a massive binary system, HD~15558 hosts a colliding wind interaction. The star was reported to be a probable non-thermal radio emitter by \citet{Bieging}, and the relativistic electrons involved in the synchrotron radio emission are most likely accelerated at the shock fronts of the wind - wind collision. The wind interaction zone is also most relevant in the context of the X-ray emission of the system. With the newly derived ephemerides, we find that the {\it ROSAT}, {\it Chandra} and {\it XMM-Newton} observations were taken at orbital phases 0.42, 0.11 and 0.46, respectively, where $\phi = 0.0$ corresponds to the periastron passage.

\subsection{HD~15570}
The optical spectrum of HD~15570 displays many peculiarities (notably strong, broad, and asymmetric H$\alpha$ emission, presence of Si\,{\sc iv} $\lambda\lambda$ 4088, 4116 and weak N\,{\sc iv} $\lambda$\,4058 emissions in addition to the usual Of emission lines of He\,{\sc ii} $\lambda$\,4686 and N\,{\sc iii} $\lambda\lambda$\,4634-4640, presence of N\,{\sc v} $\lambda\lambda$\,4604, 4620 absorptions) that led to an O4\,If$^+$ classification. A full discussion of the optical spectrum is provided by \citet{DeBecker06}. 
\citet{Willis} noted that the {\it IUE} UV spectrum of HD~15570 was intermediate between that of normal Of stars and Wolf-Rayet stars of the WN sequence, thereby suggesting an intermediate evolutionary stage. 

Several authors \citep{Hillwig,DeBecker06,DeBecker09} searched for RV variations that could reveal a spectroscopic binary. Whilst there are some low-level variations in the radial velocities, these variations are not significant and might reflect contamination of the absorption lines by some intrinsically variable wind emission. For instance, based on 68 spectra taken between September 2000 and November 2007, \citet{DeBecker09} reported a RV of $-47.1 \pm 5.8$\,km\,s$^{-1}$ for the He\,{\sc ii} $\lambda$\,4542 line. 

\citet{Polcaro} reported on strong H$\alpha$ line profile and equivalent width variability of HD~15570. They claimed variations of the EW of the H$\alpha$ emission between $-3.3$ and $-11.7$\,\AA, which translate into variations by 40\% of the mass-loss rate (assuming a smooth wind). The He\,{\sc ii} $\lambda$\,4686 emission was found to be considerably less variable \citep{Polcaro}. \citet{DeBecker06,DeBecker09} confirmed the variability of the emission lines in the spectrum of HD~15570, but at a significantly lower level than indicated by \citet{Polcaro}. 

Even though we only have four HEROS spectra of this star, they are fully consistent with the general description of the spectrum and its variability as given by \citet{DeBecker06} and \citet{DeBecker09}. For the RV of the He\,{\sc ii} $\lambda$\,4542 line (see Table\,\ref{RV15570}), we obtain $-60.3 \pm 3.8$\,km\,s$^{-1}$. Whilst this value is more negative than the one given by \citet{DeBecker09}, this difference might stem from the slightly asymmetric profile of the line when observed under different spectral resolutions.

\begin{figure*}[t!hb]
\begin{minipage}{8cm}
\begin{center}
\resizebox{8cm}{!}{\includegraphics{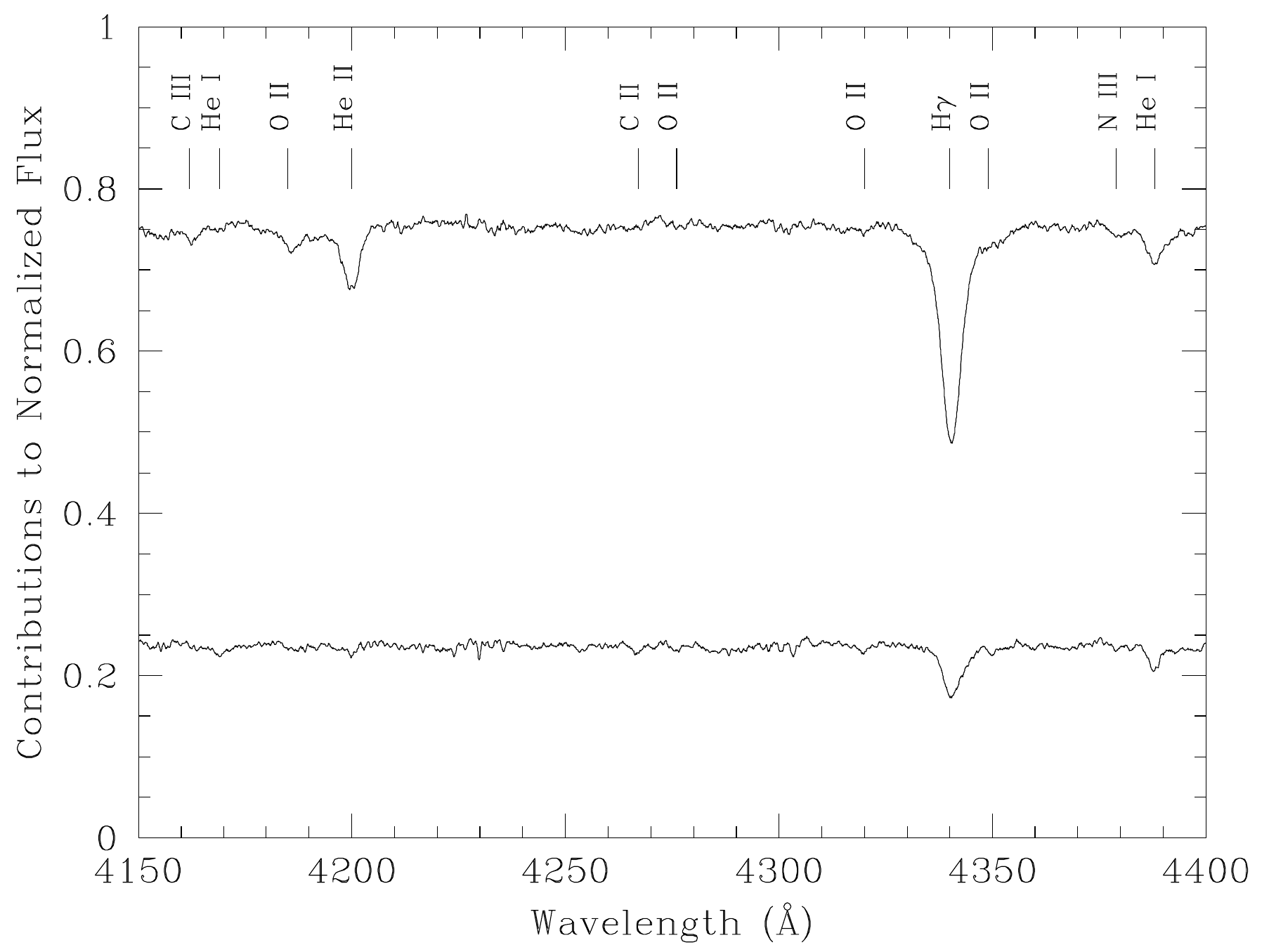}}
\end{center}
\end{minipage}
%\hfill
\begin{minipage}{8cm}
\begin{center}
\resizebox{8cm}{!}{\includegraphics{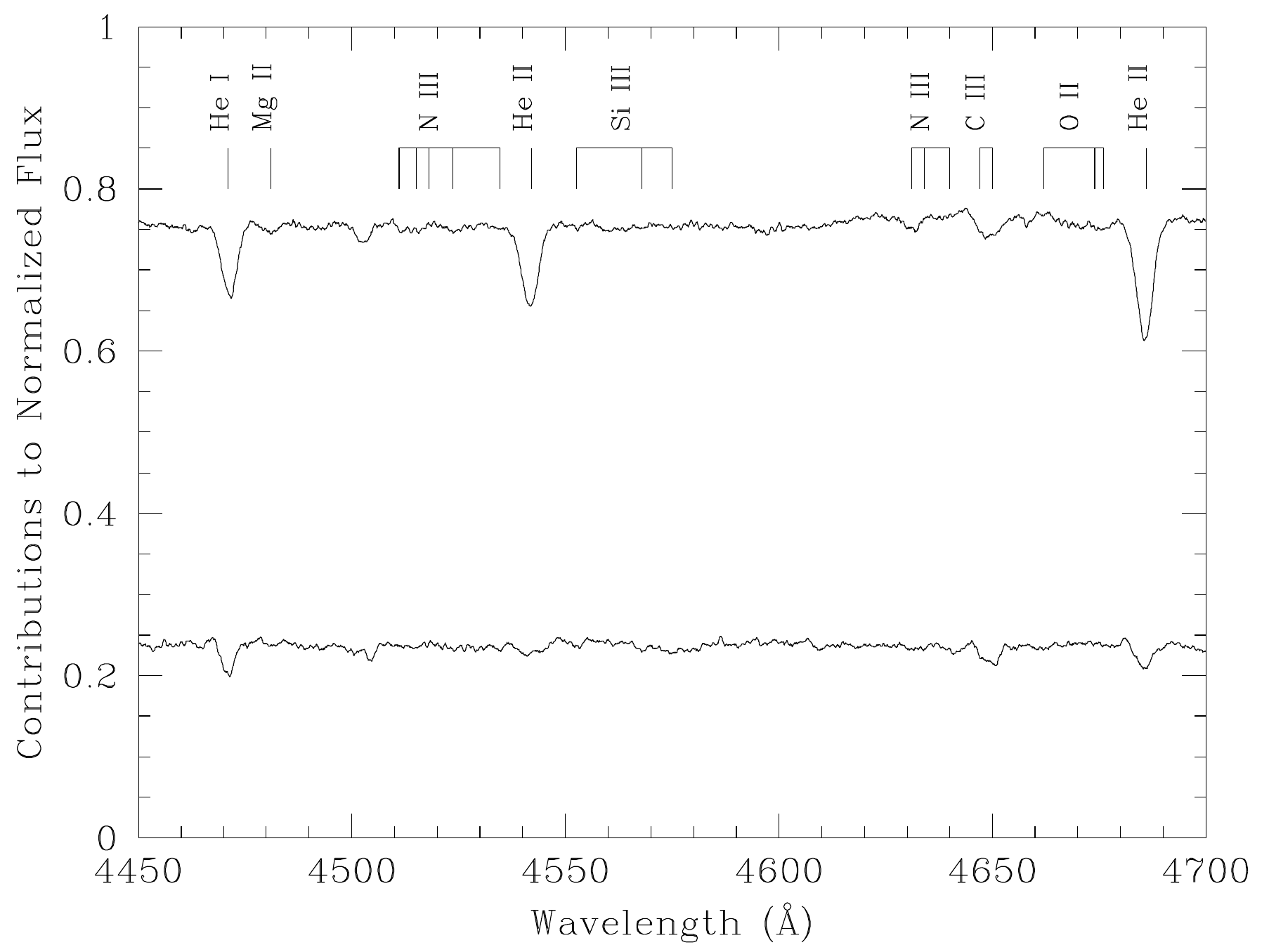}}
\end{center}
\end{minipage}
\caption{Blue spectral regions of the components of BD+60$^{\circ}$~497 obtained by disentangling of our HEROS spectra. The primary (top) and secondary (bottom) spectra are shown with their continua normalized to the relative contribution to the combined spectrum. \label{disent497}}
\end{figure*}

\citet{Bouret} and \citet{Surlan} performed spectral modelling of the FUV, UV and optical spectrum of HD~15570 using the CMFGEN \citep{CMFGEN} and PoWR \citep{PoWR} model atmosphere codes, respectively. Whilst \citet{Bouret} assumed the wind to host a distribution of optically thin clumps, \citet{Surlan} allowed the clumps to be of arbitrary optical depth. \citet{Bouret} had to assume a sub-solar phosphorus abundance to fit the P\,{\sc v} $\lambda\lambda$\,1118, 1128 resonance lines, whilst \citet{Surlan} could fit these lines with solar phosphorus abundance. \citet{Bouret} inferred $T_{\rm eff} = 38000$\,K, $\log{L/L_{\odot}} = 5.94$, and a mass-loss rate $\dot{M} = 2.19\,10^{-6}$\,M$_{\odot}$\,yr$^{-1}$. These authors also noted strongly enhanced N and strongly depleted C abundances, which most likely reflect the advanced evolutionary stage of the star. \citet{Surlan} adopted the same effective temperature, luminosity and chemical composition (except for phosphorus) and derived a mass-loss rate of $2.75\,10^{-6}$\,M$_{\odot}$\,yr$^{-1}$, in good agreement with the value of \citet{Bouret}.

\subsection{HD~15629}
\citet{Sota} proposed a spectral-type O4.5\,V((fc)) where the `c' tag indicates that the C\,{\sc iii} $\lambda\lambda$\,4647-50-51 emission lines have the same intensity as N\,{\sc iii} $\lambda$\,4634. Our two HEROS spectra suggest a slightly later spectral type (O5), but otherwise agree with the results of \citet{Sota}. We find that the Si\,{\sc iv} $\lambda$\,4089 line is filled up by emission and that the Si\,{\sc iv} $\lambda$\,4116 line is weakly in emission. Moreover, we note a group of emission lines at 6716, 6721 and 6728\,\AA\ that we tentatively identify as a blend of C\,{\sc iii} and N\,{\sc v} lines following the discussion for other stars in \citet{Walborn}. 

\citet{Repolust} analysed the optical line spectrum of HD~15629 with the FASTWIND \citep{Puls} code and inferred $T_{\rm eff} = 40500$\,K, $\log{L/L_{\odot}} = 5.60 \pm 0.13$ and  mass-loss rate $\dot{M} = 1.28\,10^{-6}$\,M$_{\odot}$\,yr$^{-1}$ (assuming a smooth wind).  
A subsequent analysis by \citet{Martins05} with the CMFGEN code accounting for wind clumping led to similar stellar parameters ($T_{\rm eff} = 41000$\,K, $\log{L/L_{\odot}} = 5.56$), but a significantly lower mass-loss rate ($3.2\,10^{-7}$\,M$_{\odot}$\,yr$^{-1}$ for a filling factor of 0.1 in the outer regions of the wind). 

\begin{figure*}[t!hb]
\begin{minipage}{8cm}
\begin{center}
\resizebox{8cm}{!}{\includegraphics{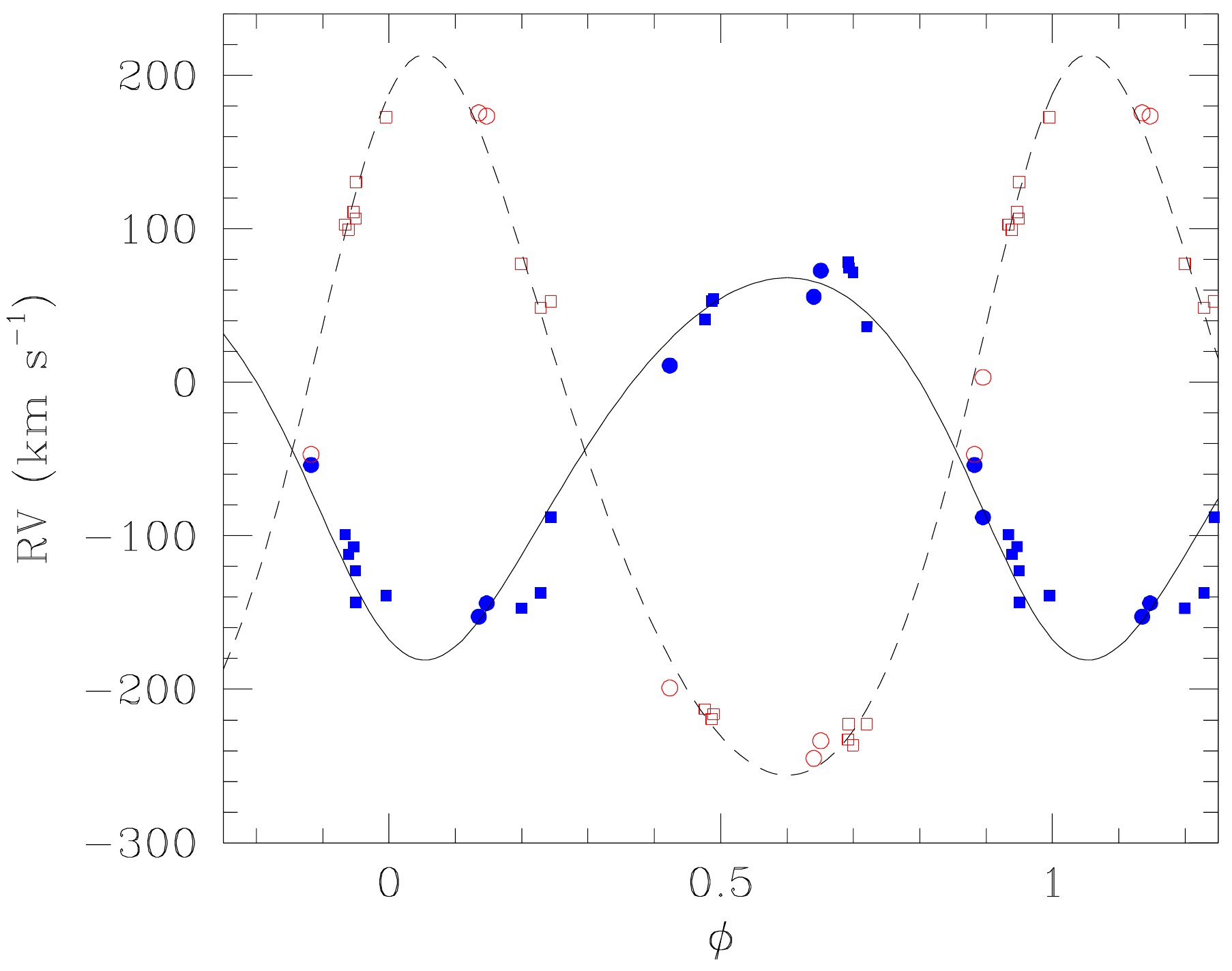}}
\end{center}
\end{minipage}
%\hfill
\begin{minipage}{8cm}
\begin{center}
\resizebox{8cm}{!}{\includegraphics{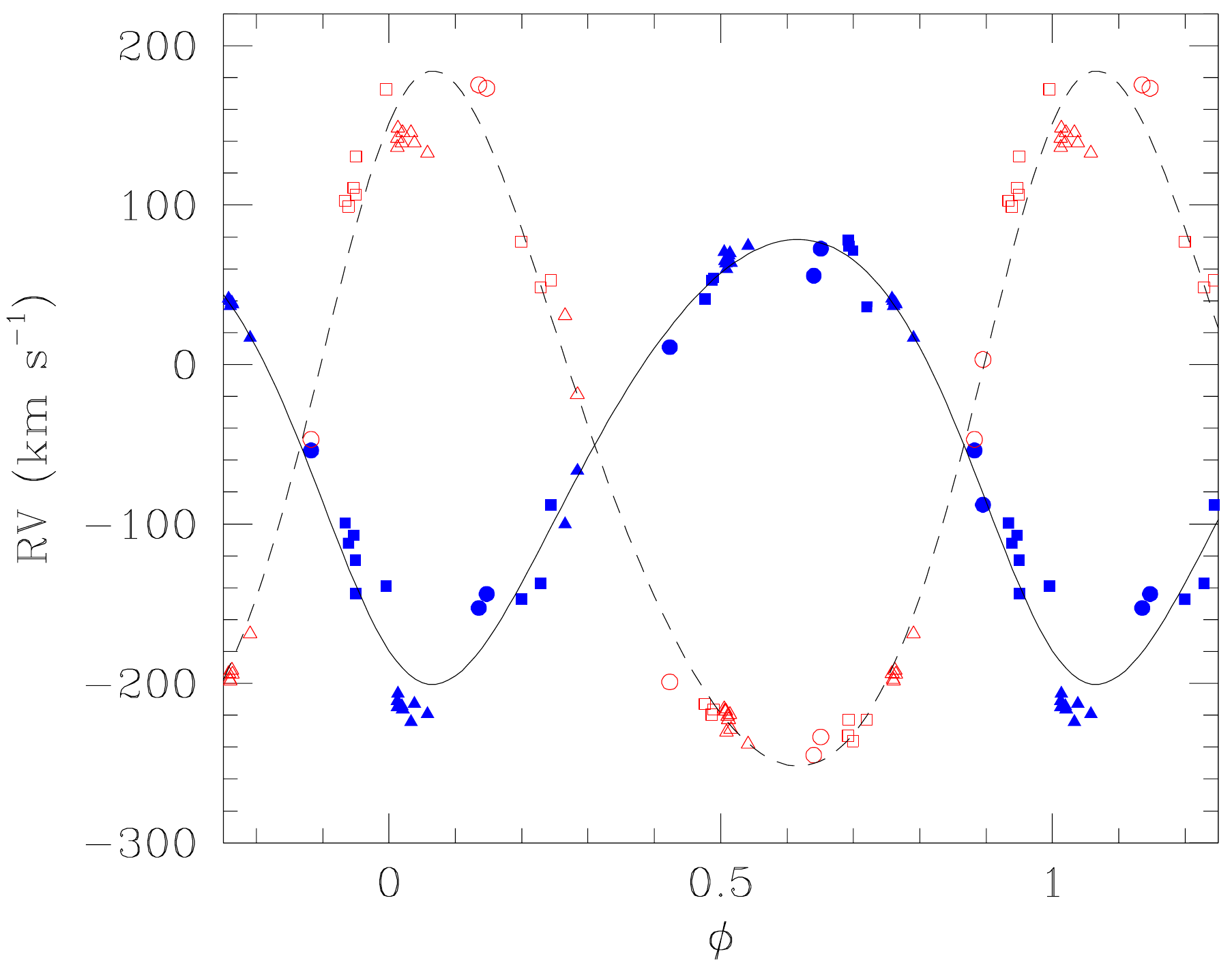}}
\end{center}
\end{minipage}
\caption{{\it Left}: new orbital solution of BD+60$^{\circ}$~497, using only RVs obtained by spectral disentangling (see text and Table\,\ref{solorb497}). Aurelie data from \citet{RDB} are shown by squares, HEROS data are given by circles. Filled and open symbols stand for the primary and secondary star, respectively. {\it Right}: same, but this time also including the RV data from \citet[][shown by triangles]{Hillwig}.\label{RVcurve497}}
\end{figure*}

Neither \citet{Hillwig} nor \citet{DeBecker06} found significant and coherent RV variability. Our two HEROS spectra broadly agree with this conclusion (see Table\,\ref{RV15629}). Therefore, there is currently no evidence for binarity of HD~15629. \cite{Peri} studied {\it WISE} images that reveal a complex layered structure, making HD~15629 a bow-shock candidate. We stress, however, that the RV of the star does not deviate significantly from that of other O-stars in IC~1805, unlike what would be expected for a runaway star.
 
\subsection{BD+60$^{\circ}$~497} 
This star was found to be a double-lined spectroscopic binary with an orbital period of 3.96\,days by \citet{RDB}. Based on a set of 16 blue spectra with a resolving power of 8000 obtained with the Aur\'elie spectrograph at the 1.52\,m telescope at Observatoire de Haute Provence, these authors classified the system as an O6.5\,V((f)) + O8.5-9.5\,V((f)). They also presented a first orbital solution assuming a circular orbit. Later on, \citet{Hillwig} combined the data of \citet{RDB} with their own spectra, which had a resolving power of 5700, to revise the orbital solution. They derived a period of $3.95863 \pm 0.00021$\,days and instead favoured an eccentric orbital solution with $e = 0.156 \pm 0.019$. 

Our new HEROS spectra clearly confirm that this system has an eccentric orbit. We applied our spectral disentangling routine, which is based on the method of \citet{GL}, to the HEROS and Aur\'elie spectra of the system and derived revised radial velocities that are listed in Table\,\ref{RV497}. The separated spectra of BD+60$^{\circ}$~497 (Fig.\,\ref{disent497}) support the previous spectral classification proposed by \citet{RDB}. Based on the \citet{Conti} criterion, we infer spectral types O6.5\,V((f)) and O8.5\,V for the primary and secondary, respectively.

Using the LOSP code with the newly derived RVs, we then derived a new orbital solution for BD+60$^{\circ}$~497 (see Fig.\,\ref{RVcurve497} and Table\,\ref{solorb497}). We considered two different sets of RV data, either restricted to the RVs obtained through spectral disentangling, or also including the RVs quoted by \citet{Hillwig}. In both cases, we find an eccentricity that matches the value proposed by \citet{Hillwig} quite well, whilst the longitude of periastron is significantly larger than the value obtained by \citet{Hillwig}. This could indicate that the system undergoes a relatively fast apsidal motion. However, by far the largest differences between the two orbital solutions concern the amplitudes of the RV curves. This also leads to somewhat different values of the mass ratio. Additional spectroscopic monitoring at high spectral resolution is probably needed to solve this problem.  

\begin{table}
\caption{New orbital solution of BD+60$^{\circ}$~497. \label{solorb497}}
\begin{center}
\begin{tabular}{l c c}
\hline
& Aur\'elie + HEROS & Aur\'elie, HEROS \\
&  data             &  + \citet{Hillwig} \\
\hline
P$_{\rm orb}$ (days) & $3.959266 \pm 0.000030$ & $3.959259 \pm 0.000024$ \\
T$_0$ (HJD) & $2\,456\,682.086 \pm 0.071$ & $2\,456\,682.037 \pm 0.069$ \\
$e$ & $0.159 \pm 0.039$  & $0.149 \pm 0.017$ \\
$\omega$ ($^{\circ}$) & $153.1 \pm 7.2$ & $148.0 \pm 6.9$ \\ 
$m_1/m_2$ & $1.88 \pm 0.09$ & $1.56 \pm 0.06$ \\
$K_1$ (km\,s$^{-1}$) & $ 124.6 \pm 4.9$ & $139.6 \pm 3.5$ \\
$K_2$ (km\,s$^{-1}$) & $ 234.7 \pm 9.2$ & $218.0 \pm 5.5$ \\
$\gamma_1$ (km\,s$^{-1}$) & $-38.8 \pm 3.6$ & $-43.4 \pm 3.3$ \\
$\gamma_2$ (km\,s$^{-1}$) & $-54.6 \pm 4.6$ & $-61.4 \pm 3.8$ \\
$m_1\,\sin^3{i}$ (M$_{\odot}$) & $11.9 \pm 1.2$ & $11.1 \pm 0.6$ \\
$m_2\,\sin^3{i}$ (M$_{\odot}$) & $6.3 \pm 0.6$ & $7.1 \pm 0.4$ \\
$a\,\sin{i}$ (R$_{\odot}$) & $27.7 \pm 0.8$ & $27.6 \pm 0.5$\\
\hline
\end{tabular}
\end{center}
\end{table}

\subsection{BD+60$^{\circ}$~498 \label{BD498}}
This star was classified as O9.7\,II-III by \citet{Sota}.  
\citet{Underhill} reported double lines in the spectrum of BD+60$^{\circ}$~498, suggesting it to be a binary. 

\begin{figure}[thb]
\begin{center}
\resizebox{8cm}{!}{\includegraphics{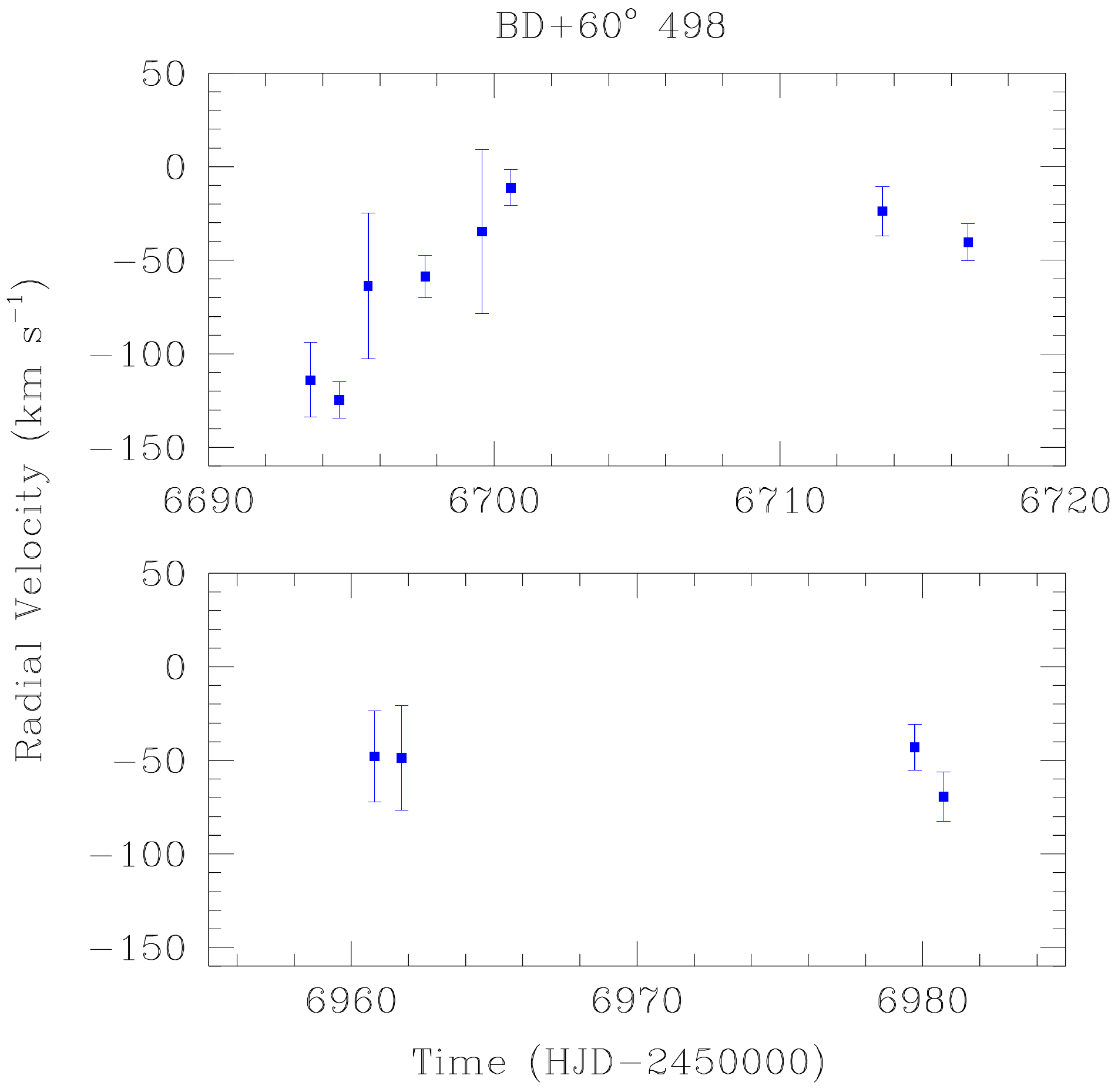}}
\end{center}
\caption{Radial velocities of BD+60$^{\circ}$~498 as measured on our HEROS data (see Table\,\ref{RV498}). The top panel corresponds to data taken in February 2014, whilst the bottom panel illustrates the data from October and November 2014. \label{BD498fig}}
\end{figure}

Our 12 HEROS spectra of this star reveal RV changes (measured on H\,{\sc i}, He\,{\sc i} and He\,{\sc ii} lines) between $-125$ and $-11$\,km\,s$^{-1}$ with a clear progression from the lowest to the least-negative value over a seven nights interval (see Table\,\ref{RV498} and Fig.\,\ref{BD498fig}). This situation suggests that this is a binary system with an orbital period of more than ten days, possibly with a rather eccentric orbit. Whilst there are some hints for a secondary spectral signature in He\,{\sc i} lines around the most extreme RVs, a far more extensive observing campaign would be needed to derive a full SB2 solution. BD+60$^{\circ}$~498 thus appears to be the third O-type binary system within IC~1805. The spectral lines are rather broad, and part of this broadening very likely stems from the blends with the companion's spectral signature. The \citet{Conti} criteria applied to the mean spectrum yield a combined spectral type O9.7 in agreement with the result of \citet{Sota}. However, the strength of the He\,{\sc ii} $\lambda$\,4686 absorption, as well as the ratio between the strengths of Si\,{\sc iv} $\lambda$\,4088 and He\,{\sc i} $\lambda$\,4143, suggest a main-sequence luminosity class instead of a giant or bright giant, as proposed by \citet{Sota}. 

\subsection{BD+60$^{\circ}$~499}
\citet{Sota} classified this star as O9.5\,V, and our HEROS spectra fully agree with this classification. 
\citet{Underhill} suggested that this is a single star. 

We have only two HEROS spectra separated by three nights. The spectra reveal very narrow lines. \citet{HG} determined a projected rotational velocity of only 18\,km\,s$^{-1}$ for this star, whilst \citet{Mimes} obtained 30\,km\,s$^{-1}$. \citet{HG} also noted a change in RV (from $-66.0$ to $-41.3$\,km\,s$^{-1}$) between their two spectra. Although our data set is clearly not sufficient to rule out binarity, we stress that our data indicate a constant RV of $-46.6$\,km\,s$^{-1}$ (Table\,\ref{RV499}). 
\subsection{BD+60$^{\circ}$~501}
\citet{RDB} provided an O7\,V((f)) spectral type, while \citet{Sota} classified this star as O7\,V(n)((f))z where the (n) qualifier stands for lines that are broadened by $v\,\sin{i} \sim 200$\,km\,s$^{-1}$ and the z tag indicates that the He\,{\sc ii} $\lambda$\,4686 line is stronger than both He\,{\sc i} $\lambda$\,4471 and He\,{\sc ii} $\lambda$\,4542. 

Neither \citet{RDB} nor \citet{Hillwig} found significant RV variations for this star, although the mean RVs obtained by these authors significantly differ. \citet{RDB} obtained a mean of $-49.9 \pm 2.5$, whilst \citet{Hillwig} instead inferred $-57.9 \pm 1.1$\,km\,s$^{-1}$. 

\begin{figure*}[htb]
\begin{minipage}{8cm}
\begin{center}
\resizebox{8cm}{!}{\includegraphics{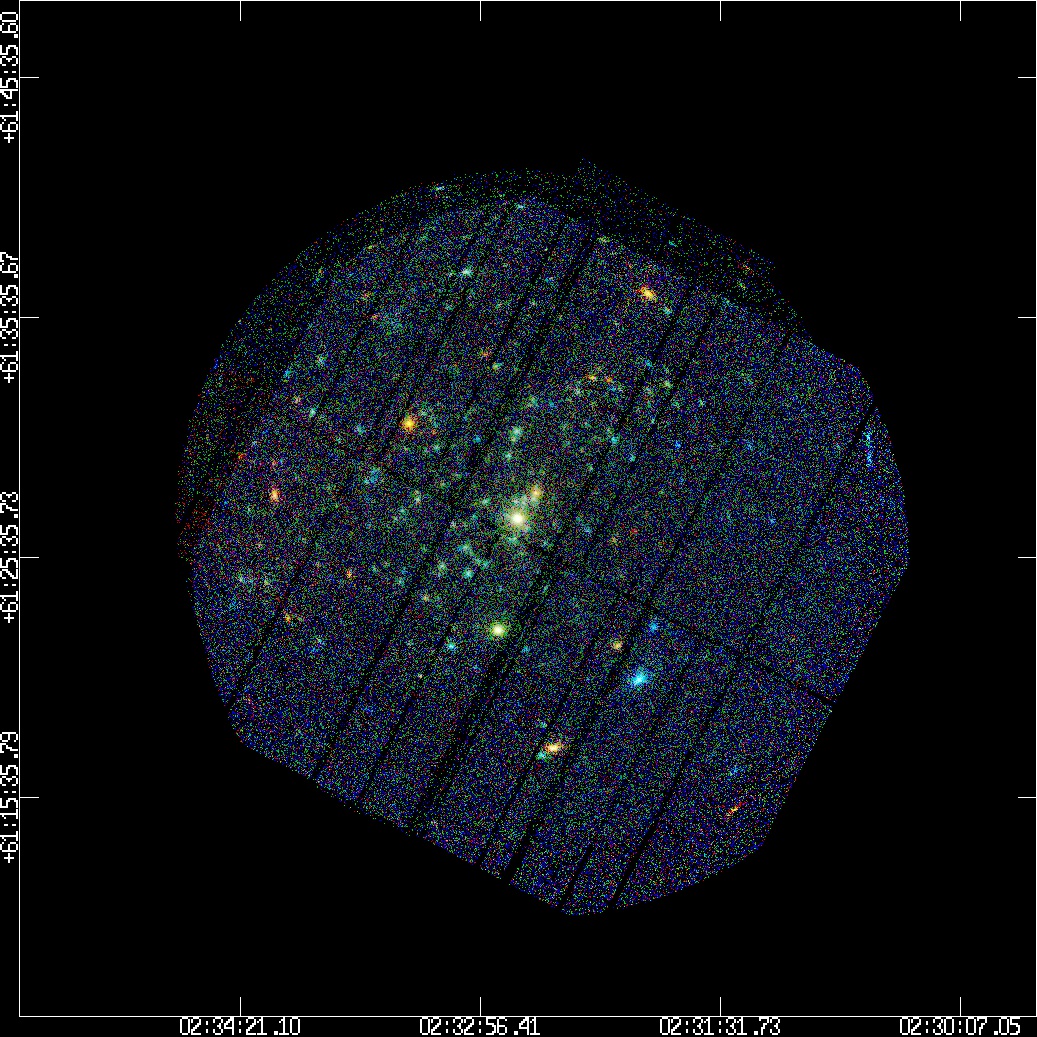}}
\end{center}
\end{minipage}
\hfill
\begin{minipage}{10cm}
\begin{center}
\resizebox{10cm}{!}{\includegraphics{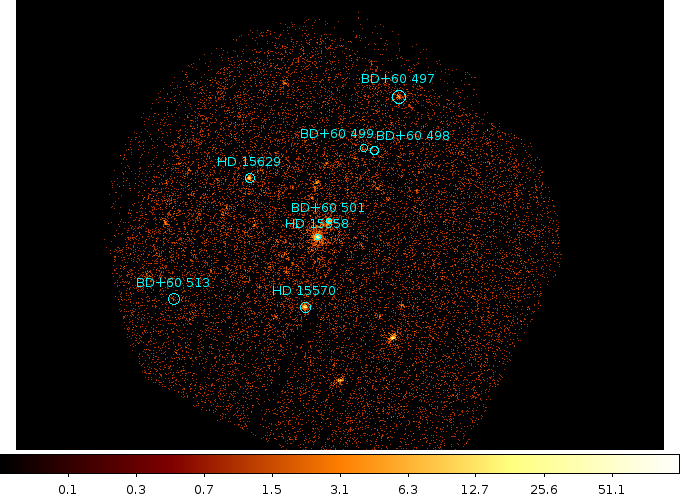}}
\end{center}
\end{minipage}
\caption{{\it Left}: energy-coded three-colour image of our {\it XMM-Newton} observation of IC~1805. Data from the three EPIC cameras were combined and exposure corrected to build this image. Red, green, and blue correspond to photon energies in the ranges $[0.5,1.0]$, $[1.0,2.0]$, and $[2.0,8.0]$\,keV, respectively. {\it Right}: combined EPIC image illustrating the location of the O-type stars in IC~1805.\label{3colour}}
\end{figure*}

\subsection{BD+60$^{\circ}$~513}
This star displays broad absorption lines in its optical spectrum indicating that it must be a fast rotator \citep{RDB,Hillwig}. Whilst some RV variations were reported, they most likely result from distortions of the broad lines by intrinsic profile variations \citep{RDB}. The star was classified as O7.5\,V((f)) by \citet{RDB} and O7\,Vnz by \citet{Sota}, where the n qualifier stands for lines that are broadened by $v\,\sin{i} \sim 300$\,km\,s$^{-1}$. 
\citet{RDB} obtained a mean RV of $-44.3 \pm 10.5$, whilst \citet{Hillwig} instead quoted $-58.6 \pm 2.8$\,km\,s$^{-1}$.
 
In February 2014 we obtained a single HEROS spectrum of the star that confirms the O7.5\,Vnz spectral type. There is an Of emission hump between 4600 and 4700\,\AA. Measuring the same spectral lines as \citet{RDB}, we obtain a heliocentric RV of $-65.7 \pm 9.7$\,km\,s$^{-1}$ where the quoted uncertainty corresponds to the dispersion among the values of the four lines. 

\section{X-ray emission in IC~1805 \label{xrays}}
Figure\,\ref{3colour} illustrates the X-ray view of IC~1805 as seen with {\it XMM-Newton}. Red, green, and blue correspond to photon energies in the ranges $[0.5,1.0]$, $[1.0,2.0]$, and $[2.0,8.0]$\,keV, respectively. The detection algorithm yields a total of 191 detected EPIC sources (see Table\,\ref{Xcat}). Except for a few soft sources, the brightest X-ray emitters appear yellow and are associated with the O-star members of the cluster. A few bright sources (XID~5, 8, 11) that are not associated with O-stars appear red in Fig.\,\ref{3colour}. These are most likely late-type foreground stars (see Sect.\,\ref{other}). Some very bright blue sources (XID~3 and 13) could be associated with background AGNs. Finally, many weak sources, most of which display a relatively hard emission (green or even blue in Fig.\,\ref{3colour}) are scattered throughout the cluster. Most of these sources are very probably pre-main-sequence (PMS) or extragalactic background sources; see below and \citet{sun16} for a full discussion of their counterparts.

\subsection{O-type stars}
Eight stars of an O-type spectral type are located inside the EPIC field of view (see Fig.\,\ref{3colour} and the objects listed in Sect.\,\ref{overview}). Two more objects inside the field of view have been classified as O-stars in the literature at some point, but this classification has been revised since then. ALS~7270 (= IC~1805 113 = LS\,I $+61^{\circ}$~277) was listed as O9.5\,Ve by \citet{SH99}. However, \citet{Negueruela} reclassified this star as B1\,Ve rather than O9.5\,Ve, explaining its non-detection in our {\it XMM-Newton} observation. Similarly, BD+60$^{\circ}$~512 is listed in the online table of \citet{Reed} as an O6 star. However, the coordinates given by \citet{Reed} and the spectral classification refer to the nearby BD+60$^{\circ}$~513 (see Sect.\,\ref{overview}). \citet{Ishida} instead classified BD+60$^{\circ}$~512 as a foreground F8\,V star. This probably explains why this star is undetected in our {\it XMM-Newton} observation. 

\begin{figure*}[htb]
\begin{minipage}{8.5cm}
\begin{center}
\resizebox{8.5cm}{!}{\includegraphics{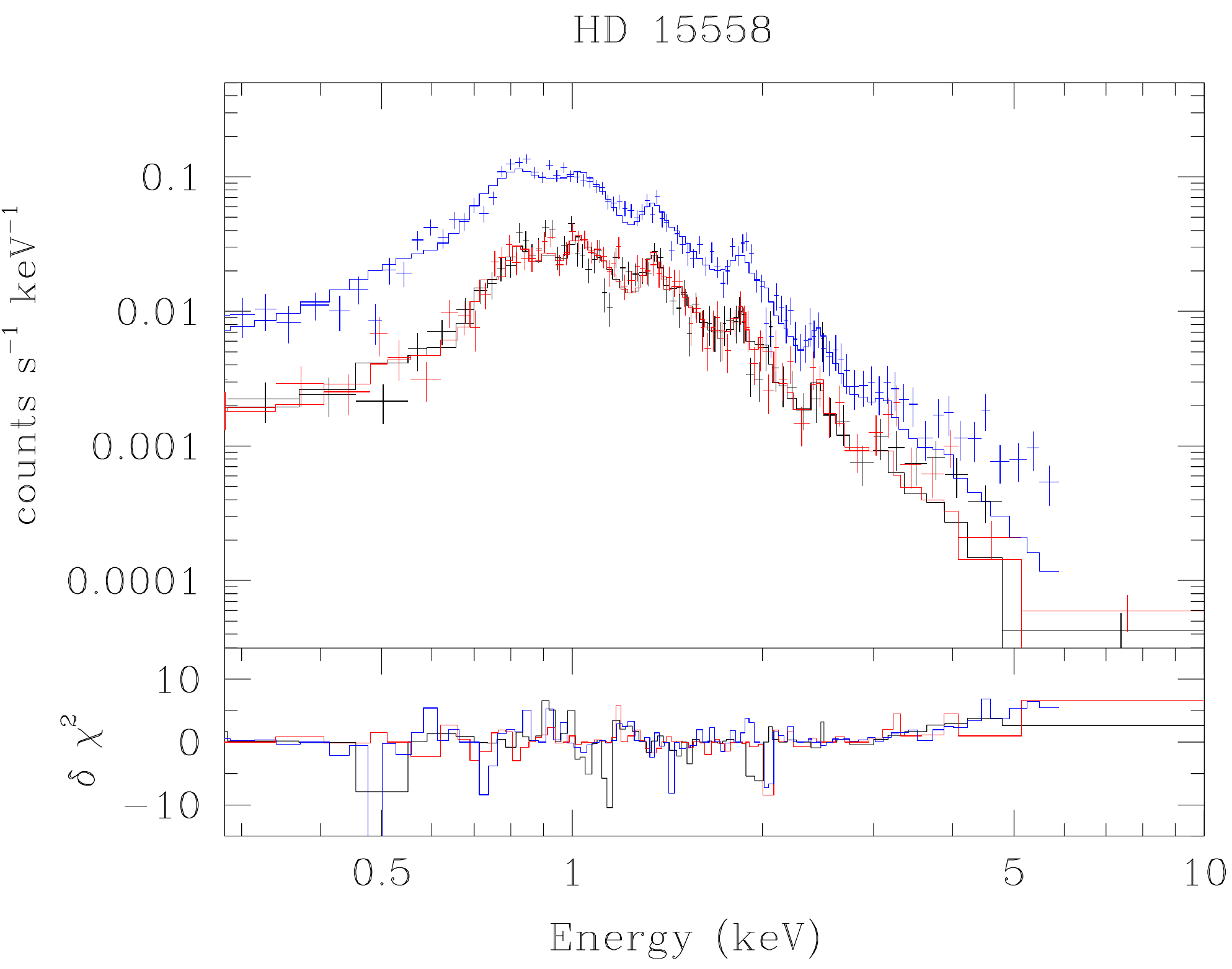}}
\end{center}
\end{minipage}
\hfill
\begin{minipage}{8.5cm}
\begin{center}
\resizebox{8.5cm}{!}{\includegraphics{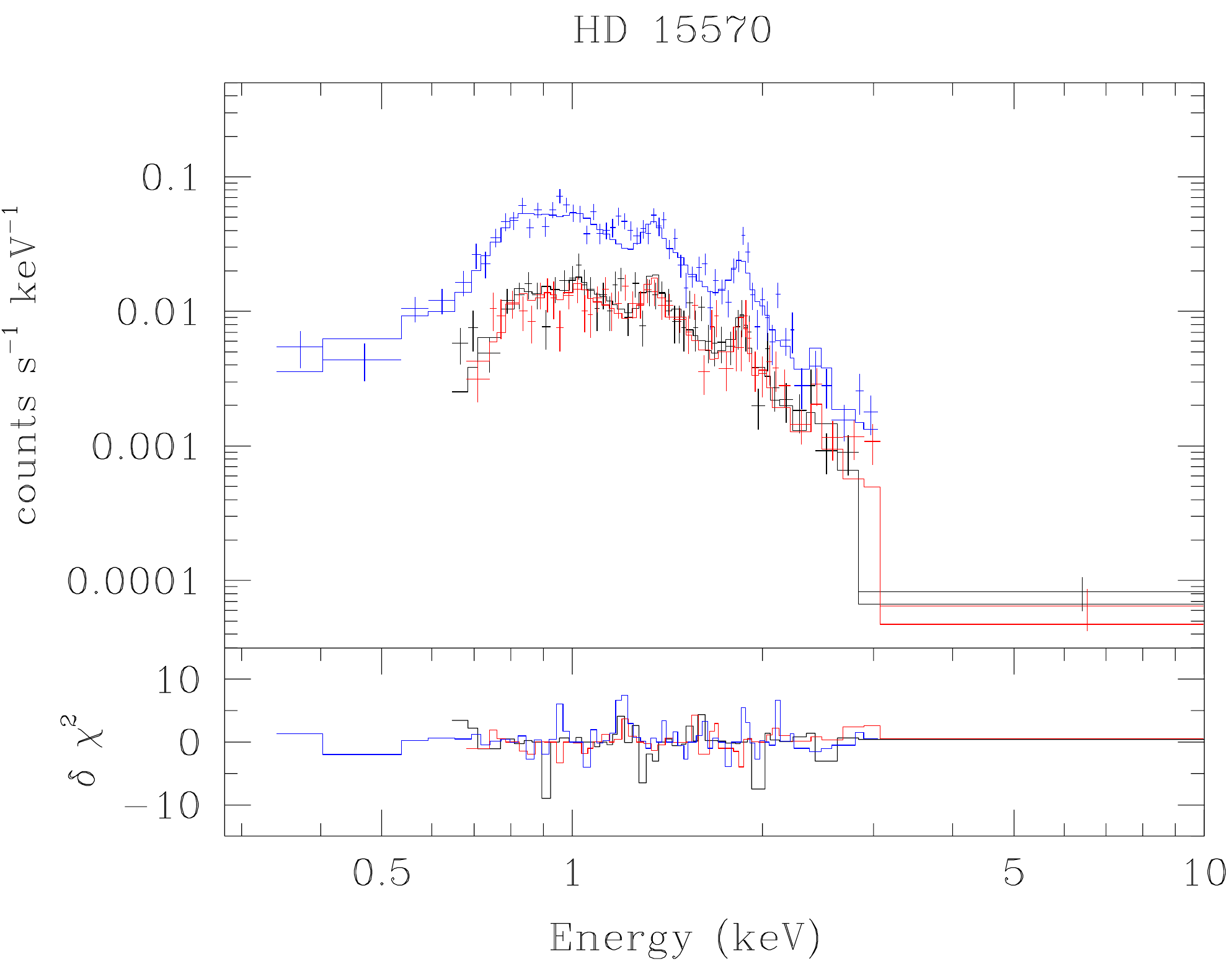}}
\end{center}
\end{minipage}
\caption{{\it Left}: comparison of the EPIC spectra of HD~15558 with the best-fit model quoted in Table\,\ref{specX}. Black, red, and blue colours refer to the spectra obtained with the EPIC-MOS1, MOS2, and pn detectors, respectively. {\it Right}: same for HD~15570.\label{SpecEPIC}}
\end{figure*}
The fits of the X-ray spectra of the O-stars were made using {\tt xspec} \citep{Arnaud} version 12.9.0i. The plasma abundances were taken to be solar \citep{Asplund}. The interstellar absorption was modelled using the T\"ubingen-Boulder ISM model \citep[{\tt tbabs},][]{Wilms}. To evaluate the interstellar neutral hydrogen column density, we adopted the photometric colours of \citet{SL} along with the intrinsic colours from \citet{MP} and the conversion between colour excess and neutral hydrogen column density of \citet{Bohlin}. X-ray spectra from massive stars can furthermore be absorbed by the ionized stellar wind. To model such an absorption, we imported the stellar wind absorption model of \citet{HD108} into {\tt xspec} as a multiplicative tabular model (hereafter labelled {\tt wind}). 
Emission from collisionally ionized equilibrium optically thin thermal plasma was modelled using the {\tt apec} models \citep{apec}. We used models computed  with ATOMDB v2.0.2.
The fits were performed for energy bins between 0.2 and 12.0\,keV. X-ray fluxes corrected for the interstellar absorption and estimates of their errors were obtained using the {\it cflux} command in {\tt xspec}. 

We tested models of the type {\tt tbabs$\times$wind$\times\sum_{i=1}^{1\,{\rm or}\,2}$apec(T$_i$)}. For low-quality spectra, the data did not allow us to constrain the value of the wind absorption component or the properties of a second plasma component. In those cases, we rather adopted a single-temperature model absorbed by the sole interstellar absorption. The results of our fits are listed in Table\,\ref{specX} and the best fits of the EPIC spectra of HD~15558 and HD~15570 are illustrated in Fig.\,\ref{SpecEPIC}.

\begin{sidewaystable*}
\caption{X-ray spectral fits of the O-type stars in IC~1805\label{specX}}
\begin{center}
%\tiny
\begin{tabular}{c c c c c c c c c c c c}
\hline
Object & Inst. &$N_H^{\rm ISM}$ & $\log{N_{\rm wind}}$ & $kT_1$ & norm$_1$ & $kT_2$ & norm$_2$ & $\chi^2_{\nu}$\,(dof) & $f_{\rm X}$ & $f_{\rm X}^{\rm un}$ & $\log{L_{\rm X}/L_{\rm bol}}$\\
\vspace*{-2mm}\\  
     &  & $(10^{22}$\,cm$^{-2}$) &             & (keV)  &  (cm$^{-5}$) & (keV)  & (cm$^{-5}$) & & \multicolumn{2}{c}{($10^{-14}$\,erg\,cm$^{-2}$\,s$^{-1}$)} &\\
\hline
\vspace*{-2mm}\\
BD+60$^{\circ}$~497 & M2 + pn & 0.48 & -- & $0.65^{+.10}_{-.05}$ & $(7.58^{+0.87}_{-1.06})\,10^{-5}$ & -- & --& $0.97 (38)$ & $5.58^{+.25}_{-.27}$ & $15.7^{+1.50}_{-2.00}$ & $-7.03 \pm 0.06$ \\
\vspace*{-2mm}\\
BD+60$^{\circ}$~498 & M2 + pn & 0.44 & -- & $0.54^{+.37}_{-.26}$ & $(9.1^{+20.5}_{-4.9})\,10^{-6}$ & -- & -- & $0.83 (2)$ & $0.65^{+.08}_{-.65}$ & $1.80^{+1.44}_{-0.64}$ & $-7.27 \pm 0.25$\\ 
\vspace*{-2mm}\\
BD+60$^{\circ}$~498 & ACIS & 0.44 & -- & $0.55^{+.31}_{-.25}$ & $(6.4^{+9.7}_{-2.5})\,10^{-6}$ & -- & -- & $0.03 (3)$ & $0.46^{+.04}_{-.12}$ & $1.29^{+0.66}_{-0.46}$ & $-7.41 \pm 0.18$ \\ 
\vspace*{-2mm}\\
BD+60$^{\circ}$~499 & M2 + pn & 0.46 & -- & $0.79^{+.24}_{-.15}$ & $(9.2^{+1.8}_{-1.8})\,10^{-6}$ & -- & -- & $1.78 (7)$ & $0.83^{+.10}_{-.11}$ & $2.00^{+0.40}_{-0.44}$ &$-7.10 \pm 0.11$ \\  
\vspace*{-2mm}\\
BD+60$^{\circ}$~499 & ACIS & 0.46 & -- & $0.62^{+.30}_{-.31}$ & $(1.30^{+2.13}_{-0.45})\,10^{-5}$ & -- & -- & $0.02 (2)$ & $1.01^{+.07}_{-.29}$ & $2.75^{+1.48}_{-0.86}$ & $-6.96 \pm 0.19$ \\  
\vspace*{-2mm}\\
BD+60$^{\circ}$~501 & EPIC & 0.42 & $21.57^{+.15}_{-.23}$ & $0.27^{+.03}_{-.03}$ & $(1.69^{+1.00}_{-0.69})\,10^{-4}$ & $0.97^{+.47}_{-.37}$ & $(2.56^{+0.74}_{-0.91})\,10^{-5}$ & $0.93 (58)$ & $3.39^{+.08}_{-.38}$ & $9.71^{+0.88}_{-0.85}$ & $-6.79 \pm 0.04$\\ 
\vspace*{-2mm}\\
BD+60$^{\circ}$~501 & ACIS & 0.42 & -- & $0.61^{+.14}_{-.26}$ & $(2.26^{+0.56}_{-0.56})\,10^{-5}$ & $1.56^{+.84}_{-.33}$ & $(1.50^{+0.65}_{-0.55})\,10^{-5}$ & $0.75 (22)$ & $3.09^{+.20}_{-.22}$ & $6.76^{+0.84}_{-0.84}$ & $-6.94 \pm 0.06$ \\ 
\vspace*{-2mm}\\
BD+60$^{\circ}$~513 & M2 + pn & 0.45 & -- & $0.71^{+.22}_{-.22}$ & $(0.99^{+0.25}_{-0.27})\,10^{-5}$ & -- & -- & $0.26 (3)$ & $0.86^{+.10}_{-.16}$ & $2.20^{+0.67}_{-0.58}$ & $-7.56 \pm 0.14$\\ 
\vspace*{-2mm}\\
HD~15558 & EPIC & 0.44 & $21.77^{+.04}_{-.05}$ & $0.33^{+.02}_{-.01}$ & $(1.09^{+0.20}_{-0.20})\,10^{-3}$ & $1.24^{+.05}_{-.05}$ & $(3.09^{+0.19}_{-0.19})\,10^{-4}$ & $1.60 (238)$ & $26.96^{+.29}_{-.60}$ & $56.9^{+1.6}_{-1.7}$ & $-6.81 \pm 0.01$ \\ 
\vspace*{-2mm}\\
HD~15558 & ACIS$^*$ & 0.44 & $21.13^{+.28}_{-1.02}$ & $0.79^{+.05}_{-.05}$ & $(1.81^{+0.74}_{-0.54})\,10^{-4}$ & $1.97^{+.17}_{-.16}$ & $(2.50^{+0.22}_{-0.27})\,10^{-4}$ & $1.49 (133)$ & $32.7^{+.40}_{-1.40}$ & $57.3^{+2.2}_{-2.2}$ & $-6.80 \pm 0.02$ \\ 
\vspace*{-2mm}\\
HD~15570 & EPIC & 0.57 & $21.95^{+.04}_{-.04}$ & $0.32^{+.05}_{-.03}$ & $(1.09^{+0.47}_{-0.37})\,10^{-3}$ & $0.96^{+.07}_{-.06}$ & $(2.80^{+0.38}_{-0.37})\,10^{-4}$ & $1.38 (153)$ & $13.91^{+.11}_{-.46}$ & $32.8^{+2.3}_{-2.2}$ & $-7.24 \pm 0.03$ \\ 
\vspace*{-2mm}\\
HD~15570 & ACIS & 0.57 & $21.97^{+.08}_{-.07}$ & $0.42^{+.15}_{-.12}$ & $(7.30^{+11.73}_{-2.92})\,10^{-4}$ & $0.95^{+.25}_{-.12}$ & $(2.69^{+1.27}_{-1.31})\,10^{-4}$ & $1.15 (96)$ & $14.59^{+.65}_{-.64}$ & $30.5^{+4.0}_{-1.4}$ & $-7.28 \pm 0.05$ \\ 
\vspace*{-2mm}\\
HD~15629 & EPIC & 0.41 & $21.49^{+.13}_{-.20}$ & $0.27^{+.01}_{-.02}$ & $(3.56^{+1.48}_{-1.10})\,10^{-4}$ & $0.87^{+.18}_{-.16}$ & $(2.87^{+1.80}_{-1.11})\,10^{-5}$ & $0.97 (95)$ & $6.32^{+.07}_{-.38}$ & $20.3^{+1.2}_{-1.1}$ & $-7.05 \pm 0.03$ \\
\vspace*{-2mm}\\
HD~15629 & ACIS & 0.41 & $21.71^{+.12}_{-.16}$ & $0.29^{+.05}_{-.03}$ & $(6.79^{+5.03}_{-3.00})\,10^{-4}$ & -- & -- & $1.22 (31)$ & $6.80^{+.57}_{-.56}$ & $21.7^{+4.4}_{-3.5}$ & $-7.01 \pm 0.08$  \\
\vspace*{-2mm}\\
\hline
\vspace*{-2mm}\\
Source b & ACIS & $0.39^{+.35}_{-.23}$ & -- & $4.0^{+10.0}_{-1.7}$ & $(3.26^{+0.97}_{-0.76})\,10^{-5}$ & -- & -- & $1.02 (9)$ & $3.86^{+.60}_{-.77}$ & $4.88^{+1.35}_{-1.48}$ \\  
\vspace*{-2mm}\\
\hline
\end{tabular}
\end{center}
\tablefoot{The ACIS-I spectrum of HD~15558 is affected by moderate pile-up. Source b is CXO~023216.25 +613312.0, i.e.\ a secondary source detected with {\it Chandra} at 4\farcs6 from the position of BD+60$^{\circ}$~499. The normalization of the {\tt apec} models corresponds to $\frac{10^{-14}\,\int\,n_e\,n_H\,dV}{d^2}$ where $d$ is the distance of the source (in cm), $n_e$ and $n_H$ are the electron and hydrogen densities of the source (in cm$^{-3}$). The fluxes $f_{\rm X}$ and $f_{\rm X}^{\rm un}$ correspond to the observed and ISM-absorption corrected fluxes in the 0.5 - 10.0\,keV energy band.}
\end{sidewaystable*}
 
Given the sizes of the extraction regions of the {\it XMM-Newton} spectra (see Sect.\,\ref{obsxmm}), we need to consider the possible contamination of the EPIC spectra of the O-type stars by other sources. 

HD~15558 is located in a very crowded region of IC~1805: \citet[][see also references therein]{MA} reported three companions within 10\farcs6 of the star. The brightest of these objects is about 2.8\,mag fainter than HD~15558 in the $z$-band. Ten sources in the {\it Chandra} data \citep{Townsley}, including the O-star binary itself, lie within the EPIC extraction region (radius of 10\arcsec\ around HD~15558). These sources are much weaker than HD~15558, however. The net number of counts of the brightest and second-brightest secondary sources is only 2.5 and 2.0\% of that of HD~15558. All other secondary sources are at least a factor of 3 weaker. 

Two sources in the ACIS data lie close to the position of BD+60$^{\circ}$~499. They are located at 1\farcs1 and 4\farcs6. The former is most likely the X-ray counterpart of the O-star\footnote{We note that these objects are located far off-axis in the ACIS data and the PSF is quite heavily elongated, thereby increasing the uncertainties on the source positions.}, whilst the latter \citep[source b = CXO\,023216.25 +613312.0,][]{Townsley} is probably an unrelated PMS star caught during a flare. Its ACIS spectrum presents a much higher plasma temperature than O-stars usually do (see last line of Table\,\ref{specX}). 

Finally, two secondary sources in the {\it Chandra} data, lie within the {\it XMM-Newton} spectral extraction region of BD+60$^{\circ}$~501 \citep{Townsley}. The brightest has a net number of ACIS counts that amounts to 18\% of that of BD+60$^{\circ}$~501 and could thus moderately contribute to the EPIC spectrum of the O-star. 

We therefore conclude that the EPIC spectra of the O-type stars are only weakly affected by contamination from secondary sources. 

\subsubsection{Comparison between different data sets}
\begin{figure}[thb]
\begin{center}
\resizebox{8cm}{!}{\includegraphics{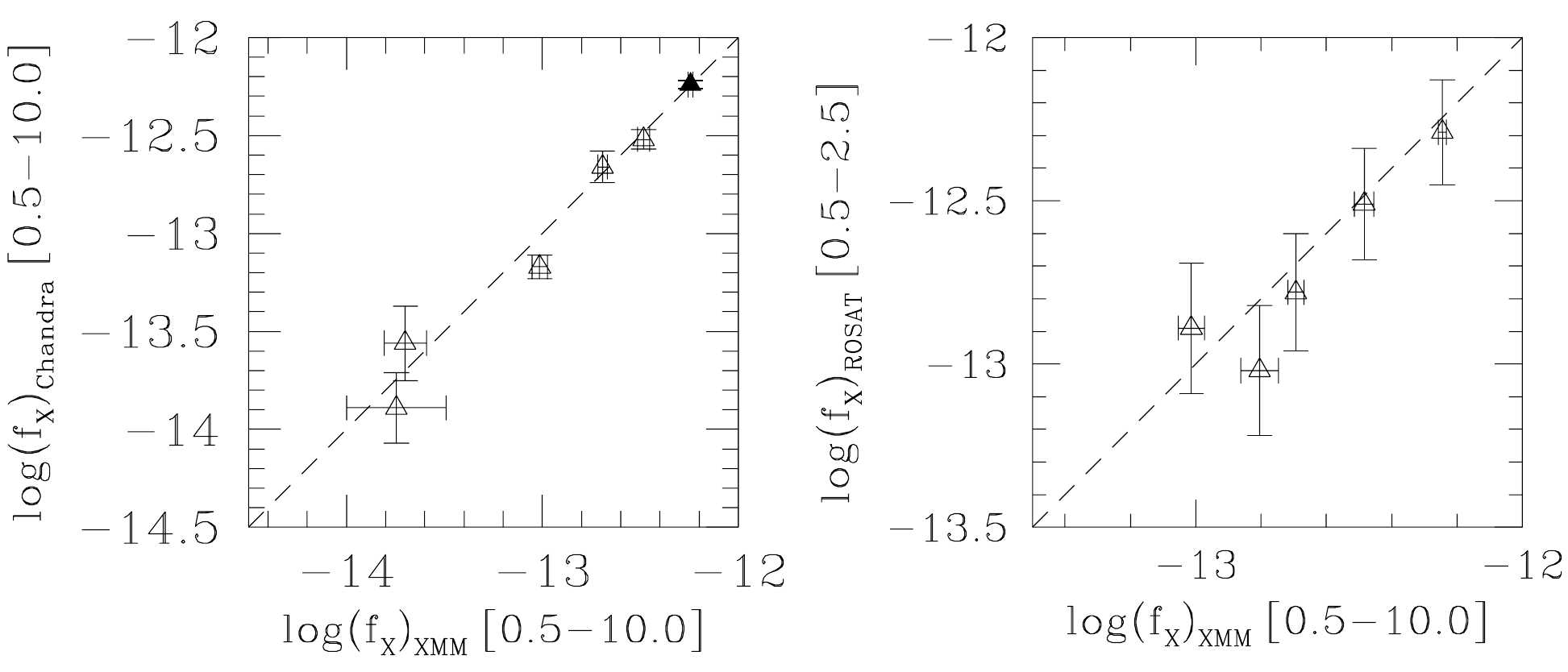}}
\end{center}
\caption{{\it Left}: comparison of the fluxes corrected for ISM absorption of O-type stars in IC~1805 in the 0.5-10\,keV energy band as determined with {\it XMM-Newton} and {\it Chandra}. The filled symbol stands for HD~15558, which is subject to mild pile-up in the {\it Chandra} data. The dashed line corresponds to the one-to-one relation. {\it Right}: comparison of the 0.5-2.5\,keV {\it ROSAT} fluxes of O-type stars versus the 0.5-10\,keV {\it XMM-Newton} fluxes.\label{Xcalibration}}
\end{figure}

Before we correlated the X-ray fluxes with the bolometric fluxes, we compared the absorption-corrected X-ray fluxes inferred from the O-type stars in common between the {\it ROSAT}, {\it Chandra} and {\it XMM-Newton} data. The results are displayed in Fig.\,\ref{Xcalibration}. The left panel shows excellent agreement between the {\it Chandra} and {\it XMM-Newton} data. When we discard HD~15558, which is subject to moderate pile-up in the ACIS data, we find an average deviation (in the sense ACIS $-$ EPIC) of $-0.03$\,dex\footnote{Including HD~15558 in the analysis does not change the conclusion in any significant way}. This result contrasts with the situation found in Cygnus\,OB2 \citep{CygOB2}, where significant differences between the two instruments were found. 
The right panel of Fig.\,\ref{Xcalibration} also indicates a good agreement within errors between the PSPC and EPIC data. The latter comparison also indicates that most of the X-ray emission of the O-type stars in IC~1805 occurs in the soft band.  

Another conclusion that can be drawn from Fig.\,\ref{Xcalibration} is that there are no large flux changes between the epochs of the various X-ray observations. Furthermore, adopting a significance level of 1\% in our variability tests (see Sect.\,\ref{obsxmm}), we found that all O-stars were compatible with a constant emission during the {\it XMM-Newton} exposure. Overall, the X-ray emission of the majority of the O-stars in IC~1805 seems constant on short and long timescales. 

\begin{figure*}[htb]
\begin{minipage}{8.5cm}
\begin{center}
\resizebox{8.5cm}{!}{\includegraphics{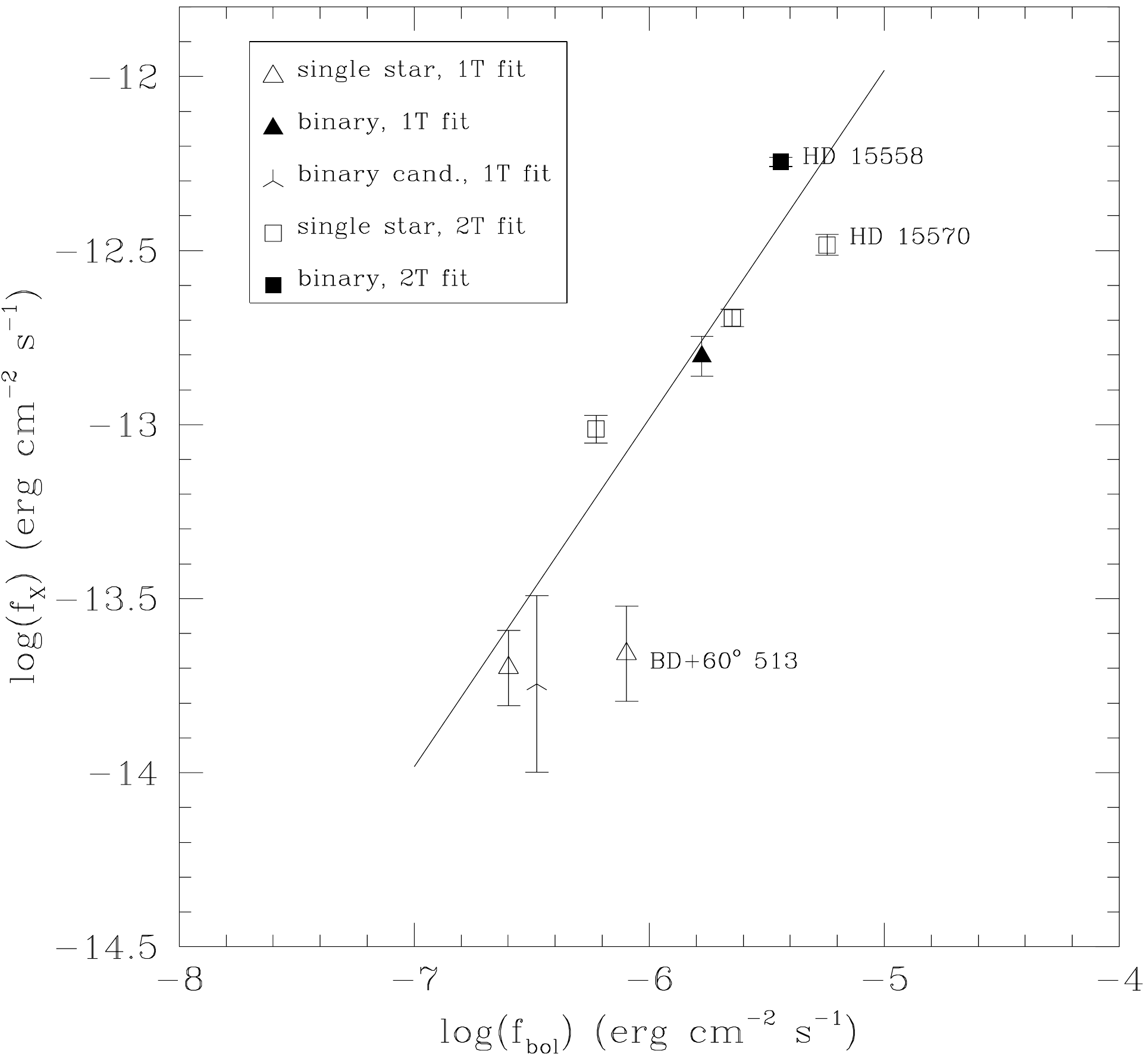}}
\end{center}
\end{minipage}
\hfill
\begin{minipage}{8.5cm}
\begin{center}
\resizebox{8.5cm}{!}{\includegraphics{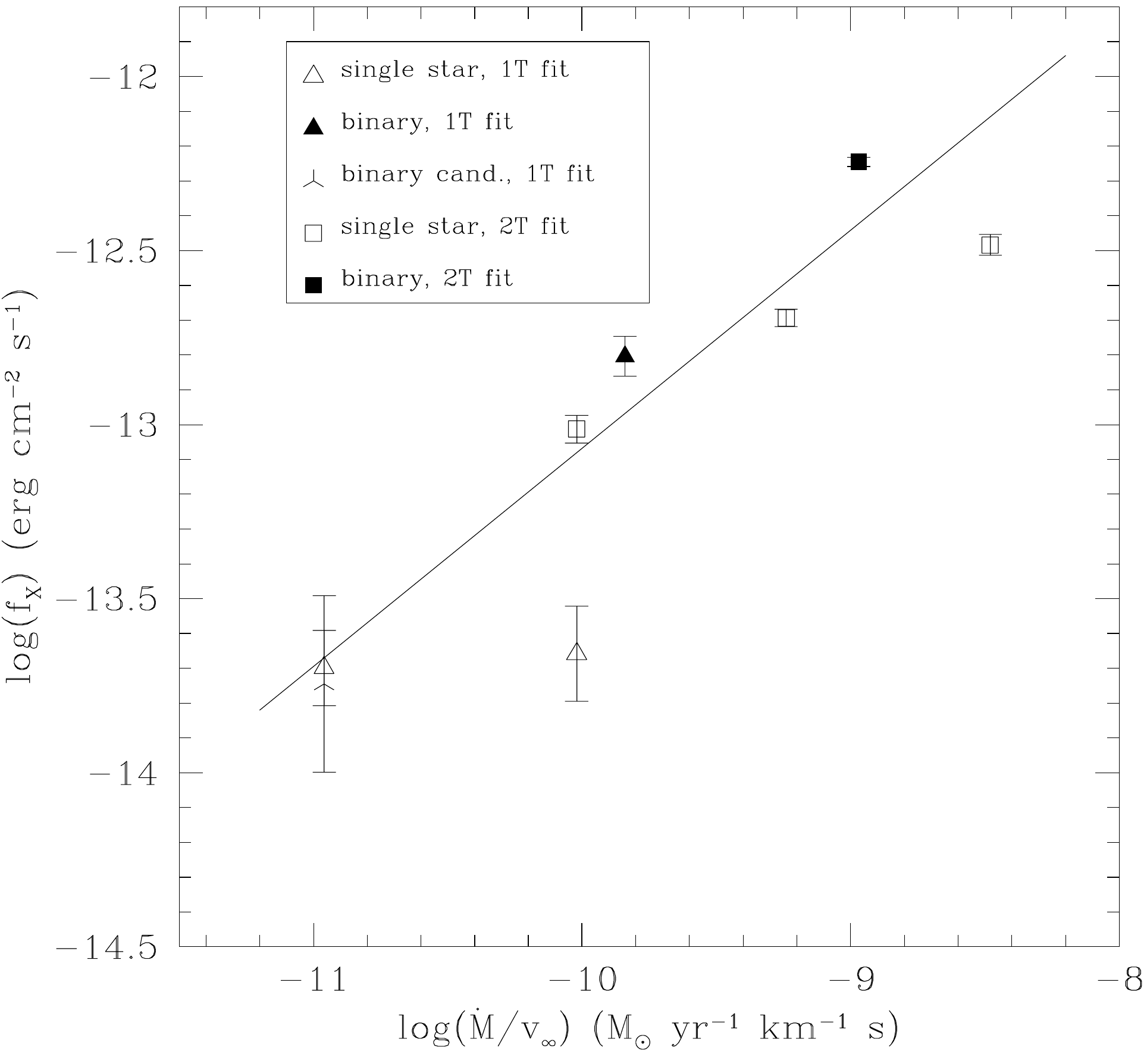}}
\end{center}
\end{minipage}
\caption{{\it Left}: logarithm of the X-ray fluxes corrected for ISM absorption of O-type stars in IC~1805 as a function of the logarithm of the bolometric fluxes. The solid line yields the best-fit scaling relation (Eq.\,\ref{eq1}) between the X-ray and bolometric fluxes considering the full sample of O-type stars, i.e.\ including binaries as well as single stars. {\it Right}: logarithm of the X-ray fluxes corrected for ISM absorption of O-type stars in IC~1805 as a function of $\log{\frac{\dot{M}}{v_{\infty}}}$. The solid lines yield the theoretical relation $\log{f_{\rm X}} \propto (1 - m)\,\log{(\frac{\dot{M}}{v_{\infty}})}$ \citep{Owocki} with $m = 0.37$ (Eq.\,\ref{eq2}). 
\label{fxvsfbol}}
\end{figure*}

\subsubsection{$L_{\rm X}/L_{\rm bol}$ relation of O-type stars}
To build the $L_{\rm X}/L_{\rm bol}$ relation of the O-type stars in IC~1805, we first need to establish their bolometric fluxes. For this purpose, we adopted the $B$ and $V$ magnitudes from \citet{SL}. The reddening $A_V$ was obtained assuming $R_V = 2.9$ \citep{SL} and using the intrinsic colours as well as bolometric corrections as a function of spectral type given by \citet{MP}. 
For the two established binary systems, HD~15558 and BD+60$^{\circ}$~497, we took the contributions from each star into account. \cite{DeBecker06} estimated that the primary of HD~15558 accounts for about 5/6 of the bolometric luminosity of the system. For BD+60$^{\circ}$~497, the studies of \citet{RDB} and \citet{Hillwig} agree about an optical brightness ratio of $\sim 0.35$ between the secondary and primary stars, and the bolometric flux was computed accordingly. Finally, for the binary candidate BD+60$^{\circ}$~498, we currently lack information on the secondary star's spectral type. For this system, we therefore assumed the bolometric correction corresponding to its combined spectral type (see Sect.\,\ref{BD498}).
    
The X-ray fluxes corrected for ISM absorption found on our {\it XMM-Newton} data are plotted as a function of the bolometric fluxes in Fig.\,\ref{fxvsfbol}. 

In some massive binary systems, colliding stellar winds provide an excess of X-ray emission in addition to the intrinsic emission generated in the winds of the individual binary components \citep[e.g.][and references therein]{GRYN}. The amount of additional X-ray emission varies tremendously from one object to the other, however, and in recent years it was found that the vast majority of massive binaries are not X-ray overluminous \citep[e.g.][]{Oskinova05,Carina,HM1,CygOB2}. From our current best knowledge of multiplicity among O-stars in IC~1805 (see Sect.\,\ref{overview}), we expect only three objects (BD+60$^{\circ}$~497, BD+60$^{\circ}$~498, and HD~15558) to host potential X-ray emission from colliding winds. 

Whilst HD~15558 is indeed the X-ray brightest object in the cluster, neither its X-ray luminosity nor its plasma temperature appear exceptional. In a wide eccentric binary such as HD~15558, the wind interaction region is likely in the adiabatic regime \citep{SBP}. As a result, the X-ray emission from the colliding wind interaction is expected to scale with the inverse of the orbital separation. In the present case, with $e = 0.42$, we expect a variation by a factor 2.4 between apastron and periastron. Our {\it XMM-Newton} observation ($\phi = 0.46$) and the {\it ROSAT} observation ($\phi = 0.42$) were both obtained near apastron when the orbital separation is largest, hence the contribution of the colliding winds should be close to its minimum value. However, the {\it Chandra} observation was taken at orbital phase $\phi = 0.11$. In principle, we might therefore obtain a rough estimate of the emission level from colliding winds by considering the variations of the X-ray flux between the {\it Chandra} and {\it XMM-Newton} observation. Unfortunately, the ACIS data are affected by pile-up, which renders such a comparison quite uncertain. At best, we note from our spectral fits that HD~15558 appears indeed brighter during the ACIS observation, but only slightly so (by 21\% in observed flux, and by less than 1\% in ISM-corrected flux), certainly not by a factor 2. Concerning the other two binaries, we note that none of them appears X-ray overluminous compared to the general trend in Fig.\,\ref{fxvsfbol}. We therefore conclude that to first order, X-ray emission from colliding winds probably does not play a major role in the properties of the X-ray emission of O-stars in IC~1805.   

Using the X-ray fluxes corrected for ISM absorption obtained from our {\it XMM-Newton} data and the estimates of the bolometric fluxes, we have fitted a simple least-squares scaling relation: 
\begin{equation}
\log{L_{\rm X}/L_{\rm bol}} = -6.98 \pm 0.20
\label{eq1}
\end{equation}

Both the average value of $\log{L_{\rm X}/L_{\rm bol}}$ and its dispersion are typical of samples of O-type stars in massive young open clusters \citep[e.g.][]{NGC6231,YN,Carina,CygOB2}. We then investigated the correlation between $f_{\rm X}$ and $\frac{\dot{M}}{v_{\infty}}$. The quantity $\frac{\dot{M}}{v_{\infty}}$ provides a measure of the mean wind density, and the cooling length of the hot shocked plasma is inversely proportional to this parameter \citep{Hillier}. Therefore, it seems quite natural that this quantity should play a role in the X-ray emission. Analysing a sample of 42 {\it ROSAT}-PSPC spectra of O-type stars, \citet{Kudritzki} empirically derived a relation $L_{\rm X} \propto \left(\frac{\dot{M}}{v_{\infty}}\right)^{-0.38}\,L_{\rm bol}^{1.34}$. Accounting for the dependence of $\frac{\dot{M}}{v_{\infty}}$ on $L_{\rm bol}$ \citep[see Eq.\ 32 of][]{Owocki}, the \citet{Kudritzki} result leads to $L_{\rm X} \propto \left(\frac{\dot{M}}{v_{\infty}}\right)^{0.67}$. Based on theoretical considerations, \citet{Owocki} predicted that the X-ray luminosity scales as $\left(\frac{\dot{M}}{v_{\infty}}\right)^{1-m}$ with $m$ in the range $[0.2,0.4]$. To check whether such a relation applies to our data, we need to estimate the wind terminal velocities and the mass-loss rates. For the former, we adopted the ratio $v_{\infty}/v_{\rm esc} = 2.6$ along with the escape velocities quoted by \citet{Muijres} for the spectral types of the stars in our sample. For the mass-loss rates, we used the values given by \citet{Muijres} as a function of spectral type, which correspond to the formalism of \citet{Vink} with $v_{\infty}/v_{\rm esc} = 2.6$. For the two double-lined spectroscopic binaries, we considered the spectral type of the primary star, while we used the combined spectral type for the binary candidate BD+60$^{\circ}$~498 (see Sect.\,\ref{overview}).  
A least-squares fit to the full sample of O-stars in IC~1805 then yields  
\begin{equation}
\log{f_{\rm X}} = (0.63 \pm 0.21)\,\log{\left(\frac{\dot{M}}{v_{\infty}}\right)_{\rm theor}}-6.81 \pm 2.10
\label{eq2}
\end{equation}

We have to stress that the values of the mass-loss rates and terminal velocities that we use here are theoretical values for a given spectral type that do not account for the effect of clumping. When we consider the two stars in our sample (HD~15570 and HD~15629) for which dedicated model atmosphere fits accounting for clumping are available in the literature  \citep{Bouret,Mimes}, we find that these clumping-corrected mass-loss rates are lower by a factor 2.9 to 4.6. When we assume, however, that the clumping correction to be applied to the \citet{Vink} mass-loss rates is essentially independent of spectral type, the overall trend in the right panel of Fig.\,\ref{fxvsfbol} should be correct. Relation\,(\ref{eq2}) nicely agrees with the empirical relation found by \citet{Kudritzki}. Our best-fit $m$ value ($0.37 \pm 0.21$) is well inside the range predicted by \citet{Owocki}. It is lower than what we found for the Cyg~OB2 stars \citep[$0.52 \pm 0.10$,][]{CygOB2}, although both values overlap within their errors. 

\subsubsection{A break in the $L_{\rm X}/L_{\rm bol}$ relation for O\,If$^+$ stars?}
The two data points that deviate most from relation\,(\ref{eq2}) are BD+60$^{\circ}$~513 at the faint end, and to a lesser extent, HD~15570 at the brighter end. If we discard these two stars, the best-fit relation becomes 
\begin{equation}
\log{f_{\rm X}} = (0.71 \pm 0.14)\,\log{\left(\frac{\dot{M}}{v_{\infty}}\right)_{\rm theor}}-5.92 \pm 1.40
\label{eq3}
\end{equation} 

The coefficients in Eqs.\,(\ref{eq2}) and (\ref{eq3}) overlap within the errors. BD+60$^{\circ}$~513, which has the highest projected rotational velocity among the O-type stars in IC~1805, is also quite faint in its $L_{\rm X}/L_{\rm bol}$ ratio. On the other hand, HD~15570, which has $\log{L_{\rm X}/L_{\rm bol}} = -7.24 \pm 0.03$, is indeed the O4\,If$^+$ star we are interested in. Does this result imply that O\,If$^+$ stars, and early-O\,If supergiants in general, are indeed X-ray faint? To answer this question, we first considered other studies of luminous O-supergiants in Galactic young open clusters. 

\citet{Carina} studied the $L_{\rm X}/L_{\rm bol}$ relation for massive stars in the Carina complex. If we focus on the earliest supergiants in their sample, we find $\log{L_{\rm X}/L_{\rm bol}} = -6.97$ for HD~93129A (O2\,If$^*$ + O3.5\,V) and $-7.21$ for Cl* Trumpler~14 MJ~257 (O3/4\,If). Generally speaking, presumably single O-stars with spectral types O2 -- O5.5 in the Carina Nebula were found to display $\log{L_{\rm X}/L_{\rm bol}} = -7.25 \pm 0.22$. Hence, there was no evidence for a significant underluminosity of the earliest O-supergiants in this cluster.  
 
\citet{HM1} studied the X-ray emission of the massive star population of the open cluster HM~1, which harbours two early-O\,If$^+$ stars, VB~4 and VB~5, which are of spectral types O4\,If$^+$ and O5\,If$^+$, respectively. Both were found to have rather low $\log{L_{\rm X}/L_{\rm bol}}$ values of $-7.4$ and $-7.3$ for VB~4 and VB~5, respectively. However, HM~1 is subject to very heavy interstellar absorption, and \citet{HM1} noted that, given the limited signal-to-noise ratio of their EPIC spectra, this large absorption could lead to an overestimate of the plasma temperature and accordingly to an underestimate of the X-ray flux for these and other stars of the same cluster. Therefore, although this result indicates a relative X-ray underluminosity of O\,If$^+$, it is not sufficient to conclude that this effect is indeed real. 

Finally, \citet{CygOB2} reported $\log{L_{\rm X}/L_{\rm bol}} = -7.03$ for the O3\,If$^*$ star Cyg~OB2 \#7, whilst the full sample of O-type stars in Cyg~OB2 (excluding the very bright, overluminous colliding wind systems Cyg~OB2 \#5, 8a and 9) yields $\log{L_{\rm X}/L_{\rm bol}} = -7.18 \pm 0.21$. Hence, the O supergiant is fully consistent with the overall relation.\\ 

Returning to the specific case of IC~1805, we examined the possible causes for the deviations seen in Fig.\,\ref{fxvsfbol}. For this purpose, we considered the recent work of \citet{Gayley}, who proposed a first attempt to estimate the X-ray emergence efficiency from the winds of all types of non-magnetic single massive stars. The emergence efficiency $\eta_{\rm X}$ is defined as the ratio between the emerging X-ray luminosity and the total excess turbulent energy flux injected into the wind. Based on first principles, \cite{Gayley} provided a simple expression of $\eta_{\rm X}$ (his Eq.\ 11) as a function of the ratio between the velocity jump needed to account for the mean X-ray photon energy and the wind terminal velocity, on the one hand, and the wind optical depth $\tau(L)$ at some specific distance $L$ in the wind\footnote{This distance scale parameter is the correlation length between the fast wind and the slower clumps. It also corresponds to the porosity length of the wind \citep{Gayley}.}, on the other hand. The latter parameter is a proxy for the mass-loss rate. We solved Eq.\ 11 of \citet{Gayley} assuming $m = 0.37$ (see Eq.\,\ref{eq2}) to estimate $\eta_{\rm X}$ as a function of $\tau(L)$. We used these results to compare the emergence efficiency for 1\,keV photons in HD~15629 ($\log{L_{\rm X}/L_{\rm bol}} = -7.05$) and HD~15570 ($\log{L_{\rm X}/L_{\rm bol}} = -7.24$). Both stars are most likely single (see Sect.\,\ref{overview}), and their stellar and wind parameters have been determined with the same model atmosphere code \citep{Martins05,Bouret}. Assuming $L = 1\,R_*$ \citep{Gayley}, we can then estimate values of $\tau(L) \sim 0.34$ for HD~15629 and $\sim 1.66$ for HD~15570. These values translate into X-ray emergence efficiencies of 35\% and 16\%, respectively. Hence, we would expect a 2.2 times (0.34\,dex) lower efficiency for the denser wind of HD~15570. Similar conclusions are obtained when we consider the mean photon energies of the {\tt apec} models in Table\,\ref{specX} instead of a photon energy of 1\,keV. Hence, the mild X-ray underluminosity of HD~15570 can very likely be traced to its specific stellar and wind parameters.  

\begin{sidewaystable*}
\caption{Spectral fits of other bright X-ray sources \label{addsrcfit}}
\tiny
\begin{center}
\begin{tabular}{c c c c c c c c c c c c c}
\hline
Name & XID & \multicolumn{5}{c}{\tt tbabs$\times$apec} & &\multicolumn{5}{c}{\tt tbabs$\times$pow} \\
\cline{3-7} \cline{9-13} \\
     &    & $N_H$ & $kT$ (keV) & norm  & $\chi^2$(dof) & $f_{\rm X}$  & & $N_H$ & $\Gamma$ & norm & $\chi^2$\,(dof) & $f_{\rm X}$  \\
& & (10$^{22}$\,cm$^{-2}$) & (keV) & (10$^{-4}$\,cm$^{-5}$) & & (10$^{-13}$\,erg\,cm$^{-2}$\,s$^{-1}$) & & (10$^{22}$\,cm$^{-2}$) & &  & & (10$^{-13}$\,erg\,cm$^{-2}$\,s$^{-1}$)\\
\hline
XMMU~J023200.4+612038 & 3    & 1.35$\pm$0.16 & 12.3$\pm$4.1  & 4.31$\pm$0.20 & 1.09(211)& 5.84$\pm$0.42 & & 1.84$\pm$0.23 & 1.86$\pm$0.14 & 1.76$\pm$0.38 & 0.95(211)& 5.51$\pm$0.34  \\
CXOU~023200.45+612039.0 &                         & 1.19$\pm$0.25 & $>$10.5       & 3.03$\pm$0.54 & 1.18(118)& 4.19$\pm$0.33 & & 1.47$\pm$0.28 & 1.59$\pm$0.22 & 0.80$\pm$0.27 & 1.13(118)& 3.98$\pm$0.40  \\
\vspace*{-2mm}\\
XMMU~J023230.2+611747 & 5    & 0.07$\pm$0.02 & 0.64$\pm$0.15 & 0.20$\pm$0.03 & 1.20(80) & 0.92$\pm$0.06 & &  &  &  & &   \\
                      &      &               &+1.65$\pm$0.26 & 0.48$\pm$0.06 &          &               & &  &  &  & &   \\
CXOU~023230.26+611748.6 &                         & 0.73$\pm$0.18 & 0.33$\pm$0.08 & 3.96$\pm$3.15 & 1.20(72) & 1.26$\pm$0.40 & & 0.39$\pm$0.12 & 3.81$\pm$0.41 & 1.09$\pm$0.31 & 1.26(74) & 1.24$\pm$0.10 \\
                      &      &               &+2.26$\pm$1.08 & 0.70$\pm$0.17 &          &               & & &  &  & &   \\
\vspace*{-2mm}\\
XMMU~J023407.5+612817 & 8    & 0.02$\pm$0.02 & 0.26$\pm$0.02 & 0.18$\pm$0.04 & 1.31(54) & 0.39$\pm$0.03 &  & &  &  & &   \\
                      &      &               &+1.20$\pm$0.17 & 0.13$\pm$0.03 &          &               &  & &  &  & &   \\
\vspace*{-2mm}\\
XMMU~J023300.0+612503 & 9    & 0.54$\pm$0.22 & 3.34$\pm$1.07 & 0.39$\pm$0.78 & 0.91(37) & 0.40$\pm$0.05 & & 0.86$\pm$0.27 & 2.48$\pm$0.39 & 0.23$\pm$0.11 & 0.86(37) & 0.39$\pm$0.06  \\
\vspace*{-2mm}\\
XMMU~J023237.2+612810 & 10   & 0.29$\pm$0.14 & 1.85$\pm$0.27 & 0.22$\pm$0.03 & 1.06(27) & 0.20$\pm$0.06 & & 0.82$\pm$0.25 & 3.63$\pm$0.76 & 0.27$\pm$0.18 & 1.07(27) & 0.19$\pm$0.03  \\
CXOU~023237.01+612811.9 &                         & 1.45$\pm$0.33 & 0.81$\pm$0.21 & 0.36$\pm$0.23 & 0.77(9)  & 0.10$\pm$0.04 & & 0.74$\pm$0.49 & 3.35$\pm$1.10 & 0.14$\pm$0.17 & 1.15(9)  & 0.12$\pm$0.05  \\
\vspace*{-2mm}\\
XMMU~J023208.0+612203 & 11   &               &               &               &          &               & & 0.78$\pm$0.28 & 5.44$\pm$1.30 & 0.44$\pm$0.34 & 1.17(32) & 0.19$\pm$0.04  \\
CXOU~023208.03+612203.6 &                         & 0.90$\pm$0.17 & 0.75$\pm$0.12 & 0.41$\pm$0.16 & 1.13(23) & 0.19$\pm$0.03 & & 0.44$\pm$0.28 & 3.63$\pm$0.99 & 0.21$\pm$0.17 & 1.05(23) & 0.23$\pm$0.04  \\
\vspace*{-2mm}\\
XMMU~J023245.8+612959 & 12   & 0.31$\pm$0.19 & 3.69$\pm$2.26 & 0.19$\pm$0.04 & 0.57(20) & 0.22$\pm$0.05 & & 0.50$\pm$0.29 & 2.19$\pm$0.53 &0.081$\pm$0.052& 0.66(20) & 0.23$\pm$0.05  \\
\vspace*{-2mm}\\
XMMU~J023155.3+612249 & 13   & 2.39$\pm$2.67 & $>3.75$       & 0.61$\pm$0.21 & 1.19(11) & 0.75$\pm$0.65 & & 2.35$\pm$2.75 & 1.33$\pm$0.86 & 0.11$\pm$0.34 & 1.19(11) & 0.77$\pm$0.10  \\
\vspace*{-2mm}\\
XMMU~J023300.9+613737 & 14   & 0.34$\pm$0.12 & 5.07$\pm$3.15 & 0.43$\pm$0.06 & 1.06(33) & 0.58$\pm$0.12 & & 0.49$\pm$0.19 & 1.97$\pm$0.38 & 0.16$\pm$0.07 & 1.07(33) & 0.59$\pm$0.11  \\
\vspace*{-2mm}\\
XMMU~J023243.0+613059 & 15   & 0.33$\pm$0.12 & 3.12$\pm$0.87 & 0.28$\pm$0.04 & 1.02(35) & 0.31$\pm$0.05 & & 0.59$\pm$0.19 & 2.47$\pm$0.36 & 0.15$\pm$0.06 & 1.02(35) & 0.31$\pm$0.04  \\
\vspace*{-2mm}\\
XMMU~J023240.7+612801 & 16   & 0.24$\pm$0.14 & 2.50$\pm$1.29 & 0.15$\pm$0.03 & 1.13(15) & 0.16$\pm$0.03 & & 0.52$\pm$0.25 & 2.73$\pm$0.73 &0.097$\pm$0.069& 1.05(15) & 0.16$\pm$0.04  \\
CXOU~023240.76+612800.2 &                         & 1.58$\pm$0.70 & 1.56$\pm$0.83 & 1.21$\pm$0.52 & 1.01(14) & 0.53$\pm$0.13 & & 1.05$\pm$0.69 & 2.64$\pm$0.64 & 0.44$\pm$0.52 & 1.35(14) & 0.61$\pm$0.17  \\
\vspace*{-2mm}\\
XMMU~J023209.2+613039 & 17   & 0.76$\pm$0.53 & 2.69$\pm$1.31 & 0.46$\pm$0.18 & 1.21(17) & 0.38$\pm$0.09 & & 1.02$\pm$0.49 & 2.60$\pm$0.59 & 0.27$\pm$0.22 & 1.29(17) & 0.39$\pm$0.11  \\
\vspace*{-2mm}\\
XMMU~J023243.3+612803 & 18   & 0.33$\pm$0.13 & 2.58$\pm$0.93 & 0.22$\pm$0.03 & 0.69(24) & 0.22$\pm$0.04 & & 0.60$\pm$0.22 & 2.59$\pm$0.49 & 0.13$\pm$0.07 & 0.79(24) & 0.23$\pm$0.04  \\
\vspace*{-2mm}\\
XMMU~J023243.7+612631 & 19   & 0.44$\pm$0.44 & 2.78$\pm$1.86 & 0.31$\pm$0.12 & 0.36(10) & 0.30$\pm$0.07 & & 0.70$\pm$0.43 & 2.55$\pm$0.67 & 0.18$\pm$0.16 & 0.39(10) & 0.31$\pm$0.09  \\
CXOU~023243.54+612632.0 &                         & 1.34$\pm$0.85 & 1.48$\pm$1.35 & 0.69$\pm$0.47 & 1.07(11) & 0.32$\pm$0.12 & & 0.95$\pm$0.59 & 2.72$\pm$0.77 & 0.29$\pm$0.33 & 1.63(11) & 0.38$\pm$0.14  \\
\vspace*{-2mm}\\
XMMU~J023317.6+612808 & 20   & 0.29$\pm$0.17 & 2.91$\pm$1.19 & 0.18$\pm$0.04 & 1.27(18) & 0.20$\pm$0.04 & & 0.67$\pm$0.30 & 2.79$\pm$0.63 & 0.13$\pm$0.09 & 1.08(18) & 0.19$\pm$0.04  \\
\vspace*{-2mm}\\
XMMU~J023254.0+612803 & 21   & 0.35$\pm$0.56 & 3.93$\pm$4.40 & 0.15$\pm$0.08 & 0.98(11) & 0.18$\pm$0.05 & & 0.56$\pm$0.58 & 2.20$\pm$0.72 &0.066$\pm$0.082& 0.98(11) & 0.18$\pm$0.06  \\
\vspace*{-2mm}\\
XMMU~J023300.9+612608 & 22   & 0.38$\pm$0.21 & 3.21$\pm$1.82 & 0.18$\pm$0.04 & 1.16(15) & 0.20$\pm$0.04 & & 0.63$\pm$0.17 & 2.43$\pm$0.60 &0.097$\pm$0.068& 1.19(15) & 0.20$\pm$0.05  \\
\hline
\end{tabular}
\end{center}
\tablefoot{For each source, all existing EPIC spectra were fitted simultaneously. When available, the results of the associated {\it Chandra} source are presented immediately below the best-fit to the {\it XMM-Newton} data. Errors correspond to 90\% confidence intervals (derived from the {\sc error} command for parameters and {\sc flux err} for fluxes) - when asymmetric, the larger error is indicated. The quoted fluxes are the observed values in the 0.5--10.\,keV energy band. The normalization of the power-law models is expressed as the number of photons\,keV$^{-1}$\,cm$^{-2}$\,s$^{-1}$ at an energy of 1\,keV, whilst the normalization of the {\tt apec} models is presented in Table\,\ref{specX}.}
\end{sidewaystable*}

\subsection{B- and A-type stars \label{Bstars}}
Of the 191 sources detected in the EPIC field of view, 13 have an optical counterpart of spectral type B, three are classified as A-type stars, and two as Herbig Ae/Be star. Spectroscopic surveys of A- and B-type stars in IC~1805 are probably far from complete. \citet{SH99}, who provide one of the most extensive spectroscopic surveys of this cluster, list 78 B-type stars in IC~1805, 49 of which lie inside our EPIC field of view. Of these 49 B-stars, 37 have a spectral type earlier than or equal to B4, whilst the remaining 12 have later spectral types. Of the 13 X-ray sources that have a B-type optical counterpart, 10 have been classified by \citet{SH99}, 9 of them are early B-type stars, whilst only 1 has a spectral type later than B4. We thus have detection rates of $\sim 24$\% for B0 - B4 stars and $\sim 8$\% for B5 - B9.5 stars. This situation is very much reminiscent of that reported by \citet{NGC6231} for the NGC~6231 cluster: a detection rate of $\sim 24$\% for B0 - B4 stars (15 objects out of 63) and a significant drop (to $\sim 7$\%, 2 objects out of 28) at later subtypes. A higher detection efficiency of 50\% (54 objects out of 108) for B-type stars up to spectral type B5 (included) was achieved in the {\it Chandra} legacy survey of Cyg~OB2 \citep{CygOB2}. This could be due to the higher sensitivity of {\it Chandra} for faint point-like sources.\\  

Three of the EPIC sources with probable A- or B-type optical counterparts were sufficiently bright to produce at least 400 EPIC counts, and we therefore extracted and analysed their spectra (see Sect.\,\ref{obsxmm}). The spectra were fitted within {\tt xspec} with either absorbed optically thin thermal plasma {\tt apec} models or absorbed power-law spectra. The results of these fits are quoted in Table \ref{addsrcfit} along with the spectral fits of other bright sources in the field of view that are discussed in Sect.\,\ref{other}. 

XMMU~J023240.7+612801 (XID~16) most probably corresponds to IC~1805 143, an early B-star with literature spectral classifications of B1\,V \citep{Massey}, B0.5\,V \citep{SH99}, or B0.3\,V \citep{HG}. The best EPIC position is at 1\farcs4 from CXOU~023240.76+612800.2. Although this separation is relatively large for a source located within 1\arcmin\ of the {\it Chandra} aimpoint, their association is highly likely. A second {\it Chandra} source (CXOU~023240.79+612758.5) lies in the {\it XMM-Newton} extraction region, but it is twice as distant and more than 50 times fainter. The absorption and X-ray flux of the target were larger during the {\it Chandra} observations: in the power-law fits, absorption doubled and flux quadrupled, while the slope did not change significantly\footnote{The usual degeneracy of the fits between a highly absorbed warm plasma and a less absorbed hot plasma prohibits clear conclusions from the thermal fits regarding changes of the absorption.}. 

The optical counterpart of XMMU~J023243.3+612803 (XID~18) is classified as a late B-type star in \citet{sun16} and as an A6 star by \citet{Wolff}. The EPIC spectra of this object are well fitted by a moderate absorption, close to the typical value for the IC~1805 cluster, and a rather high temperature. The {\it Chandra} source CXOU~023243.21+612804.5 lies at 1\farcs3 distance, which is a non-negligible separation especially since the {\it Chandra} source was not too far off-axis ($<$1\arcmin). Nevertheless, if both X-ray sources are associated, then the X-ray flux appears to have quadrupled in the {\it XMM-Newton} observation.

XMMU~J023243.7+612631 (XID~19) corresponds to IC~1805 152 classified as B3\,V by \citet{SH99} and quoted with spectral type B2.5V by \citet{sun16}. Three {\it Chandra} sources are located within the extraction region of XID~19: CXOU~023243.54+612632.0 (at 1\farcs6), CXOU~023243.39+612635.0 (at 4\farcs2), and CXOU~023243.54+612636.8 (at 5\farcs3). While the separation is non-negligible, the {\it XMM-Newton} source probably corresponds to the former source, which is 30 times brighter than each of the latter two objects. The X-ray properties (flux, absorption slightly larger than that of the cluster, moderate $\Gamma$) appear to have remained relatively constant, within the errors, between the two observations.\\
% \citet{Vasilevskis} and \citet{Sanders} quote membership probabilities of 86 and 70\%, respectively. 

A- and B-type stars are usually not expected to be strong intrinsic X-ray emitters. Except for the earliest spectral types, they lack the strong winds that produce the shocks responsible for the X-ray emission of O-type stars. At the same time, except for the latest spectral types, these stars also lack the convective outer layer where the interplay between convection and rotation leads to a dynamo mechanism able to sustain an X-ray bright corona. Observations indeed confirm that A- and B-type stars are generally X-ray dark: based on data from the {\it ROSAT} All-Sky Survey, \cite{Astars} found that stars of spectral type B1 - A9 have low, but non-zero, X-ray detection rates, roughly between 10 and 15\%. The origin of the X-ray emission of the detected sources is not clear: it could either be intrinsic to the AB stars or arise from unseen late-type companions. A similar situation holds for Herbig Ae/Be stars, where the actual origin of the X-ray emission (intrinsic to the star, produced in a magnetic star/disk interaction, or due to an unresolved late-type pre-main-sequence companion) remains a puzzle \citep{Stelzer}. However, the high incidence of binary companions \citep{Duchene} in Herbig Ae/Be stars suggests that their X-ray emission might be associated, at least to some extent, with unseen low-mass companions. 

One of the most promising scenarios for an intrinsic emission of AB stars is the magnetically confined wind shock (MCWS) model \citep{BabelMontmerle,ud-Doula}. \citet{Robrade} discussed the X-ray emission of magnetic Ap/Bp stars. This author notably found that X-ray detections are more frequent among the earliest and most luminous stars in his sample, although there is a large scatter in X-ray luminosity for a given bolometric luminosity. This situation favours an intrinsic origin of the X-ray emission of Ap/Bp stars, possibly related to the MCWS scenario as the more luminous stars have stronger winds that could more easily lead to a significant X-ray emission when interacting with the magnetic field. However, a strong (kilo-Gauss) magnetic field does not provide a sufficient condition for an intrinsic X-ray emission. In a sample of ten magnetic late-B and early-A stars, \cite{CS} failed to detect X-rays for four stars. Another argument against magnetic fields as the main explanation for the X-ray emission of B-type stars comes from the incidence rate of magnetic fields among OB stars. Sensitive spectropolarimetric surveys have  established this rate to be about 6--8\% \citep{wad13,fos15}, which is well below the X-ray detection rate. Hence, whilst magnetically confined winds could indeed explain a few percent of the X-ray detections of early B-type stars, they certainly cannot account for the full set of detections.\\

The three sources for which we were able to extract EPIC spectra display quite hard spectra, consistent with the generally high plasma temperatures found for low-mass pre-main-sequence (PMS) stars (see Sect.\,\ref{other}). Moreover, the hardening and flux increase of XID~16 between the {\it Chandra} and {\it XMM-Newton} observations may indicate a flaring event. These characteristics thus favour an interpretation of the X-ray emission arising from a late-type PMS companion and not from the AB star itself. 

\subsection{Other bright X-ray sources in the field of view \label{other}}
As explained in Sect.\,\ref{obsxmm}, we extracted EPIC spectra for 17 sources that are not O-type stars but were bright enough (they had at least 400 EPIC counts) to perform a (rough) spectral analysis. A cross-correlation with the {\it Chandra} source list of \citet{Townsley} was made for these sources, with an association considered secure whenever the separation was smaller than 1\arcsec. For those targets for which \citet{Townsley} reported at least 100 ACIS net counts (in the 0.5--8.0\,keV energy band), we also extracted the ACIS spectra (see Sect.\ref{obsacis}). The spectral fitting was made within {\tt xspec} considering optically thin thermal plasma {\tt apec} models and the possibility of non-thermal X-ray emission (represented by a power-law spectrum). Table \ref{addsrcfit} lists the results of both types of fits, and we now briefly examine each source in turn, except for the three sources that have been discussed in Sect.\,\ref{Bstars}.\\ 

The third brightest X-ray source in the field of view (XID~3) is XMMU~J023200.4+612038, which corresponds to CXOU~023200.45+612039.0 \citep{Townsley}, probably a young stellar object (YSO) according to \citet{bro13}. \citet{sun16} classify this object as a class I YSO. However, since most YSOs in IC~1805 are rather of class II, the authors caution that it is more likely for this source to be of a different nature, for example extragalactic. This possibility is indeed supported by our X-ray analysis. The best-fit absorbing column exceeds that of the cluster, and the fit by a thermal emission component yields a rather high temperature ($>$8\,keV), while the power-law model has a $\Gamma\sim2$ in the {\it XMM-Newton} data, two properties in line with an extragalactic nature. While the source appeared constant, at the 1\% significance level, during the {\it XMM-Newton} exposure, the overall flux increased by nearly 40\% in the EPIC data compared to the {\it Chandra}-ACIS data.\\ 

\begin{figure*}[htb]
\begin{minipage}{6cm}
\begin{center}
\resizebox{6cm}{!}{\includegraphics{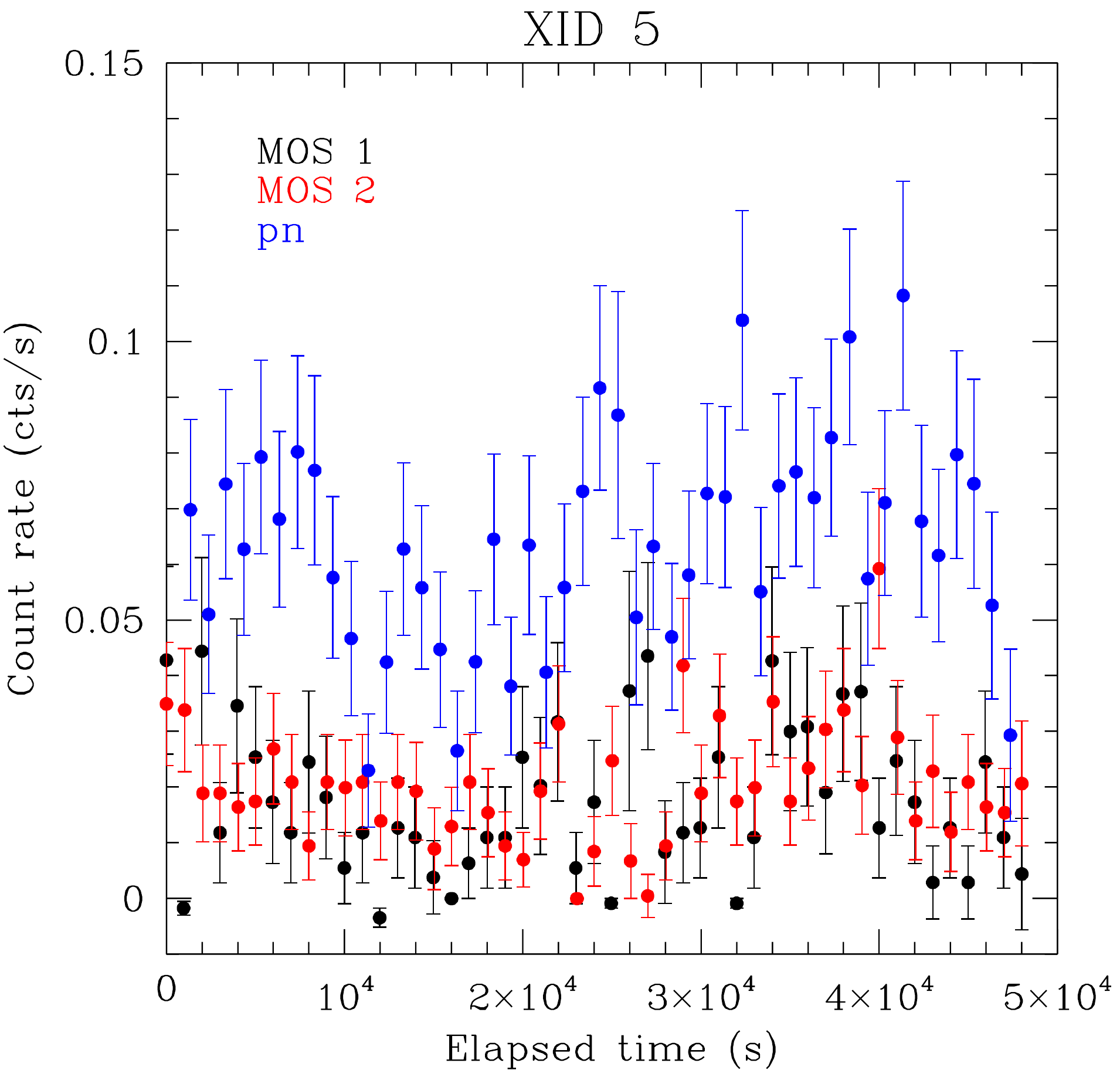}}
\end{center}
\end{minipage}
\hfill
\begin{minipage}{6cm}
\begin{center}
\resizebox{6cm}{!}{\includegraphics{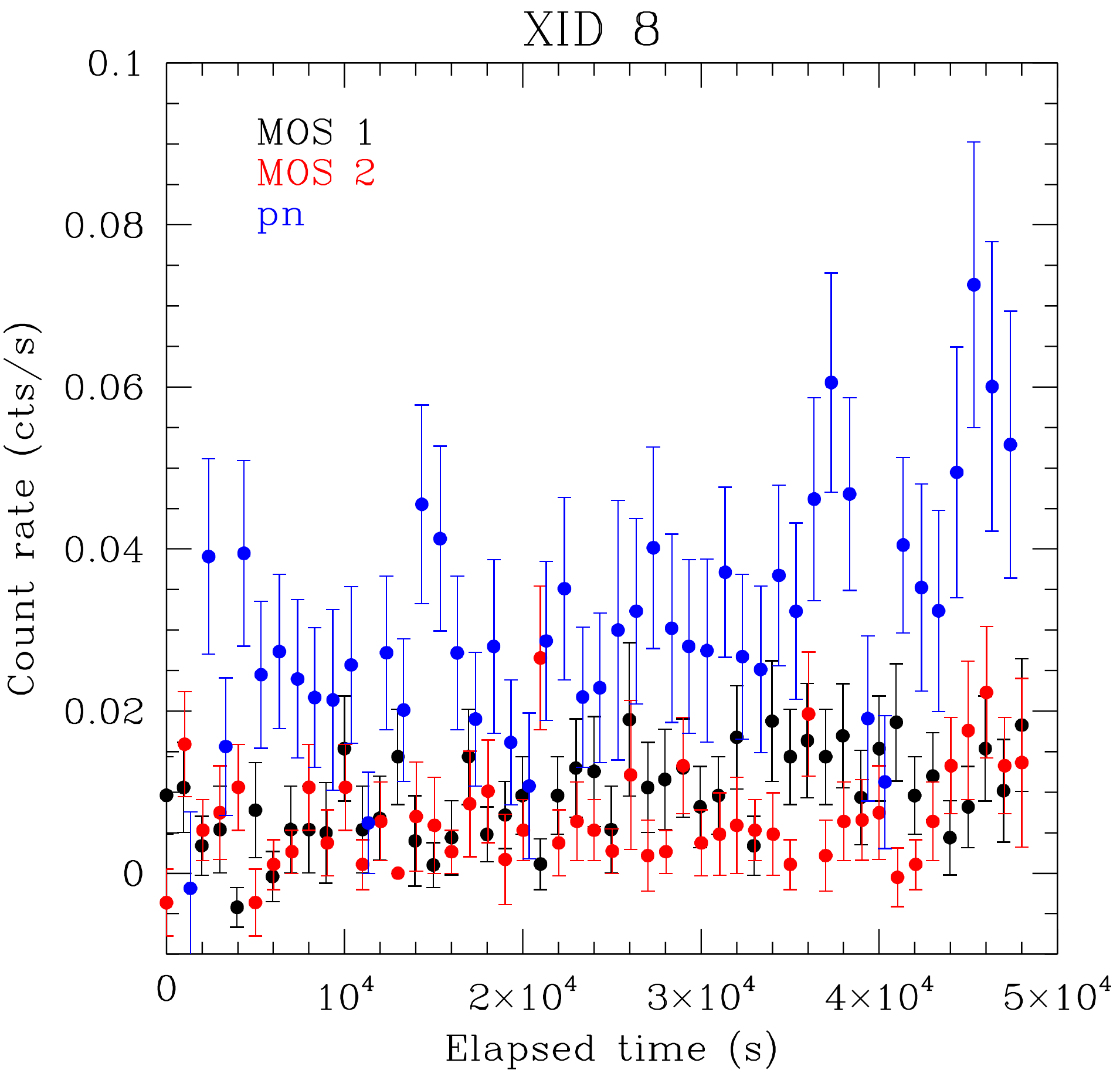}}
\end{center}
\end{minipage}
\hfill
\begin{minipage}{6cm}
\begin{center}
\resizebox{6cm}{!}{\includegraphics{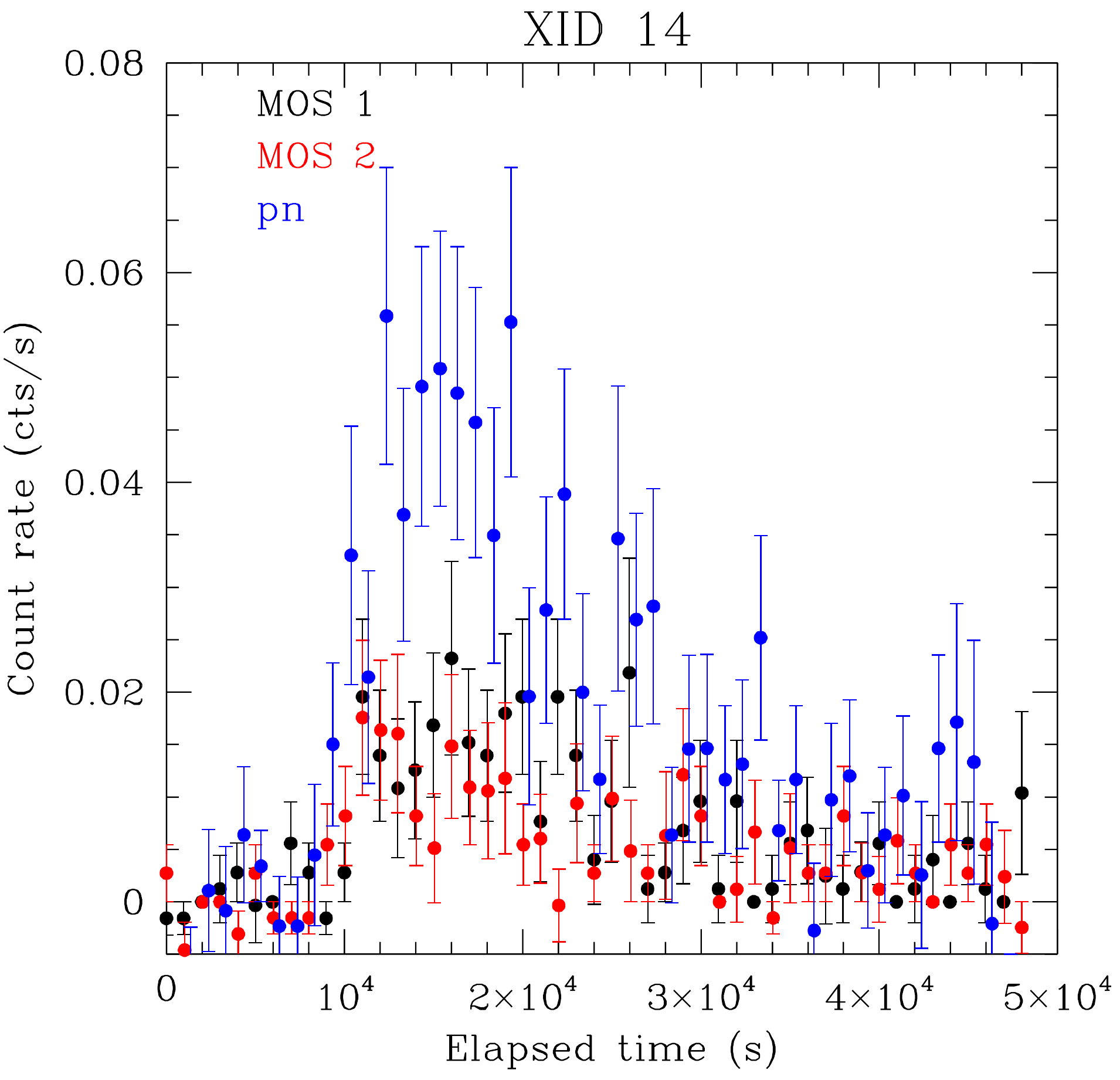}}
\end{center}
\end{minipage}
\caption{EPIC light curves of three secondary sources in the field of IC~1805: XID~5, 8, and 14, from left to right. Black, red, and blue correspond to count rates measured with the EPIC-MOS1, MOS2, and pn cameras, respectively. The light curves were extracted with time bins of 1\,ks. Time zero corresponds to the beginning of our {\it XMM-Newton} observation. 
\label{lcs}}
\end{figure*}

The spectrum of XMMU~J023230.2+611747 (XID~5) is clearly better fitted by thermal emission models, and two components are needed to achieve a good fit. This source corresponds to CXOU~023230.26+611748.6 \citep{Townsley} and IC~1805 132, an object considered as a star by \citet{bro13} and \citet{sun16}. The latter authors suggest this is a foreground F/G-type star\footnote{According to \citet{Vasilevskis} and \citet{Sanders}, the cluster membership probability of IC~1805 132 is indeed 0\%.}. The fitted temperatures are moderate, and the absorption column density is quite low, typical of nearby stellar sources. The usual trade-off between a more absorbed warm solution and a less absorbed hot solution is found, with {\it Chandra} and {\it XMM-Newton} data each converging to one of these cases, respectively (see Table\,\ref{addsrcfit}). The analysis of the {\it XMM-Newton} light curve (especially that of EPIC-pn) rejects the constancy hypothesis at the 1\% significance level, but the behaviour of the light curve appears more complex than a simple trend or a single large flare (see Fig.\,\ref{lcs}). In addition, the overall flux appeared to have decreased by about 40\% in the {\it XMM-Newton} exposure compared to the {\it Chandra} observation, although the error bar is large, hence variation is at the edge of significance.\\ 

The spectrum of XMMU~J023407.5+612817 (XID~8) is quite similar to that of the previous source, the best fit indicating two moderate temperatures with little absorption, compatible with the foreground late-type star nature pointed out by \citet{sun16}. The emission constancy is rejected at the 1\% significance level for the EPIC-pn light curve, and the EPIC-pn and EPIC-MOS1 light curves are both significantly better fitted by increasing trends than by constants (see Fig.\,\ref{lcs}). Unfortunately, this target is located outside the {\it Chandra} field of view.\\ 

The flux of XMMU~J023300.0+612503 (XID~9) appears to significantly increase during the {\it XMM-Newton} observation. Its X-ray emission appears somewhat absorbed, with either a rather steep slope for a power-law model or a very high temperature for a thermal plasma model. This source corresponds to CXOU~023259.92+612503.1 in \citet{Townsley}, whose optical/IR counterpart is classified as a young star in \citet{bro13}. \citet{sun16} also find that this counterpart lies within the PMS locus of IC~1805. With only 26 net {\it Chandra} counts \citep{Townsley}, a meaningful spectral comparison cannot be performed for this object, but we note that \citet{Townsley} attributed a photon flux of $1.62\times10^{-6}$\,ph\,cm$^{-2}$\,s$^{-1}$ with a median energy of 1.73\,keV to this source, corresponding to a flux\footnote{We note that this procedure yielded a flux close to those found in spectral fits for the six {\it Chandra} sources for which meaningful spectral information could be extracted.} of $\sim4.5\times10^{-15}$\,erg\,cm$^{-2}$\,s$^{-1}$. The target would thus have brightened by an order of magnitude between the two observations. In this context, it may be noted that constancy was also rejected at the 0.5\% significance level in the {\it Chandra} data \citep{Townsley}.\\  

Three {\it Chandra} sources are found within the {\it XMM-Newton} extraction region of XMMU~J023237.2+612810 (XID~10): CXOU~023237.01+612811.9, CXOU~023236.57+612813.4, and CXOU~023238.26+612807.6 (ordered by increasing separation). The former object clearly dominates the emission of the region in the {\it Chandra} data, as it has nearly 120 net counts, while the two other sources display fewer than sevencounts together. In addition, these two faint sources are at least twice as distant from the {\it XMM-Newton} position. While its position is shifted by 2\farcs5 compared to that of the {\it XMM-Newton} target, a quite high value for the {\it Chandra} source within 1\arcmin\ of the aimpoint, the spectral properties of the best-fit power laws are similar in both {\it Chandra} and {\it XMM-Newton} exposures, suggesting that the two objects are associated. The source flux may have doubled in the {\it XMM-Newton} observation, but this variation is again at the edge of significance in view of the large error bars. The number of counts were here too low for a meaningful light-curve analysis, but we note that a Kolmogorov-Smirnov test on the {\it Chandra} data rejected the constancy null hypothesis at the 0.5\% significance level \citep{Townsley}. \citet{sun16} find its optical/IR counterpart to lie within the PMS locus of the cluster.\\

The spectrum  of XMMU~J023208.0+612203 (XID~11) appears much better fitted by a power law than by a thermal plasma emission model. This source corresponds to CXOU~023208.03+612203.6, which is considered to be a star \citep[it appears in the PMS locus in this latter reference, but the possibility of a foreground late-type star cannot be excluded]{bro13,sun16}. \citet{Vilnius} assigned a K4\,V photometric spectral type to the optical counterpart. The source flux is similar in the {\it Chandra} and {\it XMM-Newton} observations, but we note that a Kolmogorov-Smirnov test on the {\it Chandra} data rejected the constancy null hypothesis at the $<$0.01\% significance level \citep{Townsley}.\\

XMMU~J023245.8+612959 (XID~12) has an absorption similar to that of the cluster, and it appears moderately hard ($kT\sim4$\,keV, $\Gamma\sim2$). Three {\it Chandra} sources are found within the {\it XMM-Newton} extraction region of this source: CXOU~023245.77+612959.4, CXOU~023245.84+612955.3, and CXOU~023246.78+613001.2 (in order of increasing separation). The former object clearly dominates the emission of the region in the {\it Chandra} data, as it has more than 60\,counts, while the two other sources display $\sim$7\,counts together \citep{Townsley}. Moreover, these two faint objects are at 3\farcs6 and 7\arcsec\ distance from the {\it XMM-Newton} position, respectively. The photon flux and median energy reported by \citet{Townsley} suggest a doubling of the X-ray flux in the {\it XMM-Newton} observation. \citet{sun16} find the optical/IR counterpart to lie within the PMS locus, while \citet{bro13} suggested a (young) stellar nature for the X-ray source.\\

XMMU~J023155.3+612249 (XID~13) displays a highly absorbed and hard X-ray emission. The temperature associated with a thermal model fitting is unrealistically high. This is not surprising because this target corresponds to CXOU~023155.35+612249.8, which is classified as an AGN in \citet{bro13}, a scenario reinforced by the lack of optical counterparts in the sensitive survey of \citet{sun16}. The photon flux and median energy reported by \citet{Townsley} suggest a tripling of the X-ray flux in the {\it XMM-Newton} observation.\\  

The light curve of XMMU~J023300.9+613737 (XID~14) clearly displays a moderate flare during the observation (Fig.\,\ref{lcs}). Spectral fitting yields an absorption close to that of the cluster, while the intrinsic emission is moderately hard. Only one {\it Chandra} source lies within the {\it XMM-Newton} extraction region, the young star CXOU~023300.58+613737.1, but it is offset by 2\farcs8, which is large even for an object at 10\arcmin\ off-axis angle in the ACIS data. If the two sources correspond to the same object, then the X-ray flux would have changed by a factor of 20. \citet{sun16} do not find any optical/IR counterpart to this object, while \citet{bro13} suggested a stellar nature for the X-ray source.\\

XMMU~J023243.0+613059 (XID~15) has a spectrum similar to that of the previous source. In Simbad, it appears close to 2MASS~J02324314+6130587 \citep[which has a likely YSO classification in][and a class II YSO classification in \citealt{sun16}]{bro13}. \citet{Vilnius} reported a WISE counterpart, suggesting this object to be an Ae/Be high-mass YSO (see also Sect.\,\ref{Bstars}). Its extraction region encompasses four {\it Chandra} sources: CXOU~023243.14+613058.6 (at 0\farcs75), CXOU~023243.01+613102.5 (at 3\farcs2), CXOU~023243.36+613056.3 (at 3\farcs5), and CXOU~023242.08+613105.3 (at 9\farcs5). The former and latter sources have a similar number of counts, but the latter one is quite far away from the {\it XMM-Newton} position; the two other sources are fainter by a factor of $\sim$5. If the {\it XMM-Newton} source corresponds to the closest {\it Chandra} source, then the photon flux and median energy reported by \citet{Townsley} suggest a fourfold increase of its flux.\\ 

XMMU~J023209.2+613039 (XID~17) displays a quite hard spectrum, with an absorption stronger than the typical cluster value and a high temperature/photon index. Its extraction region encompasses two {\it Chandra} sources: CXOU~023209.23+613040.8 (at 1\farcs1 - not too high considering the ACIS off-axis angle of 5\arcmin) and CXOU~023207.89+613047.4, which is five times fainter in \citet{Townsley} and at a very large angular distance of 12\farcs5 from the {\it XMM-Newton} position. The former {\it Chandra} source is a likely YSO \citep{bro13}, more precisely classified as a class II YSO by \citet{sun16}; it is also known as 2MASS~J02320924+6130404 and was found to be a WISE source probably associated with a YSO \citep{Vilnius}. Compared to {\it Chandra} observations \citep[see photon flux and median energy in][]{Townsley}, its X-ray flux has quadrupled.\\

XMMU~J023317.6+612808 (XID~20) displays an absorption compatible with that of the cluster and moderately high temperature and $\Gamma$. In its extraction region lie two {\it Chandra} sources CXOU~023317.49+612808.4 and CXOU~023316.53+612809.5, the latter being 20 times fainter and 7 times farther away from the {\it XMM-Newton} position. Despite a separation of 1\farcs2, it is therefore reasonable to consider the former {\it Chandra} source as the counterpart of the {\it XMM-Newton} source. The photon flux and median energy reported by \citet{Townsley} suggest that the object was somewhat brighter during the {\it XMM-Newton} observation. The optical/IR counterpart is a likely YSO \citep{bro13} lying in the PMS locus \citep{sun16}. \citet{Vilnius} quoted a photometric spectral type G for this object.\\ 

With a similar spectrum as the previous object, XMMU~J023254.0+612803 (XID~21) is most probably associated with CXOU~023253.96+612804.3, which lies at 1\arcsec\ (a second source in the extraction region, CXOU~023254.24+612814.6 is 30 times fainter in {\it Chandra} exposure and 10 times farther away). Its optical/IR counterpart is a young star \citep{bro13} lying in the PMS locus \citep{sun16}, and its {\it XMM-Newton} flux is somewhat higher than the one derived from the photon flux of \citet{Townsley}.\\

Again displaying a similar spectrum as the previous sources, XMMU~J023300.9+612608 (XID~22) is the probable counterpart of CXOU~023300.79+612609.6 (which was at nearly 3\arcmin\ off-axis in the {\it Chandra} data, explaining the larger separation - 1\farcs3 with the {\it XMM-Newton} position). A second source, CXOU~023300.57+612606.8, appears both eight times fainter and twice more distant, rendering an association less likely. The optical/IR counterpart is a likely YSO \citep{bro13} lying in the PMS locus \citep{sun16} and the X-ray flux is about eight times higher in the {\it XMM-Newton} observation than in the {\it Chandra} data.\\

In summary, XID~3 and 13 are probably extragalactic background sources, whilst XID~5 and 8 are probably associated with foreground late-type stars. The remaining bright X-ray sources are most probably PMS members of IC~1805. \citet{Megeath} quoted a typical distance for the W4 region of 2\,kpc. Adopting this distance and taking an average neutral hydrogen column density of $0.4\,10^{22}$\,cm$^{-2}$ along with an average plasma temperature of $kT = 2.7$\,keV (see Table\,\ref{addsrcfit}), we can convert the observed fluxes into intrinsic fluxes and thereby estimate X-ray luminosities. For the X-ray brightest PMS candidates, we found $f_{\rm X} \simeq 0.5\,10^{-13}$\,erg\,cm$^{-2}$\,s$^{-1}$ (Table\,\ref{addsrcfit}) which then translates into $L_{\rm X} \simeq 3.5\,10^{31}$\,erg\,s$^{-1}$. Whilst this is a relatively high value, it is not unusual for PMS stars in young open clusters, especially during flares \citep{NGC6231PMS,CygOB2GR}. We note also that our cross-correlation with catalogues from the literature (see Table\,\ref{Xcat}) resulted in the association of six X-ray sources with low-mass YSOs identified with WISE \citep{Vilnius}. Additional support for the conclusion that the bulk of the sources are associated with PMS stars comes from the correlation with deep, sensitive photometry \citep[see][]{sun16}.

\section{Summary and conclusions \label{conclusion}}
We have presented new X-ray observations and optical spectroscopy of the very young open cluster IC~1805. We improved the orbital solutions of the two previously known O-star binary systems, HD~15558 and BD+60$^{\circ}$~497, and found evidence that BD+60$^{\circ}$~498 is a third O-star binary. With this information, we then studied the X-ray emission of the O-star population of IC~1805. X-ray emission from colliding winds in known binary systems was found to play only a marginal role in determining the X-ray luminosity. The X-ray luminosities of O-type stars in IC~1805 follow the well-known canonical scaling relation with bolometric luminosity $\log{L_{\rm X}/L_{\rm bol}} = -6.98 \pm 0.20$. We paid particular attention to HD~15570, an O4\,If$^+$ star, to search for evidence that the dense stellar wind reduces the level of emerging X-ray emission. For this star we obtained $\log{L_{\rm X}/L_{\rm bol}} = -7.24 \pm 0.03$, which is below the scaling relation of the O-stars in the cluster, although at only 1.3 times the dispersion of the latter relation. Such a deviation is compatible with theoretical expectations based on the stellar and wind properties of the star. Considering results for other very early and luminous O-supergiants in the literature, however, we found no clear observational evidence that the most luminous O-supergiants as a class are systematically X-ray underluminous. The reason for this is the relatively small amplitude of the expected reduction effect when compared to the dispersion around the scaling relation. 

In addition to the O-type stars, our {\it XMM-Newton} X-ray data revealed many weaker sources in the field of view of IC~1805. We discussed the X-ray spectra and light curves of the brightest of these objects. Except for a few field interlopers, most of the sources are probably low-mass pre-main-sequence stars belonging to IC~1805. Finally, whilst we detected about 25\% of the spectroscopically classified B-type stars in the cluster, the properties of the detected sources are in most cases compatible with X-ray emission from otherwise undetected low-mass PMS companions, thus suggesting that the X-ray detections of B- and A-type stars in IC~1805 are due to such low-mass companions.      

\section*{Acknowledgements}
We are grateful to Prof.\ Hwankyung Sung for sharing some of his results with us before publication. Our thanks go to the referee of our paper for a swift and helpful report. We acknowledge support through an ARC grant for Concerted Research Actions, financed by the French Community of Belgium (Wallonia-Brussels Federation), from the Fonds de la Recherche Scientifique (FRS/FNRS), as well as through an XMM PRODEX contract (Belspo). The TIGRE facility is funded and operated by the universities of Hamburg, Guanajuato and Li\`ege. This research has made use of the SIMBAD database, operated at CDS, Strasbourg, France.

%\Online
\begin{appendix}
\section{List of X-ray sources detected with {\it XMM-Newton}}
Table\,\ref{Xcat} provides the full catalogue of X-ray sources detected with the EPIC instruments onboard {\it XMM-Newton} ordered by increasing right ascension. The coordinates of the sources were cross-correlated with the optical and IR catalogues of \citet{Vilnius}, \citet{Wolff}, and the SIMBAD catalogue. We adopted in each case a correlation radius of 4\arcsec, which is generally well adapted for EPIC data. For a more detailed comparison, we refer to the work of Sung et al.\ (in prep.), who present the results of a cross-correlation with their own photometric catalogue, which is much deeper than the catalogues used here. Adopting the same 4\arcsec\ radius for cross-correlation with the 2MASS \citep{2MASS} catalogue results in 72\% of the X-ray sources with a single IR counterpart and an additional 5\% with two IR counterparts.
\begin{sidewaystable*}
\caption{List of X-ray sources detected with {\it XMM-Newton} in IC~1805 \label{Xcat}}
\tiny
\begin{center}
\begin{tabular}{c c c c c c c c c c c}
\hline
Name                 & J2000.0 Coord.         & XID & \multicolumn{2}{c}{MOS1} & \multicolumn{2}{c}{MOS2} & \multicolumn{2}{c}{pn} & Optical ID \& SpT & Note \\
XMMU                & hhmmss $^{\circ}$ ' ''   &     & Ct Rate  & HR & Ct Rate & HR & Ct Rate & HR & & \\
& & & (10$^{-3}$\,ct\,s$^{-1}$) & & (10$^{-3}$\,ct\,s$^{-1}$) & (10$^{-3}$\,ct\,s$^{-1}$) & \\
\hline
J023039.8+612949 & 02:30:39.873 +61:29:49.62 & 71   &                 &                  &                 &                 &   20.45$\pm$2.37 &   0.43$\pm$0.09  \\
J023040.1+613040 & 02:30:40.176 +61:30:40.02 & 54   &                 &                  &                 &                 &   22.46$\pm$2.44 &   0.30$\pm$0.10  \\
J023051.0+613020 & 02:30:51.030 +61:30:20.33 & 170  &   1.51$\pm$0.59 &   0.30$\pm$0.36  &   1.50$\pm$0.51 &$-$0.29$\pm$0.37 &    3.53$\pm$1.12 &   0.13$\pm$0.31  \\
J023113.7+612712 & 02:31:13.771 +61:27:12.29 & 109  &   2.33$\pm$0.50 &   0.23$\pm$0.21  &   0.89$\pm$0.38 &   0.45$\pm$0.40 &    3.11$\pm$0.74 &   0.44$\pm$0.22  \\
J023116.3+612321 & 02:31:16.339 +61:23:21.69 & 172: &   1.22$\pm$0.40 &   0.27$\pm$0.32  &   0.58$\pm$0.33 &   0.75$\pm$0.42 &    2.47$\pm$1.44 &$-$1.00$\pm$0.68  \\
J023121.7+613020 & 02:31:21.778 +61:30:20.84 & 101  &   1.09$\pm$0.37 &$-$0.50$\pm$0.35  &   2.51$\pm$0.51 &   0.62$\pm$0.15 &    2.84$\pm$0.67 &$-$0.10$\pm$0.24  \\
J023122.8+613748 & 02:31:22.854 +61:37:48.74 & 121  &                 &                  &   3.20$\pm$0.61 &$-$0.99$\pm$0.13 &                  &                  \\
J023124.2+613702 & 02:31:24.275 +61:37:02.89 & 92   &                 &                  &   4.62$\pm$0.74 &$-$0.64$\pm$0.16 &                  &                  \\
J023127.7+611513 & 02:31:27.700 +61:15:13.33 & 70   &                 &                  &                 &                 &   12.10$\pm$1.60 &$-$1.00$\pm$0.13  \\
J023128.1+611644 & 02:31:28.154 +61:16:44.40 & 153  &                 &                  &                 &                 &    9.34$\pm$1.78 &   0.38$\pm$0.16  \\
J023129.2+612725 & 02:31:29.239 +61:27:25.31 & 156  &   0.31$\pm$0.21 &   0.92$\pm$0.68  &   0.61$\pm$0.23 &$-$0.92$\pm$0.32 &    2.55$\pm$0.59 &$-$0.27$\pm$0.24  \\
J023129.2+613623 & 02:31:29.282 +61:36:23.38 & 160  &                 &                  &   1.30$\pm$0.48 &   0.11$\pm$0.36 &    2.99$\pm$0.87 &$-$0.43$\pm$0.31  \\
J023138.6+613209 & 02:31:38.600 +61:32:09.53 & 141  &                 &                  &   1.46$\pm$0.33 &$-$0.45$\pm$0.23 &    3.15$\pm$0.62 &$-$0.38$\pm$0.20  \\
J023145.6+612856 & 02:31:45.632 +61:28:56.25 & 169  &   0.04$\pm$0.12 &$-$1.00$\pm$3.66  &   0.64$\pm$0.21 &   0.60$\pm$0.32 &    1.61$\pm$0.46 &   0.51$\pm$0.29  \\
J023147.0+613025 & 02:31:47.048 +61:30:25.70 & 48   &                 &                  &   2.20$\pm$0.38 &   0.60$\pm$0.15 &    5.66$\pm$0.67 &   0.46$\pm$0.11  \\
J023147.2+613207 & 02:31:47.205 +61:32:07.53 & 151  &                 &                  &   1.00$\pm$0.27 &$-$0.89$\pm$0.22 &    2.57$\pm$0.57 &$-$0.62$\pm$0.23  \\
J023150.3+613601 & 02:31:50.315 +61:36:01.01 & 45   &                 &                  &   3.10$\pm$0.54 &   0.02$\pm$0.17 &    6.37$\pm$0.87 &$-$0.43$\pm$0.14  \\
J023150.5+613257 & 02:31:50.577 +61:32:57.91 & 55   &                 &                  &   2.46$\pm$0.38 &$-$0.48$\pm$0.15 &    5.26$\pm$0.69 &$-$0.74$\pm$0.12  \\
J023150.7+613332 & 02:31:50.732 +61:33:32.00 & 191  &                 &                  &   0.75$\pm$0.26 &$-$0.55$\pm$0.32 &    2.44$\pm$0.58 &$-$0.23$\pm$0.24 & [SBJLK13]~25; f8\,V & (1) \\
J023152.4+612826 & 02:31:52.481 +61:28:26.96 & 180  &   0.38$\pm$0.18 &$-$0.60$\pm$0.42  &   0.38$\pm$0.16 &$-$1.00$\pm$0.24 &    1.27$\pm$0.36 &$-$1.00$\pm$0.24  \\
J023155.3+612249 & 02:31:55.337 +61:22:49.77 & 13   &   4.46$\pm$0.53 &   0.45$\pm$0.11  &   3.61$\pm$0.48 &   0.54$\pm$0.11 &   13.58$\pm$0.99 &   0.46$\pm$0.07  \\
J023155.3+613123 & 02:31:55.396 +61:31:23.37 & 75   &                 &                  &   1.16$\pm$0.26 &$-$0.23$\pm$0.22 &    5.14$\pm$0.84 &$-$0.41$\pm$0.16  \\
J023156.4+613235 & 02:31:56.421 +61:32:35.00 & 105  &                 &                  &   0.75$\pm$0.25 &$-$0.80$\pm$0.28 &    1.67$\pm$0.51 &$-$0.18$\pm$0.31  \\
J023156.6+613345 & 02:31:56.659 +61:33:45.38 & 93   &                 &                  &   3.16$\pm$0.45 &   0.16$\pm$0.14 &                  &                  \\
J023157.1+613643 & 02:31:57.117 +61:36:43.70 & 7    &                 &                  &  12.39$\pm$0.88 &$-$0.90$\pm$0.05 &   40.14$\pm$1.77 &$-$0.97$\pm$0.02 & BD+60$^{\circ}$~497; O6.5\,V((f))+O8.5\,V & (2) \\
J023157.2+613243 & 02:31:57.229 +61:32:43.53 & 104  &                 &                  &   1.29$\pm$0.29 &$-$0.33$\pm$0.22 &    3.64$\pm$0.60 &$-$0.45$\pm$0.16 & WISE~J023157.06+613245.4  & (*) \\
J023157.4+613348 & 02:31:57.459 +61:33:48.03 & 132: &                 &                  &   0.50$\pm$0.29 &   1.00$\pm$0.36 &    2.64$\pm$0.92 &$-$0.08$\pm$0.35  \\
J023200.0+612050 & 02:32:00.028 +61:20:50.60 & 97:  &   1.33$\pm$0.58 &   0.54$\pm$0.45  &   2.67$\pm$0.67 &   0.51$\pm$0.24 &    0.00$\pm$0.92 &                  \\
J023200.4+612038 & 02:32:00.491 +61:20:38.85 & 3    &  38.96$\pm$1.54 &   0.22$\pm$0.04  &  40.02$\pm$1.64 &   0.36$\pm$0.04 &  105.13$\pm$2.74 &   0.22$\pm$0.03  \\
J023200.7+613153 & 02:32:00.702 +61:31:53.05 & 190  &                 &                  &   0.52$\pm$0.20 &$-$0.66$\pm$0.34 &    1.72$\pm$0.45 &$-$0.48$\pm$0.26  \\
J023202.0+612648 & 02:32:02.060 +61:26:48.65 & 146  &   0.68$\pm$0.20 &$-$0.63$\pm$0.27  &   0.49$\pm$0.16 &$-$1.00$\pm$0.12 &    1.88$\pm$0.37 &$-$1.00$\pm$0.09 & IC~1805 108; g0\,V & (1) \\
J023202.7+612951 & 02:32:02.776 +61:29:51.82 & 40   &   1.76$\pm$0.30 &$-$0.03$\pm$0.17  &   1.40$\pm$0.26 &   0.07$\pm$0.19 &    3.75$\pm$0.55 &$-$0.26$\pm$0.14 & [MJD95]~J023202.42+612951.2 & \\
J023206.8+613243 & 02:32:06.884 +61:32:43.70 & 119  &                 &                  &   0.72$\pm$0.23 &$-$0.13$\pm$0.32 &    2.63$\pm$0.49 &$-$0.64$\pm$0.18 & WISE~J023206.99+613242.9 & (*) \\
J023208.0+612203 & 02:32:08.024 +61:22:03.28 & 11   &   3.34$\pm$0.45 &$-$0.80$\pm$0.11  &   3.90$\pm$0.47 &$-$0.77$\pm$0.09 &   13.52$\pm$0.96 &$-$0.92$\pm$0.05 & [SBJLK13]~53; k4\,V & (1) \\
J023208.4+613529 & 02:32:08.444 +61:35:29.80 & 187: &                 &                  &   0.50$\pm$0.20 &$-$1.00$\pm$0.24 &    2.06$\pm$0.58 &$-$0.46$\pm$0.30  \\
J023209.1+613246 & 02:32:09.151 +61:32:46.36 & 83   &                 &                  &   0.97$\pm$0.26 &$-$0.17$\pm$0.26 &    3.22$\pm$0.52 &   0.22$\pm$0.16  \\
J023209.1+612627 & 02:32:09.181 +61:26:27.64 & 86   &   0.72$\pm$0.19 &$-$0.93$\pm$0.15  &   0.80$\pm$0.19 &$-$0.99$\pm$0.12 &    2.12$\pm$0.40 &$-$0.80$\pm$0.17 & IC~1805 110; A8\,III & (3) \\
J023209.2+613039 & 02:32:09.291 +61:30:39.81 & 17   &   4.89$\pm$0.47 &   0.08$\pm$0.10  &   3.77$\pm$0.39 &$-$0.22$\pm$0.10 &    7.76$\pm$1.53 &$-$0.13$\pm$0.20 & WISE~J023209.24+613040.4 & (*) \\
J023209.6+613824 & 02:32:09.677 +61:38:24.41 & 165  &                 &                  &   1.16$\pm$0.38 &$-$0.39$\pm$0.36 &    1.36$\pm$0.53 &$-$0.86$\pm$0.36 & IC~1805 111; B2\,V & (4,**) \\
J023210.8+613307 & 02:32:10.870 +61:33:07.64 & 78   &                 &                  &   1.13$\pm$0.27 &$-$1.00$\pm$0.15 &    3.70$\pm$0.51 &$-$1.00$\pm$0.05 & BD+60$^{\circ}$~498; O9.7\,V & (2) \\
J023210.9+613100 & 02:32:10.992 +61:31:00.65 & 115  &   0.73$\pm$0.26 &$-$0.12$\pm$0.35  &   0.53$\pm$0.18 &$-$0.97$\pm$0.19 &    2.14$\pm$0.43 &$-$0.73$\pm$0.18 & [MJD95]~J023211.11+613102.8; B8 & (5) \\
J023212.5+613856 & 02:32:12.581 +61:38:56.23 & 91   &                 &                  &   1.82$\pm$0.43 &$-$0.58$\pm$0.24 &    4.17$\pm$0.82 &$-$0.66$\pm$0.20  \\
J023214.4+613336 & 02:32:14.442 +61:33:36.08 & 171  &                 &                  &   0.46$\pm$0.25 &$-$1.00$\pm$0.27 &    1.90$\pm$0.44 &$-$1.00$\pm$0.17  \\
J023216.4+613209 & 02:32:16.452 +61:32:09.30 & 161: &                 &                  &                 &                 &    1.52$\pm$0.37 &$-$1.00$\pm$0.19  \\
J023216.5+613313 & 02:32:16.578 +61:33:13.57 & 36   &                 &                  &   3.63$\pm$0.70 &$-$0.28$\pm$0.19 &    6.82$\pm$0.64 &$-$0.96$\pm$0.06 & BD+60$^{\circ}$~499; O9.5\,V & (2) \\
J023217.1+612927 & 02:32:17.129 +61:29:27.09 & 80   &   0.60$\pm$0.19 &$-$0.65$\pm$0.27  &   1.00$\pm$0.21 &$-$0.38$\pm$0.21 &    2.55$\pm$0.47 &$-$0.37$\pm$0.18  \\
J023218.2+612650 & 02:32:18.271 +61:26:50.56 & 46   &   1.03$\pm$0.23 &$-$0.02$\pm$0.23  &   0.93$\pm$0.20 &   0.30$\pm$0.21 &    3.07$\pm$0.45 &$-$0.25$\pm$0.14  \\
J023218.8+613118 & 02:32:18.885 +61:31:18.43 & 85   &   0.64$\pm$0.21 &$-$0.39$\pm$0.30  &   0.71$\pm$0.19 &$-$0.31$\pm$0.25 &    1.30$\pm$0.38 &$-$0.68$\pm$0.31  \\
\hline
\end{tabular}
\end{center}
\end{sidewaystable*}
\addtocounter{table}{-1}
\begin{sidewaystable*}
\caption{Continued}
\tiny
\begin{center}
\begin{tabular}{c c c c c c c c c c c}
\hline
Name                 & J2000.0 Coord.         & XID & \multicolumn{2}{c}{MOS1} & \multicolumn{2}{c}{MOS2} & \multicolumn{2}{c}{pn} & Optical ID \& SpT & Note \\
XMMU                & hhmmss $^{\circ}$ ' ''   &     & Ct Rate  & HR & Ct Rate & HR & Ct Rate & HR & & \\
& & & (10$^{-3}$\,ct\,s$^{-1}$) & & (10$^{-3}$\,ct\,s$^{-1}$) & (10$^{-3}$\,ct\,s$^{-1}$) & \\
\hline
J023220.2+613014 & 02:32:20.288 +61:30:14.37 & 173  &   0.32$\pm$0.15 &$-$1.00$\pm$0.36  &   0.42$\pm$0.15 &$-$0.19$\pm$0.36 &    0.95$\pm$0.33 &$-$0.38$\pm$0.34  \\
J023221.1+612718 & 02:32:21.197 +61:27:18.28 & 167  &   0.36$\pm$0.15 &$-$0.04$\pm$0.42  &   0.41$\pm$0.16 &$-$1.00$\pm$0.35 &    1.58$\pm$0.37 &$-$0.22$\pm$0.23 & [SBJLK13]~91; g5\,V & (1) \\
J023221.3+613259 & 02:32:21.302 +61:32:59.65 & 128  &                 &                  &   0.43$\pm$0.18 &$-$0.69$\pm$0.39 &    1.63$\pm$0.42 &$-$0.29$\pm$0.25  \\
J023221.7+613237 & 02:32:21.749 +61:32:37.54 & 51   &                 &                  &   1.46$\pm$0.30 &$-$0.11$\pm$0.20 &    5.82$\pm$0.60 &$-$0.52$\pm$0.09 & [SBJLK13]~93; g0\,V & (1) \\
J023222.2+612345 & 02:32:22.254 +61:23:45.44 & 124  &   0.34$\pm$0.16 &$-$1.00$\pm$0.17  &   0.28$\pm$0.15 &$-$0.35$\pm$0.53 &    1.77$\pm$0.39 &$-$0.13$\pm$0.22  \\
J023225.1+613250 & 02:32:25.106 +61:32:50.47 & 177  &                 &                  &   0.40$\pm$0.18 &$-$0.73$\pm$0.36 &    1.85$\pm$0.42 &$-$0.43$\pm$0.22  \\
J023225.3+613030 & 02:32:25.380 +61:30:30.87 & 139  &   0.54$\pm$0.18 &$-$0.20$\pm$0.32  &   0.49$\pm$0.17 &$-$0.12$\pm$0.33 &    1.23$\pm$0.36 &$-$0.52$\pm$0.27  \\
J023226.3+613111 & 02:32:26.399 +61:31:11.00 & 148  &   0.64$\pm$0.20 &$-$0.24$\pm$0.31  &   0.80$\pm$0.20 &$-$0.45$\pm$0.24 &    1.77$\pm$0.42 &$-$0.35$\pm$0.22  \\
J023229.9+612707 & 02:32:29.990 +61:27:07.31 & 120  &   0.55$\pm$0.18 &$-$0.48$\pm$0.31  &   0.55$\pm$0.17 &$-$0.65$\pm$0.28 &    1.73$\pm$0.38 &$-$0.56$\pm$0.20 & IC~1805 130; B3\,V & (3) \\
J023230.2+611747 & 02:32:30.281 +61:17:47.78 & 5    &  12.64$\pm$1.16 &$-$0.79$\pm$0.06  &  18.71$\pm$1.18 &$-$0.78$\pm$0.05 &   63.99$\pm$2.25 &$-$0.86$\pm$0.02 & IC~1805 132 & \\
J023230.7+613609 & 02:32:30.703 +61:36:09.27 & 134  &                 &                  &   0.82$\pm$0.24 &$-$0.27$\pm$0.28 &                  &                  \\
J023230.8+613206 & 02:32:30.828 +61:32:06.70 & 49   &   1.66$\pm$0.31 &   0.03$\pm$0.19  &   1.19$\pm$0.25 &   0.39$\pm$0.19 &    1.61$\pm$0.39 &   0.46$\pm$0.25  \\
J023231.9+613719 & 02:32:31.994 +61:37:19.62 & 185  &                 &                  &   0.75$\pm$0.29 &   0.23$\pm$0.37 &    2.58$\pm$0.64 &$-$0.19$\pm$0.26  \\
J023232.8+613834 & 02:32:32.881 +61:38:34.32 & 111  &                 &                  &   0.61$\pm$0.22 &$-$1.00$\pm$0.29 &    1.64$\pm$0.71 &$-$0.47$\pm$0.45 & IC~1805 134; B7\,V & (3) \\
J023233.1+612619 & 02:32:33.154 +61:26:19.41 & 66   &   0.89$\pm$0.21 &$-$0.26$\pm$0.23  &   1.13$\pm$0.27 &$-$0.58$\pm$0.20 &    1.77$\pm$0.37 &$-$0.56$\pm$0.19  \\
J023233.2+612427 & 02:32:33.277 +61:24:27.56 & 76   &   0.84$\pm$0.21 &$-$0.73$\pm$0.22  &   0.89$\pm$0.21 &$-$0.17$\pm$0.23 &    2.61$\pm$0.66 &$-$0.51$\pm$0.24 & [MJD95]~J023233.11+612427.4; a3\,V & (1) \\
J023233.5+611845 & 02:32:33.507 +61:18:45.59 & 103  &                 &                  &   1.49$\pm$0.38 &$-$0.62$\pm$0.27 &    4.39$\pm$0.78 &$-$0.20$\pm$0.18  \\
J023234.2+611728 & 02:32:34.267 +61:17:28.81 & 38   &                 &                  &   4.03$\pm$0.67 &$-$0.09$\pm$0.17 &   11.18$\pm$1.17 &$-$0.03$\pm$0.11  \\
J023236.1+612758 & 02:32:36.199 +61:27:58.97 & 61   &   0.36$\pm$0.19 &$-$0.94$\pm$0.27  &   0.83$\pm$0.25 &$-$0.22$\pm$0.28 &    1.59$\pm$0.47 &   0.01$\pm$0.30  \\
J023236.3+612824 & 02:32:36.366 +61:28:24.77 & 6    &   8.00$\pm$0.52 &$-$0.82$\pm$0.04  &   7.15$\pm$0.49 &$-$0.86$\pm$0.04 &   27.01$\pm$1.03 &$-$0.90$\pm$0.02 & BD+60$^{\circ}$~501; O7\,V(n)((f))z & (6) \\
J023236.4+612840 & 02:32:36.495 +61:28:40.99 & 42   &   0.67$\pm$0.23 &   0.13$\pm$0.35  &   0.60$\pm$0.21 &$-$0.51$\pm$0.28 &    2.60$\pm$0.49 &$-$0.67$\pm$0.14  \\
J023237.2+612810 & 02:32:37.269 +61:28:10.25 & 10   &   2.46$\pm$0.35 &$-$0.70$\pm$0.10  &   3.20$\pm$0.38 &$-$0.41$\pm$0.10 &    8.89$\pm$0.73 &$-$0.63$\pm$0.07  \\
J023237.4+613217 & 02:32:37.450 +61:32:17.42 & 59   &   1.35$\pm$0.31 &$-$0.54$\pm$0.21  &   1.51$\pm$0.27 &$-$0.43$\pm$0.17 &    3.25$\pm$0.49 &$-$0.79$\pm$0.13 & IC~1805 139; B4\,V & (3) \\
J023238.5+613204 & 02:32:38.531 +61:32:04.20 & 163  &   0.86$\pm$0.26 &$-$0.02$\pm$0.30  &   0.53$\pm$0.19 &$-$0.63$\pm$0.31 &    1.93$\pm$0.42 &$-$0.56$\pm$0.20  \\
J023239.5+612656 & 02:32:39.592 +61:26:56.94 & 29   &   1.60$\pm$0.32 &$-$0.47$\pm$0.17  &   1.61$\pm$0.29 &$-$0.46$\pm$0.16 &    4.63$\pm$0.58 &$-$0.38$\pm$0.12  \\
J023239.8+612154 & 02:32:39.898 +61:21:54.16 & 65   &   1.52$\pm$0.29 &   0.15$\pm$0.19  &   0.96$\pm$0.24 &   0.13$\pm$0.26 &    3.44$\pm$0.53 &   0.24$\pm$0.15  \\
J023240.2+612813 & 02:32:40.228 +61:28:13.28 & 28   &   1.09$\pm$0.25 &$-$0.50$\pm$0.21  &   1.33$\pm$0.27 &$-$0.53$\pm$0.17 &    3.60$\pm$0.56 &$-$0.53$\pm$0.14  \\
J023240.6+612519 & 02:32:40.680 +61:25:19.28 & 184  &   0.28$\pm$0.13 &   1.00$\pm$0.35  &   0.59$\pm$0.18 &$-$0.02$\pm$0.30 &    0.70$\pm$0.32 &   1.00$\pm$0.62  \\
J023240.7+612801 & 02:32:40.739 +61:28:01.69 & 16   &   2.21$\pm$0.34 &$-$0.50$\pm$0.13  &   2.61$\pm$0.35 &$-$0.55$\pm$0.12 &    9.95$\pm$0.73 &$-$0.56$\pm$0.06 & IC~1805 143; B0.5\,V & (3) \\
J023241.4+612554 & 02:32:41.448 +61:25:54.13 & 94   &   0.38$\pm$0.16 &$-$0.85$\pm$0.32  &   0.45$\pm$0.17 &$-$1.00$\pm$0.28 &    3.08$\pm$0.46 &$-$0.72$\pm$0.13 & [SBJLK13]~138; f6\,V & (1)  \\
J023242.0+614021 & 02:32:42.030 +61:40:21.81 & 63   &                 &                  &   2.60$\pm$0.57 &   0.12$\pm$0.21 &    6.70$\pm$1.01 &$-$0.15$\pm$0.15  \\
J023242.2+612703 & 02:32:42.267 +61:27:03.67 & 24:  &   0.02$\pm$0.30 &$-$1.00$\pm$19.7  &   0.61$\pm$0.26 &$-$0.78$\pm$0.34 &    3.56$\pm$0.74 &$-$0.67$\pm$0.17  \\
J023242.6+612721 & 02:32:42.674 +61:27:21.11 & 1    &  51.60$\pm$1.32 &$-$0.72$\pm$0.02  &  51.12$\pm$1.24 &$-$0.72$\pm$0.02 &  170.55$\pm$2.43 &$-$0.77$\pm$0.01 & HD~15558; O4.5\,III(f) & (6) \\
J023243.0+613059 & 02:32:43.096 +61:30:59.33 & 15   &   4.11$\pm$0.39 &$-$0.50$\pm$0.08  &   3.50$\pm$0.37 &$-$0.54$\pm$0.09 &   11.23$\pm$0.74 &$-$0.37$\pm$0.06 & WISE~J023243.13+613058.8; Ae/Be & (1,*) \\
J023243.1+613334 & 02:32:43.185 +61:33:34.22 & 140  &   0.80$\pm$0.22 &$-$0.63$\pm$0.27  &   0.92$\pm$0.25 &$-$0.66$\pm$0.24 &    2.01$\pm$0.86 &   0.04$\pm$0.43  \\
J023243.3+612803 & 02:32:43.340 +61:28:03.54 & 18   &   2.43$\pm$0.33 &$-$0.26$\pm$0.13  &   2.20$\pm$0.38 &$-$0.37$\pm$0.15 &    6.77$\pm$0.62 &$-$0.60$\pm$0.08 & [MJD95]~J023243.22+612804.4; A6 & (5) \\
J023243.7+612631 & 02:32:43.769 +61:26:31.82 & 19   &   3.71$\pm$0.38 &$-$0.39$\pm$0.10  &   4.40$\pm$0.41 &$-$0.33$\pm$0.09 &   11.80$\pm$1.79 &$-$0.60$\pm$0.14 & IC~1805 152; B3\,V & (3) \\
J023244.2+613040 & 02:32:44.221 +61:30:40.21 & 41   &   1.35$\pm$0.25 &$-$0.19$\pm$0.18  &   1.52$\pm$0.27 &$-$0.65$\pm$0.17 &    4.11$\pm$0.52 &$-$0.73$\pm$0.11 & [SBJLK13]~145; g8\,IV & (1) \\
J023244.3+612722 & 02:32:44.301 +61:27:22.59 & 31:  &   3.66$\pm$0.72 &$-$0.33$\pm$0.17  &   3.36$\pm$0.54 &$-$0.55$\pm$0.12 &    6.17$\pm$1.16 &$-$0.37$\pm$0.15 & IC~1805 153; B2 & (7) \\
J023245.6+612631 & 02:32:45.688 +61:26:31.20 & 30   &   0.67$\pm$0.23 &$-$0.78$\pm$0.25  &   1.06$\pm$0.25 &$-$0.32$\pm$0.23 &    2.66$\pm$0.55 &$-$0.83$\pm$0.17  \\
J023245.8+612959 & 02:32:45.860 +61:29:59.04 & 12   &   2.20$\pm$0.29 &$-$0.29$\pm$0.13  &   1.97$\pm$0.28 &$-$0.32$\pm$0.14 &    7.03$\pm$0.59 &$-$0.37$\pm$0.08  \\
J023246.6+612729 & 02:32:46.673 +61:27:29.30 & 39   &   0.25$\pm$0.25 &   1.00$\pm$1.39  &   0.72$\pm$0.40 &$-$0.40$\pm$0.49 &    3.47$\pm$0.57 &$-$0.33$\pm$0.15  \\
J023247.1+613754 & 02:32:47.195 +61:37:54.77 & 166  &                 &                  &   0.34$\pm$0.22 &$-$0.32$\pm$0.67 &    1.74$\pm$0.50 &$-$0.85$\pm$0.25  \\
J023248.7+614050 & 02:32:48.719 +61:40:50.10 & 131  &                 &                  &   1.99$\pm$0.50 &$-$0.34$\pm$0.27 &                  &                  \\
J023248.7+612425 & 02:32:48.750 +61:24:25.54 & 175  &   0.31$\pm$0.16 &   0.10$\pm$0.51  &   0.63$\pm$0.20 &$-$0.27$\pm$0.31 &    1.33$\pm$0.36 &$-$0.09$\pm$0.27  \\
J023249.3+613615 & 02:32:49.325 +61:36:15.87 & 144  &                 &                  &   0.62$\pm$0.23 &$-$0.71$\pm$0.35 &    2.74$\pm$0.56 &$-$0.68$\pm$0.21  \\
J023249.5+612242 & 02:32:49.547 +61:22:42.07 & 2    &  24.93$\pm$0.96 &$-$0.76$\pm$0.03  &  26.43$\pm$1.12 &$-$0.65$\pm$0.03 &   83.05$\pm$1.91 &$-$0.83$\pm$0.01 & HD~15570; O4\,If$^+$ & (2) \\
J023250.3+613341 & 02:32:50.351 +61:33:41.92 & 56   &   1.81$\pm$0.31 &$-$0.81$\pm$0.14  &   1.69$\pm$0.28 &$-$0.50$\pm$0.15 &    3.96$\pm$0.53 &$-$0.75$\pm$0.12  \\
\hline
\end{tabular}
\end{center}
\end{sidewaystable*}
\addtocounter{table}{-1}
\begin{sidewaystable*}
\caption{Continued}
\tiny
\begin{center}
\begin{tabular}{c c c c c c c c c c c}
\hline
Name                 & J2000.0 Coord.         & XID & \multicolumn{2}{c}{MOS1} & \multicolumn{2}{c}{MOS2} & \multicolumn{2}{c}{pn} & Optical ID \& SpT & Note \\
XMMU                & hhmmss $^{\circ}$ ' ''   &     & Ct Rate  & HR & Ct Rate & HR & Ct Rate & HR & & \\
& & & (10$^{-3}$\,ct\,s$^{-1}$) & & (10$^{-3}$\,ct\,s$^{-1}$) & (10$^{-3}$\,ct\,s$^{-1}$) & \\
\hline
J023254.0+612803 & 02:32:54.072 +61:28:03.71 & 21   &   1.41$\pm$0.25 &$-$0.52$\pm$0.16  &   1.91$\pm$0.28 &$-$0.16$\pm$0.14 &    6.31$\pm$0.56 &$-$0.42$\pm$0.08  \\
J023254.2+613411 & 02:32:54.239 +61:34:11.98 & 43   &   0.85$\pm$0.23 &$-$0.88$\pm$0.16  &   1.23$\pm$0.26 &$-$0.65$\pm$0.18 &    3.58$\pm$0.49 &$-$1.00$\pm$0.06  \\
J023254.2+612527 & 02:32:54.266 +61:25:27.73 & 32   &   2.28$\pm$0.31 &$-$0.18$\pm$0.13  &   1.91$\pm$0.30 &$-$0.18$\pm$0.16 &    5.97$\pm$1.11 &$-$0.44$\pm$0.18  \\
J023256.4+612534 & 02:32:56.466 +61:25:34.49 & 57   &   1.03$\pm$0.23 &$-$0.83$\pm$0.16  &   0.88$\pm$0.23 &   0.17$\pm$0.26 &    2.92$\pm$0.51 &$-$0.69$\pm$0.14  \\
J023256.7+613040 & 02:32:56.711 +61:30:40.62 & 58   &   1.27$\pm$0.24 &   0.13$\pm$0.19  &   1.14$\pm$0.24 &   0.33$\pm$0.20 &    3.17$\pm$0.52 &   0.18$\pm$0.17  \\
J023257.2+612108 & 02:32:57.290 +61:21:08.55 & 145  &                 &                  &   0.77$\pm$0.25 &$-$0.35$\pm$0.32 &    2.34$\pm$0.49 &   0.22$\pm$0.21  \\
J023257.9+612726 & 02:32:57.964 +61:27:26.21 & 44   &   1.60$\pm$0.25 &$-$0.20$\pm$0.16  &   1.54$\pm$0.26 &$-$0.29$\pm$0.16 &    1.12$\pm$1.04 &   0.27$\pm$0.91 & IC~1805 166; B2\,V & (3) \\
J023259.1+612556 & 02:32:59.182 +61:25:56.50 & 158  &   0.50$\pm$0.18 &$-$0.71$\pm$0.31  &   0.53$\pm$0.19 &$-$0.36$\pm$0.35 &    1.25$\pm$0.39 &$-$0.78$\pm$0.26  \\
J023300.0+612503 & 02:33:00.016 +61:25:03.08 & 9    &   3.19$\pm$0.47 &$-$0.38$\pm$0.14  &   3.72$\pm$0.39 &$-$0.23$\pm$0.10 &   12.24$\pm$0.77 &$-$0.29$\pm$0.06  \\
J023300.8+612704 & 02:33:00.800 +61:27:04.90 & 183  &   0.27$\pm$0.15 &$-$1.00$\pm$0.47  &   0.22$\pm$0.13 &$-$1.00$\pm$0.35 &    1.41$\pm$0.36 &$-$0.59$\pm$0.24  \\
J023300.8+612912 & 02:33:00.868 +61:29:12.00 & 154  &   0.48$\pm$0.16 &   0.53$\pm$0.33  &   0.68$\pm$0.19 &   0.83$\pm$0.22 &                  &                  \\
J023300.9+612608 & 02:33:00.903 +61:26:08.68 & 22   &   2.12$\pm$0.29 &$-$0.39$\pm$0.13  &   2.17$\pm$0.30 &$-$0.37$\pm$0.13 &    6.85$\pm$0.61 &$-$0.33$\pm$0.08  \\
J023300.9+613737 & 02:33:00.969 +61:37:37.82 & 14   &   6.42$\pm$0.67 &$-$0.26$\pm$0.10  &   4.93$\pm$0.61 &$-$0.12$\pm$0.13 &   15.76$\pm$1.15 &$-$0.36$\pm$0.07  \\
J023301.5+613134 & 02:33:01.502 +61:31:34.08 & 88   &   0.57$\pm$0.18 &$-$0.17$\pm$0.31  &   1.12$\pm$0.23 &$-$0.01$\pm$0.21 &    2.31$\pm$0.80 &   0.11$\pm$0.35  \\
J023302.9+612605 & 02:33:02.926 +61:26:05.82 & 87   &   0.93$\pm$0.21 &   0.47$\pm$0.23  &   0.48$\pm$0.19 &   0.28$\pm$0.40 &    1.51$\pm$0.37 &   0.84$\pm$0.25  \\
J023303.8+612909 & 02:33:03.866 +61:29:09.29 & 178  &   0.46$\pm$0.16 &$-$0.91$\pm$0.19  &   0.46$\pm$0.19 &$-$0.00$\pm$0.41 &    1.57$\pm$0.38 &$-$0.59$\pm$0.24 &  WISE~J023303.82+612909.6  & (*) \\
J023304.6+613343 & 02:33:04.600 +61:33:43.11 & 176  &   0.63$\pm$0.22 &$-$0.67$\pm$0.35  &   0.00$\pm$0.09 &                 &    0.86$\pm$0.30 &$-$1.00$\pm$0.22  \\
J023305.1+612249 & 02:33:05.181 +61:22:49.44 & 118  &   0.22$\pm$0.35 &   0.11$\pm$1.56  &   1.05$\pm$0.30 &$-$0.62$\pm$0.22 &    2.13$\pm$0.56 &$-$0.73$\pm$0.25  \\
J023305.2+612706 & 02:33:05.282 +61:27:06.94 & 35   &   1.07$\pm$0.22 &$-$0.44$\pm$0.19  &   1.32$\pm$0.25 &$-$0.17$\pm$0.19 &    3.81$\pm$0.50 &$-$0.69$\pm$0.11 & IC~1805 175; B2.5\,V & (3) \\
J023305.6+612201 & 02:33:05.696 +61:22:01.79 & 23   &                 &                  &   2.21$\pm$0.62 &$-$0.33$\pm$0.28 &   12.20$\pm$0.85 &$-$0.19$\pm$0.07  \\
J023306.1+613731 & 02:33:06.136 +61:37:31.49 & 62   &   1.80$\pm$0.40 &$-$0.33$\pm$0.22  &   1.34$\pm$0.34 &$-$0.81$\pm$0.23 &    2.82$\pm$0.95 &$-$0.39$\pm$0.35  \\
J023307.2+613609 & 02:33:07.215 +61:36:09.31 & 168  &   0.67$\pm$0.27 &$-$0.60$\pm$0.37  &   0.82$\pm$0.27 &   0.00$\pm$0.33 &    2.43$\pm$0.52 &$-$0.41$\pm$0.22  \\
J023307.8+614043 & 02:33:07.825 +61:40:43.84 & 157: &   0.90$\pm$0.35 &$-$0.21$\pm$0.40  &   1.12$\pm$0.43 &   0.06$\pm$0.38 &                  &                  \\
J023308.9+612847 & 02:33:08.981 +61:28:47.78 & 110  &   0.46$\pm$0.18 &$-$0.52$\pm$0.36  &   0.43$\pm$0.16 &$-$0.71$\pm$0.33 &    1.83$\pm$0.37 &$-$0.91$\pm$0.16 & [MJD95]~J023308.90+612845.7; b8\,IV, Ae/Be & (1) \\
J023309.0+612521 & 02:33:09.085 +61:25:21.41 & 25   &   1.70$\pm$0.30 &$-$0.71$\pm$0.14  &   2.38$\pm$0.32 &$-$0.58$\pm$0.12 &    6.12$\pm$0.59 &$-$0.47$\pm$0.09 & [SBJLK13]~206; g5\,V & (1) \\
J023310.2+612500 & 02:33:10.208 +61:25:00.59 & 79   &   0.74$\pm$0.21 &   0.03$\pm$0.28  &   0.38$\pm$0.18 &$-$0.42$\pm$0.50 &    2.03$\pm$0.41 &$-$0.24$\pm$0.20  \\
J023310.3+613207 & 02:33:10.352 +61:32:07.91 & 164  &   0.38$\pm$0.17 &$-$0.54$\pm$0.45  &   0.20$\pm$0.14 &$-$0.31$\pm$0.71 &    1.18$\pm$0.32 &$-$0.92$\pm$0.20 & WISE~J023310.03+613208.9 & (*) \\
J023310.3+614106 & 02:33:10.378 +61:41:06.94 & 50   &   3.91$\pm$0.68 &$-$0.05$\pm$0.18  &   5.31$\pm$0.80 &   0.09$\pm$0.15 &                  &                  \\
J023310.4+612402 & 02:33:10.483 +61:24:02.89 & 117  &   0.37$\pm$0.16 &$-$1.00$\pm$0.21  &   0.59$\pm$0.19 &$-$0.27$\pm$0.31 &    1.44$\pm$0.37 &$-$0.37$\pm$0.24  \\
J023311.1+613018 & 02:33:11.139 +61:30:18.39 & 34   &   1.67$\pm$0.27 &$-$0.56$\pm$0.14  &   1.23$\pm$0.24 &$-$0.70$\pm$0.16 &    3.46$\pm$0.47 &$-$0.40$\pm$0.13 & [SBJLK13]~212; k-m\,V & (1) \\
J023311.5+613114 & 02:33:11.577 +61:31:14.86 & 133  &   0.23$\pm$0.15 &   0.02$\pm$0.67  &   0.62$\pm$0.20 &$-$0.95$\pm$0.23 &    2.37$\pm$0.45 &   0.02$\pm$0.19 & WISE~J023311.42+613114.9  & (*) \\
J023312.0+613058 & 02:33:12.069 +61:30:58.13 & 150  &   0.39$\pm$0.15 &$-$1.00$\pm$0.16  &   0.61$\pm$0.18 &$-$1.00$\pm$0.08 &    1.90$\pm$0.40 &$-$0.97$\pm$0.13 & IC~1805 184; f5\,V & (1) \\
J023312.9+613130 & 02:33:12.974 +61:31:30.94 & 152  &   0.65$\pm$0.20 &$-$1.00$\pm$0.26  &   0.45$\pm$0.19 &$-$0.67$\pm$0.33 &    1.72$\pm$0.41 &$-$0.28$\pm$0.23  \\
J023313.5+613152 & 02:33:13.517 +61:31:52.33 & 130  &   0.48$\pm$0.19 &$-$0.49$\pm$0.37  &   0.39$\pm$0.18 &$-$0.57$\pm$0.40 &    1.22$\pm$0.39 &$-$0.66$\pm$0.33  \\
J023314.6+613011 & 02:33:14.658 +61:30:11.79 & 129  &   0.47$\pm$0.18 &   0.05$\pm$0.39  &   0.66$\pm$0.20 &$-$0.23$\pm$0.29 &    1.46$\pm$0.38 &$-$0.41$\pm$0.24  \\
J023314.8+612402 & 02:33:14.802 +61:24:02.34 & 27   &   1.25$\pm$0.25 &$-$0.63$\pm$0.18  &   1.98$\pm$0.31 &$-$0.74$\pm$0.12 &    5.19$\pm$0.56 &$-$0.66$\pm$0.10 & IC~1805 187; b-f & (1) \\
J023315.4+612604 & 02:33:15.489 +61:26:04.80 & 123  &   0.30$\pm$0.16 &$-$0.29$\pm$0.51  &   0.63$\pm$0.19 &$-$0.64$\pm$0.28 &    1.10$\pm$0.32 &$-$0.99$\pm$0.16  \\
J023315.7+613144 & 02:33:15.787 +61:31:44.92 & 64   &   0.92$\pm$0.54 &   0.39$\pm$0.61  &   1.51$\pm$0.36 &$-$0.24$\pm$0.23 &    4.66$\pm$0.55 &$-$0.56$\pm$0.11  \\
J023316.1+612045 & 02:33:16.193 +61:20:45.92 & 102  &                 &                  &   1.21$\pm$0.34 &$-$0.36$\pm$0.28 &    6.81$\pm$1.20 &$-$0.66$\pm$0.17  \\
J023317.6+612808 & 02:33:17.669 +61:28:08.39 & 20   &   2.43$\pm$0.33 &$-$0.54$\pm$0.12  &   2.75$\pm$0.33 &$-$0.70$\pm$0.09 &    5.76$\pm$0.57 &$-$0.39$\pm$0.09 & [SBJLK13]~228; g &  \\
J023317.8+612827 & 02:33:17.858 +61:28:27.38 & 69   &   0.36$\pm$0.17 &$-$0.04$\pm$0.47  &   0.52$\pm$0.20 &$-$0.55$\pm$0.33 &    2.25$\pm$0.44 &$-$0.81$\pm$0.17  \\
J023320.3+612208 & 02:33:20.356 +61:22:08.49 & 126  &                 &                  &   0.76$\pm$0.25 &$-$0.13$\pm$0.33 &    2.63$\pm$0.52 &$-$0.40$\pm$0.20 & [MJD95]~J023320.18+612211.0; F8 & (1) \\
J023320.7+613117 & 02:33:20.733 +61:31:17.89 & 4    &  13.26$\pm$0.71 &$-$1.00$\pm$0.01  &  11.75$\pm$0.67 &$-$0.98$\pm$0.02 &   48.83$\pm$1.46 &$-$0.99$\pm$0.01 & HD~15629; O5\,V((fc)) & (2) \\
J023321.6+613014 & 02:33:21.698 +61:30:14.12 & 125  &   0.10$\pm$0.16 &$-$1.00$\pm$1.84  &   0.56$\pm$0.20 &$-$0.76$\pm$0.27 &    2.24$\pm$0.43 &$-$0.61$\pm$0.19  \\
J023322.5+612509 & 02:33:22.545 +61:25:09.14 & 73   &   1.03$\pm$0.24 &   0.46$\pm$0.24  &   0.59$\pm$0.24 &$-$0.08$\pm$0.41 &    1.59$\pm$0.40 &   0.34$\pm$0.25  \\
J023322.8+612740 & 02:33:22.881 +61:27:40.49 & 60   &   0.93$\pm$0.22 &$-$0.33$\pm$0.23  &   1.05$\pm$0.24 &$-$0.31$\pm$0.22 &    2.82$\pm$0.45 &$-$0.67$\pm$0.15  \\
J023323.6+612443 & 02:33:23.643 +61:24:43.10 & 47   &   1.50$\pm$0.28 &$-$0.22$\pm$0.19  &   1.46$\pm$0.30 &$-$0.18$\pm$0.21 &    3.69$\pm$0.51 &$-$0.33$\pm$0.13  \\
J023325.0+613553 & 02:33:25.090 +61:35:53.94 & 114  &   0.54$\pm$0.27 &   0.24$\pm$0.48  &   0.99$\pm$0.30 &$-$0.20$\pm$0.31 &    2.15$\pm$0.57 &$-$0.23$\pm$0.27  \\
\hline
\end{tabular}
\end{center}
\end{sidewaystable*}
\addtocounter{table}{-1}
\begin{sidewaystable*}
\caption{Continued}
\tiny
\begin{center}
\begin{tabular}{c c c c c c c c c c c}
\hline
Name                 & J2000.0 Coord.         & XID & \multicolumn{2}{c}{MOS1} & \multicolumn{2}{c}{MOS2} & \multicolumn{2}{c}{pn} & Optical ID \& SpT & Note \\
XMMU                & hhmmss $^{\circ}$ ' ''   &     & Ct Rate  & HR & Ct Rate & HR & Ct Rate & HR & & \\
& & & (10$^{-3}$\,ct\,s$^{-1}$) & & (10$^{-3}$\,ct\,s$^{-1}$) & (10$^{-3}$\,ct\,s$^{-1}$) & \\
\hline
J023325.2+612719 & 02:33:25.280 +61:27:19.57 & 33   &   1.12$\pm$0.24 &$-$0.60$\pm$0.17  &   1.26$\pm$0.25 &$-$0.29$\pm$0.19 &    2.33$\pm$0.42 &$-$0.43$\pm$0.16  \\
J023326.4+613531 & 02:33:26.482 +61:35:31.95 & 186: &   0.39$\pm$0.23 &   0.79$\pm$0.49  &   1.05$\pm$0.34 &$-$0.60$\pm$0.35 &    1.69$\pm$0.55 &   0.10$\pm$0.32  \\
J023328.6+612527 & 02:33:28.658 +61:25:27.82 & 189  &   0.48$\pm$0.20 &$-$0.57$\pm$0.38  &   0.62$\pm$0.22 &$-$0.27$\pm$0.37 &    1.21$\pm$0.33 &$-$0.89$\pm$0.21  \\
J023330.2+612855 & 02:33:30.263 +61:28:55.34 & 116  &   0.47$\pm$0.19 &$-$0.92$\pm$0.25  &   0.00$\pm$0.07 &                 &    1.76$\pm$0.38 &$-$0.43$\pm$0.21 & IC~1805 200; B4\,V & (3) \\
J023331.2+612922 & 02:33:31.251 +61:29:22.91 & 113  &   0.79$\pm$0.24 &$-$0.54$\pm$0.25  &   1.14$\pm$0.27 &$-$0.63$\pm$0.21 &    0.79$\pm$0.47 &$-$0.28$\pm$0.62  \\
J023332.7+612514 & 02:33:32.735 +61:25:14.37 & 107  &   0.59$\pm$0.20 &$-$0.56$\pm$0.32  &   1.08$\pm$0.26 &$-$0.46$\pm$0.24 &    1.38$\pm$0.38 &$-$0.79$\pm$0.26  \\
J023332.8+613545 & 02:33:32.815 +61:35:45.09 & 90   &   1.14$\pm$0.32 &$-$0.73$\pm$0.25  &   1.07$\pm$0.29 &$-$1.00$\pm$0.22 &    3.25$\pm$0.63 &$-$0.85$\pm$0.19  \\
J023333.0+612917 & 02:33:33.097 +61:29:17.21 & 67   &   1.06$\pm$0.26 &$-$0.18$\pm$0.24  &   2.27$\pm$0.36 &   0.10$\pm$0.16 &    2.62$\pm$0.46 &   0.05$\pm$0.18  \\
J023333.4+612855 & 02:33:33.421 +61:28:55.37 & 100  &   1.16$\pm$0.27 &   0.52$\pm$0.24  &   1.14$\pm$0.28 &   0.16$\pm$0.25 &    1.93$\pm$0.44 &   0.25$\pm$0.23  \\
J023334.4+612703 & 02:33:34.492 +61:27:03.99 & 135  &   0.27$\pm$0.18 &   1.00$\pm$0.46  &   0.61$\pm$0.20 &$-$0.92$\pm$0.21 &    1.95$\pm$0.43 &   0.01$\pm$0.22  \\
J023334.5+613840 & 02:33:34.596 +61:38:40.27 & 84   &   1.41$\pm$0.40 &$-$0.82$\pm$0.25  &   1.55$\pm$0.37 &$-$0.96$\pm$0.15 &    2.60$\pm$0.77 &$-$1.00$\pm$0.11  \\
J023334.6+611921 & 02:33:34.624 +61:19:21.91 & 182  &                 &                  &   1.87$\pm$0.53 &   0.94$\pm$0.12 &    3.23$\pm$0.74 &   0.59$\pm$0.22  \\
J023335.0+612811 & 02:33:35.012 +61:28:11.24 & 181  &   0.61$\pm$0.22 &   0.22$\pm$0.38  &   0.90$\pm$0.28 &   0.39$\pm$0.31 &    0.71$\pm$0.32 &   1.00$\pm$0.44  \\
J023335.4+612852 & 02:33:35.472 +61:28:52.99 & 72   &   1.30$\pm$0.28 &$-$0.40$\pm$0.21  &   1.65$\pm$0.32 &$-$0.05$\pm$0.19 &    3.69$\pm$0.51 &$-$0.42$\pm$0.13  \\
J023336.2+613637 & 02:33:36.254 +61:36:37.31 & 112  &   0.81$\pm$0.32 &$-$0.51$\pm$0.38  &   1.57$\pm$0.39 &$-$0.62$\pm$0.26 &    3.05$\pm$0.66 &$-$1.00$\pm$0.19  \\
J023338.3+613103 & 02:33:38.328 +61:31:03.57 & 99   &   1.02$\pm$0.30 &$-$0.18$\pm$0.29  &   1.63$\pm$0.34 &   0.11$\pm$0.20 &    3.70$\pm$0.57 &$-$0.54$\pm$0.15  \\
J023341.5+612502 & 02:33:41.505 +61:25:02.47 & 68   &                 &                  &   0.81$\pm$0.58 &$-$1.00$\pm$0.96 &    5.87$\pm$0.65 &$-$0.70$\pm$0.10  \\
J023344.2+612440 & 02:33:44.232 +61:24:40.88 & 179  &                 &                  &   0.71$\pm$0.27 &   0.49$\pm$0.35 &    1.35$\pm$0.45 &$-$0.46$\pm$0.35  \\
J023346.0+612754 & 02:33:46.077 +61:27:54.25 & 147  &   0.71$\pm$0.23 &$-$0.57$\pm$0.27  &   0.93$\pm$0.27 &$-$0.56$\pm$0.28 &    2.84$\pm$0.87 &$-$0.94$\pm$0.26  \\
J023347.7+612453 & 02:33:47.769 +61:24:53.37 & 155  &   0.73$\pm$0.29 &$-$0.56$\pm$0.35  &   0.48$\pm$0.26 &   1.00$\pm$0.19 &    1.51$\pm$0.45 &   1.00$\pm$0.29  \\
J023348.5+612836 & 02:33:48.591 +61:28:36.48 & 174  &   0.58$\pm$0.22 &$-$0.86$\pm$0.23  &   0.07$\pm$0.11 &$-$0.07$\pm$1.62 &    2.46$\pm$0.51 &$-$0.57$\pm$0.21  \\
J023351.6+612213 & 02:33:51.691 +61:22:13.54 & 77   &                 &                  &   1.71$\pm$0.40 &$-$0.61$\pm$0.24 &    5.83$\pm$0.84 &$-$0.27$\pm$0.14  \\
J023351.8+613355 & 02:33:51.857 +61:33:55.36 & 74   &   2.35$\pm$0.45 &$-$0.31$\pm$0.19  &   1.76$\pm$0.43 &$-$0.46$\pm$0.26 &    6.34$\pm$0.83 &$-$0.62$\pm$0.13  \\
J023352.1+612525 & 02:33:52.115 +61:25:25.80 & 127  &   0.55$\pm$0.25 &$-$0.03$\pm$0.45  &   0.72$\pm$0.30 &$-$0.67$\pm$0.39 &    1.96$\pm$0.50 &$-$0.71$\pm$0.28  \\
J023352.2+613737 & 02:33:52.234 +61:37:37.72 & 96   &   1.12$\pm$0.37 &$-$0.70$\pm$0.34  &   2.08$\pm$0.47 &$-$0.71$\pm$0.22 &                  &                  \\
J023352.6+612615 & 02:33:52.667 +61:26:15.06 & 142  &   0.71$\pm$0.26 &$-$0.60$\pm$0.30  &   0.89$\pm$0.28 &$-$0.45$\pm$0.32 &    0.66$\pm$0.45 &$-$0.39$\pm$0.68  \\
J023353.0+613717 & 02:33:53.040 +61:37:17.01 & 137  &   0.59$\pm$0.34 &$-$0.16$\pm$0.59  &   2.53$\pm$0.61 &   0.09$\pm$0.23 &                  &                  \\
J023353.7+612151 & 02:33:53.766 +61:21:51.80 & 108  &                 &                  &   1.64$\pm$0.48 &   0.42$\pm$0.25 &    3.66$\pm$0.72 &   0.53$\pm$0.18  \\
J023354.4+613145 & 02:33:54.436 +61:31:45.95 & 26   &   2.84$\pm$0.64 &$-$0.45$\pm$0.22  &   3.62$\pm$0.52 &$-$0.38$\pm$0.15 &    9.18$\pm$0.90 &$-$0.35$\pm$0.10  \\
J023357.4+612626 & 02:33:57.484 +61:26:26.81 & 122  &   0.78$\pm$0.27 &$-$0.73$\pm$0.28  &   0.60$\pm$0.26 &$-$0.53$\pm$0.48 &    1.85$\pm$0.51 &$-$0.53$\pm$0.27 & [MJD95]~J023357.20+612627.9; B5 & (1) \\
J023359.8+613215 & 02:33:59.824 +61:32:15.58 & 82   &   0.56$\pm$0.35 &$-$0.86$\pm$0.44  &   1.40$\pm$0.37 &$-$0.65$\pm$0.27 &    5.48$\pm$0.76 &$-$0.55$\pm$0.13  \\
J023402.6+612310 & 02:34:02.631 +61:23:10.34 & 37   &                 &                  &   2.13$\pm$0.41 &$-$0.95$\pm$0.11 &    7.71$\pm$0.89 &$-$1.00$\pm$0.09 & BD+60$^{\circ}$\,513; O7\,Vnz & (6) \\
J023403.5+613323 & 02:34:03.573 +61:33:23.02 & 52   &   1.88$\pm$0.42 &$-$0.07$\pm$0.22  &   2.51$\pm$0.53 &   0.42$\pm$0.18 &    5.16$\pm$0.83 &   0.17$\pm$0.16  \\
J023405.4+612940 & 02:34:05.426 +61:29:40.80 & 138  &   1.48$\pm$0.37 &   0.18$\pm$0.25  &   0.83$\pm$0.35 &   1.00$\pm$0.29 &    2.45$\pm$0.62 &   0.40$\pm$0.25  \\
J023407.5+612817 & 02:34:07.576 +61:28:17.89 & 8    &   8.48$\pm$0.72 &$-$0.90$\pm$0.05  &   7.13$\pm$0.69 &$-$0.90$\pm$0.07 &   29.91$\pm$1.47 &$-$0.95$\pm$0.03  \\
J023407.9+612936 & 02:34:07.984 +61:29:36.49 & 95   &   0.83$\pm$0.32 &$-$0.75$\pm$0.28  &   1.09$\pm$0.34 &$-$0.50$\pm$0.32 &    2.81$\pm$0.59 &$-$0.88$\pm$0.18 & [MJD95]~J023407.72+612933.8 & \\
J023410.3+612438 & 02:34:10.385 +61:24:38.83 & 53   &   1.75$\pm$0.43 &$-$0.84$\pm$0.18  &   2.30$\pm$0.52 &$-$0.42$\pm$0.24 &    4.65$\pm$0.78 &$-$0.78$\pm$0.17 & IC~1805 237; A0 & (5) \\
J023412.6+612611 & 02:34:12.631 +61:26:11.78 & 98   &   1.60$\pm$0.40 &$-$0.42$\pm$0.23  &   1.63$\pm$0.44 &$-$0.12$\pm$0.28 &    2.46$\pm$0.60 &$-$0.78$\pm$0.25  \\
J023414.7+613027 & 02:34:14.748 +61:30:27.73 & 106  &   0.79$\pm$0.34 &   0.12$\pm$0.44  &   1.96$\pm$0.51 &   0.34$\pm$0.23 &    3.35$\pm$0.71 &$-$0.02$\pm$0.21  \\
J023415.6+612442 & 02:34:15.616 +61:24:42.89 & 143  &   0.67$\pm$0.32 &$-$0.81$\pm$0.32  &   1.33$\pm$0.51 &   0.52$\pm$0.30 &    4.02$\pm$0.80 &$-$0.51$\pm$0.20  \\
J023416.5+612741 & 02:34:16.545 +61:27:41.33 & 159  &   0.52$\pm$0.26 &$-$1.00$\pm$0.25  &   1.49$\pm$0.47 &   0.26$\pm$0.29 &    2.49$\pm$0.64 &$-$0.42$\pm$0.27  \\
J023419.0+612446 & 02:34:19.065 +61:24:46.68 & 81   &   2.71$\pm$0.55 &$-$0.29$\pm$0.20  &   1.62$\pm$0.47 &$-$0.59$\pm$0.33 &    7.80$\pm$1.02 &$-$0.32$\pm$0.13  \\
J023420.2+613532 & 02:34:20.244 +61:35:32.12 & 162  &   1.25$\pm$0.43 &$-$0.34$\pm$0.37  &   0.66$\pm$0.38 &$-$0.25$\pm$0.62 &                  &                  \\
J023421.6+612342 & 02:34:21.667 +61:23:42.99 & 188  &                 &                  &   2.22$\pm$0.62 &   0.60$\pm$0.21 &    3.33$\pm$0.87 &   0.36$\pm$0.25  \\
J023426.9+613307 & 02:34:26.934 +61:33:07.15 & 136  &   1.60$\pm$0.64 &   0.06$\pm$0.40  &   1.63$\pm$0.49 &$-$0.10$\pm$0.31 &                  &                  \\
J023428.3+613341 & 02:34:28.334 +61:33:41.67 & 89   &   2.31$\pm$0.95 &   0.55$\pm$0.36  &   4.31$\pm$0.79 &   0.94$\pm$0.07 &                  &                  \\
J023431.0+613037 & 02:34:31.094 +61:30:37.78 & 149  &   1.71$\pm$0.52 &$-$0.47$\pm$0.31  &   0.94$\pm$0.38 &$-$0.53$\pm$0.45 &                  &                 & IC~1805 260; B2\,V & (3) \\
\hline
\end{tabular}
\end{center}
\tablefoot{XID indicates the source number ordered by decreasing flux. An XID followed by a colon indicates sources that are doubtful because they fall (partially) in a detector gap or because they fall onto the wings of a bright source. HR is the hardness ratio defined as $(H-S)/(H+S)$ where $S$ and $H$ refer to the count rates in the soft (0.4 -- 2.0\,keV) and hard (2.0 -- 10.0\,keV) energy bands. IC~1805, [MJD95], [SBJLK13] numbers refer to the catalogues of \citet{Vasilevskis}, \citet{Massey} and \citet{Vilnius}, respectively. Small letters for the spectral type indicate photometric spectral classifications given by \citet{Vilnius}. The numerical notes in the last column refer to the bibliographic reference for spectral types: (1) \citet{Vilnius}, (2) this work, (3) \citet{SH99}, (4) \citet{Antipin}, (5) \citet{Wolff}, (6) \citet{Sota}, and (7) \citet{Ishida}. An asterisk identifies {\it WISE} sources that were classified as YSO candidates by \citet{Vilnius}. Finally, the double asterisk identifies the Algol-type eclipsing binary V~1166\,Cas (= IC~1805 111 = LS\,I $+61^{\circ}$~275) which has an orbital period of 1.3210\,d \citep{Antipin}.} % 74% membershi$\log{L_{\rm X}/L_{\rm bol}} = -7.03$p prob (VSvA), 53% Sanders 
\end{sidewaystable*}

\newpage
\section{Newly determined radial velocities \label{Journalopt}}
In this section we provide the journal of the new optical spectra of the O-type stars obtained with the HEROS spectrograph at the 1.2~m TIGRE telescope along with the RVs inferred from these data.

\begin{table}[h]
\caption{RVs of the He\,{\sc ii} $\lambda$\,4542 line as measured on the HEROS spectra of HD~15558. Typical uncertainties are 5\,km\,s$^{-1}$.\label{RV15558}}
\begin{center}
\begin{tabular}{c c}
\hline
HJD$-2\,450\,000$ & RV$_1$ \\
                  & km\,s$^{-1}$ \\
\hline 
    6512.867 & $-26.1$ \\
    6518.819 & $-35.8$ \\
    6533.823 & $-22.5$ \\
    6537.853 & $-21.6$ \\
    6560.899 & $-14.3$ \\
    6566.885 & $-16.3$ \\
    6575.830 & $-12.1$ \\
    6580.770 & $-19.3$ \\
    6581.901 &  $-5.6$ \\
    6584.809 &  $-8.4$ \\
    6592.667 & $-11.4$ \\
    6606.784 & $-12.5$ \\
    6640.688 & $-21.5$ \\
    6662.591 & $-31.2$ \\
    6689.592 & $-66.9$ \\
    6690.645 & $-65.4$ \\
\hline
\end{tabular}
\end{center}
\end{table}

\begin{table}[h]
\caption{Newly determined RVs of HD~15570\label{RV15570}}
\begin{center}
\begin{tabular}{c c c c}
\hline
HJD$-2\,450\,000$ & \multicolumn{3}{c}{RV (km\,s$^{-1}$)}\\
& He\,{\sc ii} $\lambda$\,4542 & N\,{\sc iii} $\lambda$\,4634 & N\,{\sc iii} $\lambda$\,4641 \\
\hline 
 6509.918 & $-61.0$ & $-69.0$ & $-73.3$ \\ % 4633.184 4639.886
 6519.877 & $-65.2$ & $-58.9$ & $-74.5$ \\ % 4633.339 4639.867
 6680.564 & $-56.6$ & $-59.3$ & $-72.2$ \\ % 4633.333 4639.903
 6958.815 & $-58.3$ & $-61.6$ & $-70.5$ \\ % 4633.298 4639.928
\hline
Mean & $-60.3 \pm 3.8$ & $-62.2 \pm 4.7$ & $-72.6 \pm 1.7$ \\
\hline
\end{tabular}
\end{center}
\end{table}

\begin{table}[h]
\caption{Newly determined RVs of HD~15629\label{RV15629}}
\begin{center}
\begin{tabular}{c c c c}
\hline
HJD$-2\,450\,000$ & \multicolumn{3}{c}{RV (km\,s$^{-1}$)}\\
& He\,{\sc ii} $\lambda$\,4542 & N\,{\sc iii} $\lambda$\,4634 & N\,{\sc iii} $\lambda$\,4641 \\
\hline 
 6680.658 & $-53.8$ & $-58.5$ & $-79.1$ \\ % 4470.515 -> -66.8 
 6958.757 & $-52.7$ & $-78.7$ & $-78.2$ \\ % 4470.540 -> -65.2
\hline
Mean & $-53.3 \pm 0.8$ & $-68.6 \pm 14.3$ & $-78.7 \pm 0.6$ \\
\hline
\end{tabular}
\end{center}
\end{table}

\begin{table}
\caption{Newly determined RVs of BD+60$^{\circ}$~497 \label{RV497}}
\begin{center}
\begin{tabular}{c c c c}
\hline
HJD$-2\,450\,000$ & RV$_1$ & RV$_2$ & Instrument \\
                  & km\,s$^{-1}$ & km\,s$^{-1}$ & \\
\hline 
    2520.644 & $-107.1$   &   $110.7$ & A \\ 
    2523.600 &  $ 74.4$   &  $-222.8$ & A \\
    2524.552 &  $-99.3$   &   $102.7$ & A \\
    2527.555 &   $78.3$   &  $-232.7$ & A \\
    2528.533 & $-112.1$   &    $99.2$ & A \\
    2529.562 & $-147.3$   &    $77.0$ & A \\
    2531.543 &  $ 71.5$   &  $-236.4$ & A \\
    2532.534 & $-122.6$   &   $106.6$ & A \\
    2533.638 & $-137.3$   &    $48.4$ & A \\
    2916.583 & $-143.5$   &   $130.4$ & A \\
    2918.667 &   $40.9$   &  $-212.9$ & A \\
    2919.632 &   $36.2$   &  $-222.7$ & A \\
    2922.677 &   $54.4$   &  $-216.3$ & A \\
    2925.665 &  $-87.9$   &    $52.7$ & A \\
    2928.641 & $-139.1$   &   $172.5$ & A \\
    2934.545 &   $53.0$   &  $-219.4$ & A \\
    6681.572 &  $-53.9$   &   $-47.0$ & H \\
    6685.581 &  $-88.0$   &    $ 3.2$ & H \\
    6686.579 & $-143.9$   &   $173.4$ & H \\
    6687.672 &   $10.9$   &  $-199.1$ & H \\
    6688.573 &  $ 72.7$   &  $-233.6$ & H \\
    6682.573 & $-152.7$   &   $175.5$ & H \\
    6684.571 &   $55.7$   &  $-245.0$ & H \\
\hline
\end{tabular}
\tablefoot{The RVs were determined through cross-correlation in the disentangling process. The A and H letters in the last column indicate Aur\'elie and HEROS data, respectively.}
\end{center}
\end{table}

\begin{table}
\caption{Newly determined RVs of BD+60$^{\circ}$~498 \label{RV498}}
\begin{center}
\begin{tabular}{c c c}
\hline
HJD$-2\,450\,000$ & RV & $\sigma$(RV) \\
                  & km\,s$^{-1}$ & km\,s$^{-1}$\\
\hline 
    6693.578 & $-114.0$ & 19.9 \\
    6694.581 & $-124.7$ &  9.7 \\
    6695.586 & $-63.7$ & 38.8 \\
    6697.590 & $-58.6$ & 11.2 \\
    6699.569 & $-34.7$ & 43.9 \\
    6700.585 & $-11.2$ &  9.6 \\
    6713.594 & $-23.7$ & 13.2 \\
    6716.596 & $-40.4$ &  9.9 \\
    6960.821 & $-47.9$ & 24.3 \\
    6961.758 & $-48.7$ & 28.0 \\
    6979.719 & $-42.9$ & 12.4 \\   
    6980.733 & $-69.4$ & 13.2 \\
\hline
\end{tabular}
\tablefoot{The RVs are the mean of the values measured on the He\,{\sc i} $\lambda\lambda$\,4026, 4121, 4144, 4388, 4471, 4713, H$\gamma$, H$\beta$ and He\,{\sc ii} $\lambda$\,4686 lines.}
\end{center}
\end{table}
\begin{table}
\caption{Newly determined RVs of BD+60$^{\circ}$~499 \label{RV499}}
\begin{center}
\begin{tabular}{c c c}
\hline
HJD$-2\,450\,000$ & RV & $\sigma$(RV) \\
                  & km\,s$^{-1}$ & km\,s$^{-1}$\\
\hline 
 6696.589 & $-45.9$ & 5.2 \\
 6699.628 & $-47.3$ & 4.9 \\
\hline
\end{tabular}
\tablefoot{The RVs are the mean of the values measured on the He\,{\sc i} $\lambda\lambda$\,4026, 4121, 4144, 4388, 4471, 4713, H$\gamma$, H$\beta$ and He\,{\sc ii} $\lambda$\,4686 lines.}
\end{center}
\end{table}
\begin{table}
\caption{Newly determined RV of BD+60$^{\circ}$~513 \label{RV513}}
\begin{center}
\begin{tabular}{c c c}
\hline
HJD$-2\,450\,000$ & RV & $\sigma$(RV) \\
                  & km\,s$^{-1}$ & km\,s$^{-1}$\\
\hline 
 6698.600 & $-65.7$ & 9.7 \\
\hline
\end{tabular}
\tablefoot{The RV is the mean of the values measured on the He\,{\sc i} $\lambda$\,4471, He\,{\sc ii} $\lambda\lambda$\,4542, 4686 and H$\beta$ lines.}
\end{center}
\end{table}

\end{appendix}

\begin{thebibliography}{}
\bibitem[Antipin(2008)]{Antipin}
Antipin, S.V. 2008, Peremennye Zvezdy Prilozhenie 8,, 19A
\bibitem[Arnaud(1996)]{Arnaud} 
Arnaud, K.A.\ 1996, in {\it Astronomical Data Analysis Software and Systems V}, eds.\ G.\ Jacoby, \& J.\ Barnes, ASP Conf.\ Series, 101, 17
\bibitem[Asplund et al.(2009)]{Asplund}
Asplund, M., Grevesse, N., Sauval, A.J., \& Scott, P.\ 2009, ARA\&A, 47, 481
\bibitem[Babel \& Montmerle(1997)]{BabelMontmerle} 
Babel, J., \& Montmerle, T.\ 1997, ApJ, 485, L29
\bibitem[Baumgardt et al.(2000)]{Baumgardt}
Baumgardt, H., Dettbarn, C., \& Wielen, R.\ 2000, A\&AS, 146, 251
\bibitem[Bergh\"ofer et al.(1997)]{Berghoefer} 
Bergh\"ofer, T.W., Schmitt, J.H.M.M., Danner, R., \& Cassinelli, J.P.\ 1997, A\&A, 322, 167
\bibitem[Bieging et al.(1989)]{Bieging}
Bieging, J.H., Abbott, D.C., \& Churchwell, E.B.\ 1989, ApJ, 340, 518
\bibitem[Bohlin et al.(1978)]{Bohlin} 
Bohlin, R.C., Savage, B.D., \& Drake, J.F.\ 1978, ApJ, 224, 132
\bibitem[Bouret et al.(2012)]{Bouret}
Bouret, J.-C., Hillier, D.J., Lanz, T., \& Fullerton, A.W.\ 2012, A\&A, 544, A67
\bibitem[Broos et al.(2013)]{bro13} 
Broos, P.S., Getman, K.V., Povich, M.S., et al.\ 2013, ApJS, 209, 32 
\bibitem[Conti(1973)]{Conti} 
Conti, P.S.\ 1973, ApJ, 179, 181
\bibitem[Conti et al.(1995)]{Conti95}
Conti, P.S., Hanson, M.M., Morris, P.W., Willis, A.J., \& Fossey, S.J.\ 1995, ApJ, 445, L35
\bibitem[Czesla \& Schmitt(2007)]{CS}
Czesla, S., \& Schmitt, J.H.M.M.\ 2007, A\&A, 465, 493
\bibitem[De Becker(2013)]{DeBecker13}
De Becker, M.\ 2013, New Astronomy, 25, 7
\bibitem[De Becker et al.(2006)]{DeBecker06}
De Becker, M., Rauw, G., Manfroid, J., \& Eenens, P.\ 2006, A\&A, 456, 1121
\bibitem[De Becker et al.(2009)]{DeBecker09}
De Becker, M., Rauw, G., \& Linder, N.\ 2009, ApJ, 704, 964
\bibitem[Duch\^ene(2015)]{Duchene}
Duch\^ene, G.\ 2015, Ap\&SS, 355, 291
\bibitem[Feigelson et al.(2013)]{Feigelson}
Feigelson, E.D., Townsley, L.K., Broos, P.S., et al.\ 2013, ApJS, 209, 26
\bibitem[Feldmeier et al.(1997)]{Feldmeier} 
Feldmeier, A., Puls, J., \& Pauldrach, A.W.A.\ 1997, A\&A, 322, 878
\bibitem[Fossati et al.(2015)]{fos15} 
Fossati, L., Castro, N., Sch\"oller, M., et al.\ 2015, A\&A, 582, A45
\bibitem[Garmany \& Massey(1981)]{GM}
Garmany, C.D., \& Massey, P.\ 1981, PASP, 93, 500
\bibitem[Gayley(2016)]{Gayley}
Gayley, K.G.\ 2016, AdSpR, 58, 719 
\bibitem[Gonz\'alez \& Levato(2006)]{GL} 
Gonz\'alez, J.F., \& Levato, H.\ 2006, A\&A, 448, 283
\bibitem[Gosset et al.(2001)]{Gosset} 
Gosset, E., Royer, P., Rauw, G., Manfroid, J., \& Vreux, J.-M.\ 2001, MNRAS, 327, 435
\bibitem[Gosset et al.(2005)]{Gosset2} 
Gosset, E., Naz\'e, Y., Claeskens, J.-F., et al.\ 2005, A\&A, 492, 685 
\bibitem[Hamann \& Gr\"afener(2004)]{PoWR}
Hamann, W.-R., \& Gr\"afener, G.\ 2004, A\&A, 427, 697
\bibitem[Harnden et al.(1979)]{Harnden} 
Harnden, F.,R.,Jr., Branduardi, G., Gorenstein, P., et al.\ 1979, ApJ, 234, L51
\bibitem[Heck et al.(1985)]{HMM} 
Heck, A., Manfroid, J., \& Mersch, G.\ 1985, A\&AS, 59, 63
\bibitem[Hillier \& Miller(1998)]{CMFGEN}
Hillier, D.J., \& Miller, D.L.\ 1998, ApJ, 496, 407
\bibitem[Hillier et al.(1993)]{Hillier}
Hillier, D.J., Kudritzki, R.P., Pauldrach, A.W., et al.\ 1993, A\&A, 276, 117 
\bibitem[Hillwig et al.(2006)]{Hillwig} 
Hillwig, T.C., Gies, D.R., Bagnuolo, W.G., et al.\ 2006, ApJ, 639, 1069
\bibitem[Hinkle et al.(2000)]{Hinkle}
Hinkle, K., Wallace, L., Valenti, J., \& Harmer, D.\ 2000, Visible and Near Infrared Atlas of the Arcturus Spectrum 3727-9300\,\AA, eds.\ K.\ Hinkle, L.\ Wallace, J.\ Valenti, \& D.\ Harmer, San Francisco: ASP 
\bibitem[Huang \& Gies(2006)]{HG} 
Huang, W., \& Gies, D.R.\ 2006, ApJ, 648, 580
\bibitem[Ignace et al.(2000)]{Ignace}
Ignace, R., Oskinova, L.M., \& Foullon, C.\ 2000, MNRAS, 318, 214
\bibitem[Ishida(1970)]{Ishida}
Ishida, K.\ 1970, PASP, 22, 277
\bibitem[Jansen et al.(2001)]{Jansen} 
Jansen, F., Lumb, D., Altieri, B., et al.\ 2001, A\&A, 365, L1
\bibitem[Kudritzki et al.(1996)]{Kudritzki}
Kudritzki, R.P., Palsa, R., Feldmeier, A., Puls, J., \& Pauldrach, A.W.\ 1996, in {\it R\"ontgenstrahlung from the Universe}, eds.\ H.U.\ Zimmermann, J.\ Tr\"umper, \& H.\ Yorke, MPE Rep.\ 263, 9
\bibitem[Lindgren(1976)]{lin76}
Lindgren, B.W.\ 1976, Statistical Theory - Third Edition, Mc Millan Pub.\ (New York)
\bibitem[Ma\'{\i}z Apell\'aniz(2010)]{MA}
Ma\'{\i}z Apell\'aniz, J.\ 2010, A\&A, 518, A1
\bibitem[Martins \& Plez(2006)]{MP} 
Martins, F., \& Plez, B.\ 2006, A\&A, 457, 637
\bibitem[Martins et al.(2005)]{Martins05}
Martins, F., Schaerer, D., Hillier, D.J., et al.\ 2005, A\&A, 441, 735
\bibitem[Martins et al.(2015)]{Mimes}
Martins, F., Herv\'e, A., Bouret, J.-C., et al.\ 2015, A\&A, 575, A34
\bibitem[Massey et al.(1995)]{Massey}
Massey, P., Johnson, K.E., \& Degioia-Eastwood, K.\ 1995, ApJ, 454, 151
\bibitem[Megeath et al.(2008)]{Megeath}
Megeath, S.T., Townsley, L.K., Oey, M.S., \& Tieftrunk, A.,R.\ 2008, in {\it Handbook of Star Forming Regions Vol. I}, ed.\ Bo Reipurth, Astronomical Society of the Pacific, 264
\bibitem[Mernier \& Rauw(2013)]{M17}
Mernier, F., \& Rauw, G.\ 2013, New Astronomy, 20, 42
\bibitem[Mittag(2010)]{Mittag}
Mittag, M., Hempelmann, A., Gonz\'alez-P\'erez, J.N., \& Schmitt, J.H.M.M.\ 2010, AdAst, 101502
\bibitem[Muijres et al.(2012)]{Muijres}
Muijres, L.E., Vink, J.S., de Koter, A., M\"uller, P.E., \& Langer, N.\ 2012, A\&A, 537, A37
\bibitem[Naz\'e(2009)]{YN} 
Naz\'e, Y.\ 2009, A\&A, 506, 1055
\bibitem[Naz\'e et al.(2004)]{HD108} 
Naz\'e, Y., Rauw, G., Vreux, J.-M., \& De Becker, M.\ 2004, A\&A, 417, 667
\bibitem[Naz\'e et al.(2011)]{Carina} 
Naz\'e, Y., Broos, P.S., Oskinova, L., et al.\ 2011, ApJS, 194, 7 
\bibitem[Naz\'e et al.(2013a)]{HM1} 
Naz\'e, Y., Rauw, G., Sana, H., \& Corcoran, M.F.\ 2013a, A\&A, 555, A83
\bibitem[Naz\'e et al.(2013b)]{zetaPup2} 
Naz\'e, Y., Oskinova, L.M., \& Gosset, E.\ 2013b, ApJ, 763, 143 
\bibitem[Naz\'e et al.(2014)]{YNBfield} 
Naz\'e, Y., Petit, V., Rinbrand, M., et al.\ 2014, ApJS, 215, 10
\bibitem[Negueruela et al.(2004)]{Negueruela}
Negueruela, I., Steele, I.A., \& Bernabeu, G.\ 2004, AN, 325, 749
\bibitem[Oskinova(2005)]{Oskinova05}
Oskinova, L.M.\ 2005, MNRAS, 361, 679
\bibitem[Oskinova(2015)]{OskinovaWR} 
Oskinova, L.M.\ 2015, in {\it Wolf-Rayet Stars}, eds.\ W.-R.\ Hamann, A.\ Sander, H.\ Todt, 295 
\bibitem[Oskinova et al.(2003)]{Oskinova} 
Oskinova, L.M., Ignace, R., Hamann, W.-R., Pollock, A.M.T., \& Brown, J.C.\ 2003, A\&A, 402, 755
\bibitem[Owocki et al.(2013)]{Owocki} 
Owocki, S.P., Sundqvist, J.O., Cohen, D.H., \& Gayley, K.G.\ 2013, MNRAS, 429, 3379
\bibitem[Peri et al.(2012)]{Peri}
Peri, C.S., Benaglia, P., Brookes, D.P., Stevens, I.R., \& Isequilla, N.L.\ 2012, A\&A, 538, A108
\bibitem[Pittard \& Parkin(2010)]{PP} 
Pittard, J.M., \& Parkin, E.R.\ 2010, MNRAS, 403, 1657
\bibitem[Polcaro et al.(2003)]{Polcaro}
Polcaro, V.F., Viotti, R., Norci, L., et al.\ 2003, in {\it International Conference on magnetic fields in O, B and A stars}, eds.\ L.A.\ Balona, H.F.\ Henrichs, \& T.\ Medupe, ASP Conf. Series, 305, 377
\bibitem[Puls et al.(2005)]{Puls}
Puls, J., Urbaneja, M.A., Venero, R., et al.\ 2005, A\&A, 435, 669
\bibitem[Rauw(2011)]{CygOB2GR}
Rauw, G.\ 2011, A\&A, 536, A31
\bibitem[Rauw \& De Becker(2004)]{RDB} 
Rauw, G., \& De Becker, M.\ 2004, A\&A, 421, 693
\bibitem[Rauw \& Naz\'e(2016)]{GRYN} 
Rauw, G., \& Naz\'e, Y.\ 2016, AdSpR, 58, 761
\bibitem[Rauw et al.(2015)]{CygOB2} 
Rauw, G., Naz\'e, Y., Wright, N.J., et al.\ 2015, ApJS, 221, 1
\bibitem[Reed(2005)]{Reed}
Reed, B.C.\ 2005, AJ, 130, 1652
\bibitem[Repolust et al.(2004)]{Repolust}
Repolust, T., Puls, J., \& Herrero, A.\ 2004, A\&A, 415, 349
\bibitem[Robrade(2016)]{Robrade} 
Robrade, J.\ 2016, AdSpR, 58, 727
\bibitem[Sana et al.(2006a)]{SGR} 
Sana, H., Gosset, E., \& Rauw, G.\ 2006a, MNRAS, 371, 67
\bibitem[Sana et al.(2006b)]{NGC6231} 
Sana, H., Rauw, G., Naz\'e, Y., Gosset, E., \& Vreux, J.-M.\ 2006b, MNRAS, 372, 661 
\bibitem[Sana et al.(2007)]{NGC6231PMS} 
Sana, H., Rauw, G., Sung, H., Gosset, E., \& Vreux, J.-M.\ 2007, MNRAS, 377, 945
\bibitem[Sanders(1972)]{Sanders}
Sanders, W.L.\ 1972, A\&A, 16, 58
\bibitem[Schmitt et al.(2014)]{Schmitt}
Schmitt, J.H.M.M., Schr\"oder, K.-P., Rauw, G., et al.\ 2014, AN 335, 787
\bibitem[Schr\"oder \& Schmitt(2007)]{Astars}
Schr\"oder, C., \& Schmitt, J.H.M.M.\ 2007, A\&A, 475, 677
\bibitem[Sciortino et al.(1990)]{Sciortino} 
Sciortino, S., Vaiana, G.S., Harnden, F.R.Jr., et al.\ 1990, ApJ, 361, 621
\bibitem[Shi \& Hu(1999)]{SH99}
Shi, H.M., \& Hu, J.Y.\ 1999, A\&AS, 136, 313
\bibitem[Skrutzkie et al.(2006)]{2MASS}
Skrutskie, M.F., Cutri, R.M., Stiening, R., et al.\ 2006, AJ, 131, 1163
\bibitem[Smith \& Brickhouse(2001)]{apec} 
Smith, R.K., \& Brickhouse, N.S.\ 2001, ApJ, 556, L91
\bibitem[Sota et al.(2011)]{Sota}
Sota, A., Ma\'{\i}z Apell\'aniz, J., Walborn, N.R., et al.\ 2011, ApJS, 193, 24
\bibitem[Stelzer et al.(2009)]{Stelzer}
Stelzer, B., Robrade, J., Schmitt, J.H.M.M., \& Bouvier, J.\ 2009, A\&A, 493, 1109
\bibitem[Stevens et al.(1992)]{SBP} 
Stevens, I.R., Blondin, J.M., \& Pollock, A.M.T.\ 1992, ApJ, 386, 265
\bibitem[Strai\v{z}ys et al.(2013)]{Vilnius}
Strai\v{z}ys, V., Boyle, R.P., Jnusz, R., Laugalys, V., \& Kazlauskas, A.\ 2013, A\&A, 554, A3
\bibitem[Str\"uder et al.(2001)]{pn} 
Str\"uder, L., Briel, U., Dennerl, K., et al.\ 2001, A\&A, 365, L18
\bibitem[Sung \& Lee(1995)]{SL}
Sung, H., \& Lee, S.-W.\ 1995, Journal of the Korean Astronomical Society, 28, 119 
\bibitem[\v{S}urlan et al.(2013)]{Surlan}
\v{S}urlan, B., Hamann, W.-R., Aret, A., et al.\ 2013, A\&A, 559, A130
\bibitem[Townsley et al.(2014)]{Townsley}
Townsley, L.K., Broos, P.S., Garmire, G.P., et al.\ 2014, ApJS, 213, 1
\bibitem[Turner et al.(2001)]{MOS} 
Turner, M.J.L., Abbey, A., Arnaud, M., et al.\ 2001, A\&A, 365, L27
\bibitem[ud-Doula \& Owocki(2002)]{ud-Doula} 
ud-Doula, A., \& Owocki, S.P.\ 2002, ApJ, 576, 413
\bibitem[ud-Doula \& Naz\'e(2016)]{ud-DoulaYN} 
ud-Doula, A., \& Naz\'e, Y.\ 2016, AdSpR, 58, 680 
\bibitem[Underhill(1967)]{Underhill} 
Underhill, A.B.\ 1967, in {\it Determination of Radial Velocities and their Applications}, IAU Symp., 30, 167
\bibitem[Vasilevskis et al.(1965)]{Vasilevskis} 
Vasilevskis, S., Sanders, W.L., \& Van Altena, W.F.\ 1965, AJ, 70, 806
\bibitem[Vink et al.(2001)]{Vink}
Vink, J.S., de Koter, A., \& Lamers, H.J.G.L.M.\ 2001, A\&A, 369, 574
\bibitem[Wade et al.(2014)]{wad13} 
Wade, G.~A., Grunhut, J., Alecian, E., et al.\ 2014, in {\it Magnetic Fields throughout Stellar Evolution}, IAU Symp., 302, 265  
\bibitem[Walborn(2001)]{Walborn}
Walborn, N.R.\ 2001, in {\it Eta Carinae and Other Mysterious Stars: The Hidden Opportunities of Emission Spectroscopy}, eds.\ T.R.\ Gull, S.\ Johannson, \& K.\ Davidson, ASP Conference Series, 242, 217
\bibitem[Wessolowski(1996)]{Wessolowski} 
Wessolowski, U.\ 1996, in {\it R\"ontgenstrahlung from the Universe}, eds.\ H.U.\ Zimmermann, J.\ Tr\"umper, \& H.\ Yorke, MPE Rep., 263, 75
\bibitem[Willis \& Stickland(1980)]{Willis} 
Willis, A.J., \& Stickland, D.J.\ 1980, MNRAS, 190, 27
\bibitem[Wilms et al.(2000)]{Wilms}
Wilms, J., Allen, A., \& McCray, R.\ 2000, ApJ, 542, 914
\bibitem[Wolfe et al.(1967)]{WHS} 
Wolfe, R.H.Jr., Horak, H.G., \& Storer, N.W.\ 1965, in {\it Modern Astrophysics. A Memorial to Otto Struve}, ed.\ M.\ Hack, (New York: Gordon \& Breach), 251
\bibitem[Wolff et al.(2011)]{Wolff}
Wolff, S.C., Strom, S.E., \& Rebull, L.M.\ 2011, ApJ, 726, 19
\bibitem[Sung et al.(in prep.)]{sun16}
\end{thebibliography}
\end{document}